\input harvmac
\input amssym



\overfullrule=0pt

\font\manual=manfnt \def\dbend{\lower3.5pt\hbox{\manual\char127}}

\def\ie{{\it i.e.}}
\def\eg{{\it e.g.}}
\def\cf{{\it c.f.}}

\def\etc{{\it etc}}

\def\sst{\scriptscriptstyle}

\def\frac#1#2{{#1\over#2}}
\def\coeff#1#2{{\textstyle{#1\over #2}}}
\def\half{\frac12}
\def\hf{{\textstyle\half}}
\def\d{\partial}
\def\p{\partial}

\def\inbar{\,\vrule height1.5ex width.4pt depth0pt}
\def\IR{\relax{\rm I\kern-.18em R}}
\def\IC{\relax\hbox{$\inbar\kern-.3em{\rm C}$}}
\def\IQ{\relax\hbox{$\inbar\kern-.3em{\rm Q}$}}
\def\IH{\relax{\rm I\kern-.18em H}}
\def\IN{\relax{\rm I\kern-.18em N}}
\def\IP{\relax{\rm I\kern-.18em P}}
\font\cmss=cmss10
\font\cmsss=cmss10 at 7pt
\def\IZ{\relax\ifmmode\mathchoice
{\hbox{\cmss Z\kern-.4em Z}}{\hbox{\cmss Z\kern-.4em Z}}
{\lower.9pt\hbox{\cmsss Z\kern-.4em Z}}
{\lower1.2pt\hbox{\cmsss Z\kern-.4em Z}}\else{\cmss Z\kern-.4em
Z}\fi}
\def\Z{{\IZ}}

\catcode`\@=11
\def\slash#1{\mathord{\mathpalette\c@ncel{#1}}}
\def\underrel#1\over#2{\mathrel{\mathop{\kern\z@#1}\limits_{#2}}}

\def\lessapprox{{\buildrel{<}\over{\scriptstyle\sim}}}
\catcode`\@=12
%

%
%
\def\ket#1{|#1\rangle}

\def\vev#1{\langle#1\rangle}
\def\det{{\rm det}}

\def\Tr{{\rm Tr}}
\def\mod{{\rm mod}}
\def\sinh{{\rm sinh}}   
\def\cosh{{\rm cosh}}   
\def\tanh{{\rm tanh}}

\def\det{{\rm det}}

%
%
 \def\CA{{\cal A}}
 
\def\CC{{\cal C}} 

 \def\CG{{\cal G}}
 \def\CH{{\cal H}}
\def\II{{\cal I}} \def\CI{{\cal I}}

\def\MM{{\cal M}} \def\CM{{\cal M}}
 \def\CN{{\cal N}}
 \def\CO{{\cal O}}

 \def\CR{{\cal R}}
 \def\CS{{\cal S}}
 \def\CT{{\cal T}}

 \def\CW{{\cal W}}

%
%
\def\unlockat{\catcode`\@=11}
\def\lockat{\catcode`\@=12}
\unlockat
\def\newsec#1{\global\advance\secno by1\message{(\the\secno. #1)}
\global\subsecno=0\global\subsubsecno=0\eqnres@t\noindent
{\bf\the\secno. #1}
\writetoca{{\secsym} {#1}}\par\nobreak\medskip\nobreak}
\global\newcount\subsecno \global\subsecno=0
\def\subsec#1{\global\advance\subsecno
by1\message{(\secsym\the\subsecno. #1)}
\ifnum\lastpenalty>9000\else\bigbreak\fi\global\subsubsecno=0
\noindent{\it\secsym\the\subsecno. #1}
\writetoca{\string\quad {\secsym\the\subsecno.} {#1}}
\par\nobreak\medskip\nobreak}
\global\newcount\subsubsecno \global\subsubsecno=0
\def\subsubsec#1{\global\advance\subsubsecno by1
\message{(\secsym\the\subsecno.\the\subsubsecno. #1)}
\ifnum\lastpenalty>9000\else\bigbreak\fi
\noindent\quad{\secsym\the\subsecno.\the\subsubsecno.}{#1}
\writetoca{\string\qquad{\secsym\the\subsecno.\the\subsubsecno.}{#1}}
\par\nobreak\medskip\nobreak}
\def\subsubseclab#1{\DefWarn#1\xdef
#1{\noexpand\hyperref{}{subsubsection}%
{\secsym\the\subsecno.\the\subsubsecno}%
{\secsym\the\subsecno.\the\subsubsecno}}%
\writedef{#1\leftbracket#1}\wrlabeL{#1=#1}}
\lockat
%
\def\sym{{\sl Sym}}
\def\cag{{\CA_\gamma}}
\def\dto{ D(2,1\vert \alpha) }
\def\sd{{\bf s}}
\def\ld{{\bf l}}

\def\ppd{{+\dot +}}
\def\mmd{{-\dot -}}
\def\pmd{{+\dot -}}
\def\mpd{{-\dot +}}
\def\pmpmd{{\pm\dot \pm}}

\def\pmmpd{{\pm\dot \mp}}
\def\mppmd{{\mp\dot \pm}}
\def\mdmd{\dot - \dot -}
\def\mm{--}
\def\shf{{\sst-\frac12}}
\def\bar{\overline}



\def\CS{{\cal S}}

\def\SCh{{\rm SCh}}
\def\tl{\tilde }

\def\cag{{\CA_{\gamma}} }
\def\IH{{\bf H}}
\def\SCh{{\rm SCh} }

\def\syms{ {\rm Sym}^N(\CS)  }
\def\dto{ D(2,1\vert \alpha) }
\def\ap{A^{(+)} }
\def\am{A^{(-)} }

\def\imt{{\rm Im}\tau}


\noblackbox
\overfullrule=0pt
\def\Title#1#2{\rightline{#1}\ifx\answ\bigans\nopagenumbers\pageno0\vskip1in
\else\pageno1\vskip.8in\fi \centerline{\titlefont #2}\vskip .5in}

\font\cmss=cmss10 \font\cmsss=cmss10 at 7pt
%

\let\includefigures=\iftrue
\newfam\black
\includefigures
\input epsf
\def\figin{\epsfcheck\figin}\def\figins{\epsfcheck\figins}
\def\epsfcheck{\ifx\epsfbox\UnDeFiNeD
\message{(NO epsf.tex, FIGURES WILL BE IGNORED)}
\gdef\figin##1{\vskip2in}\gdef\figins##1{\hskip.5in}
\else\message{(FIGURES WILL BE INCLUDED)}%
\gdef\figin##1{##1}\gdef\figins##1{##1}\fi}
\def\DefWarn#1{}
\def\figinsert{\goodbreak\midinsert}
\def\ifig#1#2#3{\DefWarn#1\xdef#1{fig.~\the\figno}
\writedef{#1\leftbracket fig.\noexpand~\the\figno}%
\figinsert\figin{\centerline{#3}}\medskip\centerline{\vbox{\baselineskip12pt
\advance\hsize by -1truein\noindent\footnotefont{\bf Fig.~\the\figno:} #2}}
\bigskip\endinsert\global\advance\figno by1}
\else
\def\ifig#1#2#3{\xdef#1{fig.~\the\figno}
\writedef{#1\leftbracket fig.\noexpand~\the\figno}%
\global\advance\figno by1}
\fi

\font\cmss=cmss10 \font\cmsss=cmss10 at 7pt

\def\IB{\relax\hbox{$\inbar\kern-.3em{\rm B}$}}
\def\IC{\relax\hbox{$\inbar\kern-.3em{\rm C}$}}
\def\IQ{\relax\hbox{$\inbar\kern-.3em{\rm Q}$}}
\def\ID{\relax\hbox{$\inbar\kern-.3em{\rm D}$}}
\def\IE{\relax\hbox{$\inbar\kern-.3em{\rm E}$}}
\def\IF{\relax\hbox{$\inbar\kern-.3em{\rm F}$}}
\def\IG{\relax\hbox{$\inbar\kern-.3em{\rm G}$}}
\def\IGa{\relax\hbox{${\rm I}\kern-.18em\Gamma$}}
\def\IH{\relax{\rm I\kern-.18em H}}
\def\IK{\relax{\rm I\kern-.18em K}}
\def\IL{\relax{\rm I\kern-.18em L}}
\def\IP{\relax{\rm I\kern-.18em P}}
\def\IR{\relax{\rm I\kern-.18em R}}
\def\Z{\relax\ifmmode\mathchoice
{\hbox{\cmss Z\kern-.4em Z}}{\hbox{\cmss Z\kern-.4em Z}}
{\lower.9pt\hbox{\cmsss Z\kern-.4em Z}}
{\lower1.2pt\hbox{\cmsss Z\kern-.4em Z}}\else{\cmss Z\kern-.4em
Z}\fi}
\def\II{\relax{\rm I\kern-.18em I}}

\def\S{{\bf S}}

\def\R{\IR}

\def\hf{{1\over 2}}

\def\CA {{\cal A}}

\def\CC {{\cal C}}

\def\CG {{\cal G}}
\def\CH {{\cal H}}
\def\CI {{\cal I}}

\def\CM {{\cal M}}
\def\CN {{\cal N}}
\def\CO {{\cal O}}

\def\CR {{\cal R}}
\def\CS {{\cal S}}
\def\CT {{\cal T}}

\def\CW {{\cal W}}


\def\p{\partial}


\def\zb {\bar{z}}


\def\Tr{{\rm Tr}}

\def\vol{{\rm vol}}

\def\p{\partial}

\def\inbar{\,\vrule height1.5ex width.4pt depth0pt}

\def\l{{\ell}}

\def\a{\alpha}

\def\g{\gamma}
\def\d{\delta}

\def\la{\lambda}
\def\th{\theta}
\def\s{\sigma}

\def\bar{\overline}

\def\example#1{\bgroup\narrower\footnotefont\baselineskip\footskip\bigbreak
\hrule\medskip\nobreak\noindent {\bf Example}. {\it #1\/}\par\nobreak}
\def\endexample{\medskip\nobreak\hrule\bigbreak\egroup}


\lref\ManinHN{
Y.~I.~Manin and M.~Marcolli,
``Holography principle and arithmetic of algebraic curves,''
Adv.\ Theor.\ Math.\ Phys.\  {\bf 5}, 617 (2002)
[arXiv:hep-th/0201036].}

\lref\STVP{A.~Sevrin, W.~Troost and A.~Van Proeyen,
``Superconformal Algebras In Two-Dimensions With N=4,''
Phys.\ Lett.\ B {\bf 208}, 447 (1988).}

\lref\SpindelSR{
P.~Spindel, A.~Sevrin, W.~Troost and A.~Van Proeyen,
``Extended Supersymmetric Sigma Models On Group Manifolds. 1. The Complex
Structures,''
Nucl.\ Phys.\ B {\bf 308}, 662 (1988);
%
A.~Sevrin, W.~Troost, A.~Van Proeyen and P.~Spindel,
``Extended Supersymmetric Sigma Models On Group Manifolds. 2. Current
Algebras,''
Nucl.\ Phys.\ B {\bf 311}, 465 (1988);
%
A.~Van Proeyen,
``Realizations Of N=4 Superconformal Algebras On Wolf Spaces,''
Class.\ Quant.\ Grav.\  {\bf 6}, 1501 (1989).
}

\lref\Schoutens{K.~Schoutens,
``O(N) Extended Superconformal Field Theory In Superspace,''
Nucl.\ Phys.\ B {\bf 295}, 634 (1988).}

\lref\Srep{K.~Schoutens,
``A Nonlinear Representation Of The D = 2 SO(4) Extended
Superconformal Algebra,'' Phys.\ Lett.\ B {\bf 194}, 75 (1987).}

\lref\SScomments{A.~Schwimmer and N.~Seiberg,
``Comments On The N=2, N=3, N=4 Superconformal Algebras In Two-Dimensions,''
Phys.\ Lett.\ B {\bf 184}, 191 (1987).}

\lref\DST{F.~Defever, S.~Schrans and K.~Thielmans,
``Moding Of Superconformal Algebras,''
Phys.\ Lett.\ B {\bf 212}, 467 (1988).}

\lref\KS{A.~Klemm and M.~G.~Schmidt,
``Orbifolds By Cyclic Permutations Of Tensor Product Conformal
Field Theories,'' Phys.\ Lett.\ B {\bf 245}, 53 (1990);
J.~Fuchs, A.~Klemm and M.~G.~Schmidt,
``Orbifolds By Cyclic Permutations In Gepner Type Superstrings
And In The Corresponding Calabi-Yau Manifolds,''
Annals Phys.\  {\bf 214}, 221 (1992).}

\lref\ElitzurMM{S.~Elitzur, O.~Feinerman, A.~Giveon and D.~Tsabar,
``String theory on AdS(3) x S(3) x S(3) x S(1),''
Phys.\ Lett.\ B {\bf 449}, 180 (1999); hep-th/9811245.}

\lref\deBoerRH{J.~de Boer, A.~Pasquinucci and K.~Skenderis,
``AdS/CFT dualities involving large 2d N = 4 superconformal symmetry,''
Adv.\ Theor.\ Math.\ Phys.\  {\bf 3}, 577 (1999); hep-th/9904073.}

\lref\GPTVP{M.~Gunaydin, J.~L.~Petersen, A.~Taormina and A.~Van Proeyen,
``On The Unitary Representations Of A Class Of N=4 Superconformal Algebras,''
Nucl.\ Phys.\ B {\bf 322}, 402 (1989).}

\lref\PT{J.~L.~Petersen and A.~Taormina,
``Characters Of The N=4 Superconformal Algebra With Two Central Extensions,''
Nucl.\ Phys.\ B {\bf 331}, 556 (1990).}
%

\lref\BMN{D.~Berenstein, J.~M.~Maldacena and H.~Nastase,
``Strings in flat space and pp waves from N = 4 super Yang Mills,''
JHEP {\bf 0204}, 013 (2002); hep-th/0202021.}

\lref\AGS{R.~Argurio, A.~Giveon and A.~Shomer,
``Superstrings on AdS(3) and symmetric products,''
JHEP {\bf 0012}, 003 (2000), hep-th/0009242;
``String theory on AdS(3) and symmetric products,''
Fortsch.\ Phys.\  {\bf 49}, 409 (2001), hep-th/0012117.}

\lref\LM{O.~Lunin and S.~D.~Mathur,
``Three-point functions for M(N)/S(N) orbifolds with N = 4 supersymmetry,''
Commun.\ Math.\ Phys.\  {\bf 227}, 385 (2002); hep-th/0103169.}

\lref\Gomis{J.~Gomis, L.~Motl and A.~Strominger,
``pp-wave / CFT(2) duality,'' JHEP {\bf 0211}, 016 (2002)
[arXiv:hep-th/0206166].}

\lref\brown{ J.~D.~Brown and M.~Henneaux,
 ``Central Charges In The Canonical Realization Of Asymptotic Symmetries: An
Example From Three-Dimensional Gravity,'' Commun.\ Math.\ Phys.\
{\bf 104}, 207 (1986).
}

\lref\preskill{ J.~Preskill, P.~Schwarz, A.~D.~Shapere, S.~Trivedi
and F.~Wilczek, ``Limitations on the statistical description of
black holes,'' Mod.\ Phys.\ Lett.\ A {\bf 6}, 2353 (1991).
}

\lref\Lunin{O.~Lunin and S.~D.~Mathur,
``Correlation Functions For M(N)/S(N) Orbifolds,''
Int.\ J.\ Mod.\ Phys.\ A {\bf 16S1C}, 967 (2001).}

\lref\BVW{N.~Berkovits, C.~Vafa and E.~Witten,
``Conformal field theory of AdS background with Ramond-Ramond flux,''
JHEP {\bf 9903}, 018 (1999), hep-th/9902098.}

\lref\LP{M.~Luscher and K.~Pohlmeyer,
``Scattering Of Massless Lumps And Nonlocal Charges In
The Two-Dimensional Classical Nonlinear Sigma Model,''
Nucl.\ Phys.\ B {\bf 137}, 46 (1978).}

\lref\Luscher{M.~Luscher,
``Quantum Nonlocal Charges And Absence Of Particle Production
In The Two-Dimensional Nonlinear Sigma Model,''
Nucl.\ Phys.\ B {\bf 135}, 1 (1978).}

\lref\BPR{I.~Bena, J.~Polchinski and R.~Roiban,
``Hidden symmetries of the AdS(5) x S**5 superstring,'' hep-th/0305116.}

\lref\DNW{L.~Dolan, C.~R.~Nappi and E.~Witten,
``A relation between approaches to integrability in superconformal
Yang-Mills theory,'' hep-th/0308089.}

\lref\Bowcock{P.~Bowcock,
``Exceptional superconformal algebras,''
Nucl.\ Phys.\ B {\bf 381}, 415 (1992), hep-th/9202061.}

\lref\FL{E.~S.~Fradkin and V.~Y.~Linetsky,
``Results of the classification of superconformal algebras in
two-dimensions,'' Phys.\ Lett.\ B {\bf 282}, 352 (1992), hep-th/9203045.}

\lref\DKSS{S.~Deger, A.~Kaya, E.~Sezgin and P.~Sundell,
``Spectrum of D = 6, N = 4b supergravity on AdS(3) x S(3),''
Nucl.\ Phys.\ B {\bf 536}, 110 (1998), hep-th/9804166.}

\lref\LPS{H.~Lu, C.~N.~Pope and E.~Sezgin,
``SU(2) reduction of six-dimensional (1,0) supergravity,''
hep-th/0212323.}

\lref\LPSb{H.~Lu, C.~N.~Pope and E.~Sezgin,
``Yang-Mills-Chern-Simons supergravity,'' hep-th/0305242.}

\lref\APT{G.~Arutyunov, A.~Pankiewicz and S.~Theisen,
``Cubic couplings in D = 6 N = 4b supergravity on AdS(3) x S(3),''
Phys.\ Rev.\ D {\bf 63}, 044024 (2001), hep-th/0007061.}

\lref\Mathur{S.~D.~Mathur,
``Gravity on AdS(3) and flat connections in the boundary CFT,''
hep-th/0101118.}

\lref\fragm{J.~M.~Maldacena, J.~Michelson and A.~Strominger,
``Anti-de Sitter fragmentation,'' JHEP {\bf 9902}, 011 (1999);
hep-th/9812073.}

\lref\SW{N.~Seiberg and E.~Witten,
``The D1-D5 system and singular CFT,'' JHEP {\bf 9904}, 017 (1999);
hep-th/9903224.}

\lref\WittenDS{
E.~Witten,
``Supersymmetric index of three-dimensional gauge theory,''
arXiv:hep-th/9903005.
}

\lref\LarsenDH{
F.~Larsen and E.~J.~Martinec,
``Currents and moduli in the (4,0) theory,''
JHEP {\bf 9911}, 002 (1999)
[arXiv:hep-th/9909088].
}

\lref\AV{B.~S.~Acharya and C.~Vafa,
``On domain walls of N = 1 supersymmetric Yang-Mills in four dimensions,''
hep-th/0103011.}

\lref\Mnorms{Greg's notes {\it Mnorms.tex}}

\lref\PS{J.~Polchinski, M.J.~Strassler,
``The String Dual of a Confining Four-Dimensional Gauge Theory,''
hep-th/0003136.}

\lref\PolchinskiRQ{
J.~Polchinski,
``String Theory. Vols. 1\&2,''
Cambridge Univ. Press, 1998.
}

\lref\Drinfeld{V.G.~Drinfeld,
``Hopf algebras and the quantum Yang-Baxter equation,''
Soviet Math. Dokl. {\bf 32} (1985) 254;
``Quantum Groups,'' Proceedings Int. Cong. Math. Berkeley (1986) 798;
``A new realization of Yangians and quantum affine algebras,''
Soviet Math. Dokl. {\bf 36} (1988) 212.}

\lref\FRT{L.D.~Faddeev, N.Yu.~Reshetikhin, L.A.~Takhtajan,
``Quantization of Lie groups and Lie algebras,''
Leningrad Math. J. {\bf 1} (1990) 193.}

\lref\AACFR{D.~Arnaudon, J.~Avan, N.~Crampe, L.~Frappat and E.~Ragoucy,
``R-matrix presentation for (super)-Yangians Y(g),''
J. Math. Phys. {\bf 44} (2003) 302.}

\lref\IMP{K.~Ito, J.~O.~Madsen and J.~L.~Petersen,
``Free field representations and screening operators for
the N=4 doubly extended superconformal algebras,''
Phys.\ Lett.\ B {\bf 292}, 298 (1992); hep-th/9207010.}

\lref\IMadsen{K.~Ito and J.~O.~Madsen,
``Hamiltonian reduction and classical extended superconformal algebras,''
Phys.\ Lett.\ B {\bf 283}, 223 (1992); hep-th/9202058.}

\lref\Wadhm{E.~Witten,
``Sigma Models And The ADHM Construction Of Instantons,''
J. Geom. Phys. {\bf 15} (1995) 215.}

\lref\Dadhm{M.~Douglas, ``Gauge Fields and D-branes,''
J. Geom. Phys. {\bf 28} (1998) 255.}

\lref\GGPT{
J.~P.~Gauntlett, G.~W.~Gibbons, G.~Papadopoulos and P.~K.~Townsend,
``Hyper-Kaehler manifolds and multiply intersecting branes,''
Nucl.\ Phys.\ B {\bf 500} (1997) 133; [arXiv:hep-th/9702202].}

\lref\OT{N.~Ohta, P.K.~Townsend,
``Supersymmetry of M-Branes at Angles,'' Phys.Lett. {\bf B418} (1998) 77.}

\lref\PTgrassm{G.~Papadopoulos, A.~Teschendorff,
``Grassmannians,Calibrations and Five-Brane Intersections,''
Class.Quant.Grav. {\bf 17} (2000) 2641.}

\lref\NS{H.~Nicolai and H.~Samtleben,
``Kaluza-Klein supergravity on AdS(3) x S(3),'' hep-th/0306202.}

\lref\OPT{H.~Ooguri, J.~L.~Petersen and A.~Taormina,
``Modular Invariant Partition Functions for the Doubly Extended
N=4 Superconformal Algebras,'' Nucl.\ Phys.\ B {\bf 368}, 611 (1992).}

\lref\PTone{J.~L.~Petersen and A.~Taormina,
``Characters Of The N=4 Superconformal Algebra With Two Central Extensions,''
Nucl.\ Phys.\ B {\bf 331}, 556 (1990).}

\lref\PTtwo{J.~L.~Petersen and A.~Taormina,
``Characters Of The N=4 Superconformal Algebra With Two Central
Extensions: 2. Massless Representations,''
Nucl.\ Phys.\ B {\bf 333}, 833 (1990).}

\lref\HL{R.~Harvey and H.B.~Lawson, Jr., ``Calibrated geometries",
Acta Math. {\bf 148} (1982) 47.}

\lref\AV{B.~S.~Acharya and C.~Vafa,
``On domain walls of N = 1 supersymmetric Yang-Mills in four dimensions,''
hep-th/0103011.}

\lref\GST{S.~Gukov, J.~Sparks and D.~Tong,
``Conifold transitions and fivebrane condensation in M-theory
on Spin(7)  manifolds,'' Class.\ Quant.\ Grav.\  {\bf 20} (2003) 665.}

\lref\Lmoduli{N.~D.~Lambert, ``Moduli and brane intersections,''
Phys.\ Rev.\ D {\bf 67}, 026006 (2003) 026006.}

\lref\deBoerIP{
J.~de Boer,
``Six-dimensional supergravity on S**3 x AdS(3) and 2d conformal field
theory,''
Nucl.\ Phys.\ B {\bf 548}, 139 (1999)
[arXiv:hep-th/9806104];
F.~Larsen,
``The perturbation spectrum of black holes in N = 8 supergravity,''
Nucl.\ Phys.\ B {\bf 536}, 258 (1998)
[arXiv:hep-th/9805208].
}

\lref\ElitzurMM{
S.~Elitzur, O.~Feinerman, A.~Giveon and D.~Tsabar,
``String theory on AdS(3) x S(3) x S(3) x S(1),''
Phys.\ Lett.\ B {\bf 449}, 180 (1999), [arXiv:hep-th/9811245].}

\lref\MaldacenaHW{J.~M.~Maldacena and H.~Ooguri,
``Strings in AdS(3) and SL(2,R) WZW model. I,''
J.\ Math.\ Phys.\  {\bf 42}, 2929 (2001), [arXiv:hep-th/0001053].}

\lref\MartinecCF{E.~J.~Martinec and W.~McElgin,
``String theory on AdS orbifolds,''
JHEP {\bf 0204}, 029 (2002), [arXiv:hep-th/0106171].}

\lref\Sommov{L.~Sommovigo,
``Penrose limit of $AdS_3\times S^3\times S^3\times S^1$
and its associated  sigma-model,''
JHEP {\bf 0307} (2003) 035, hep-th/0305151.}

\lref\MaldacenaBW{
J.~M.~Maldacena and A.~Strominger,
``AdS(3) black holes and a stringy exclusion principle,''
JHEP {\bf 9812}, 005 (1998)
[arXiv:hep-th/9804085].
}

\lref\AspinwallEV{
P.~S.~Aspinwall,
``Resolution of orbifold singularities in string theory,''
arXiv:hep-th/9403123.
}

\lref\GiveonNS{
A.~Giveon, D.~Kutasov and N.~Seiberg,
``Comments on string theory on AdS(3),''
Adv.\ Theor.\ Math.\ Phys.\  {\bf 2}, 733 (1998)
[arXiv:hep-th/9806194].
}

\lref\AharonyDP{
O.~Aharony, M.~Berkooz and E.~Silverstein,
``Non-local string theories on $AdS_3 \times \S^3$
and stable  non-supersymmetric backgrounds,''
Phys.\ Rev.\ D {\bf 65}, 106007 (2002)
[arXiv:hep-th/0112178].
E.~Witten,
``Multi-trace operators, boundary conditions, and AdS/CFT correspondence,''
arXiv:hep-th/0112258.
M.~Berkooz, A.~Sever and A.~Shomer,
``Double-trace deformations, boundary conditions and spacetime
singularities,'' JHEP {\bf 0205}, 034 (2002)
[arXiv:hep-th/0112264].
}

\lref\AharonyPA{
O.~Aharony, M.~Berkooz and E.~Silverstein,
``Multiple-trace operators and non-local string theories,''
JHEP {\bf 0108}, 006 (2001)
[arXiv:hep-th/0105309].
}

\lref\DixonQV{
L.~J.~Dixon, D.~Friedan, E.~J.~Martinec and S.~H.~Shenker,
``The Conformal Field Theory Of Orbifolds,''
Nucl.\ Phys.\ B {\bf 282}, 13 (1987).
%
S.~Hamidi and C.~Vafa,
``Interactions On Orbifolds,''
Nucl.\ Phys.\ B {\bf 279}, 465 (1987).
%
M.~A.~Bershadsky and A.~O.~Radul,
``Conformal Field Theories With Additional Z(N) Symmetry,''
Sov.\ J.\ Nucl.\ Phys.\  {\bf 47}, 363 (1988)
[Yad.\ Fiz.\  {\bf 47}, 575 (1988)].
}

\lref\MartinecCF{
E.~J.~Martinec and W.~McElgin,
``String theory on AdS orbifolds,''
JHEP {\bf 0204}, 029 (2002)
[arXiv:hep-th/0106171].
}

\lref\MartinecXQ{
E.~J.~Martinec and W.~McElgin,
``Exciting AdS orbifolds,''
JHEP {\bf 0210}, 050 (2002)
[arXiv:hep-th/0206175].
}

\lref\WittenYU{
E.~Witten,
``On the conformal field theory of the Higgs branch,''
JHEP {\bf 9707}, 003 (1997)
[arXiv:hep-th/9707093].
}

\lref\DixonBG{
L.~J.~Dixon,
``Some World Sheet Properties Of Superstring Compactifications,
On Orbifolds And Otherwise,''
{\it Lectures given at the 1987 ICTP Summer Workshop
in High Energy Phsyics and Cosmology, Trieste, Italy, Jun 29 - Aug 7, 1987}
}

\lref\GreenQU{
M.~B.~Green and N.~Seiberg,
``Contact Interactions In Superstring Theory,''
Nucl.\ Phys.\ B {\bf 299}, 559 (1988).
}

\lref\DijkgraafGF{
R.~Dijkgraaf,
``Instanton strings and hyperKaehler geometry,''
Nucl.\ Phys.\ B {\bf 543}, 545 (1999)
[arXiv:hep-th/9810210].
}

\lref\LuninYV{
O.~Lunin and S.~D.~Mathur,
``Correlation functions for M(N)/S(N) orbifolds,''
Commun.\ Math.\ Phys.\  {\bf 219}, 399 (2001)
[arXiv:hep-th/0006196];
``Three-point functions for M(N)/S(N) orbifolds with N = 4 supersymmetry,''
Commun.\ Math.\ Phys.\  {\bf 227}, 385 (2002)
[arXiv:hep-th/0103169].
}

\lref\BanksDD{
T.~Banks, M.~R.~Douglas, G.~T.~Horowitz and E.~J.~Martinec,
``AdS dynamics from conformal field theory,''
arXiv:hep-th/9808016.
}

\lref\AchucarroGM{
A.~Achucarro and P.~K.~Townsend,
``Extended Supergravities In D = (2+1) As Chern-Simons Theories,''
Phys.\ Lett.\ B {\bf 229}, 383 (1989).
}

\lref\IzquierdoJZ{
J.~M.~Izquierdo and P.~K.~Townsend,
Class.\ Quant.\ Grav.\  {\bf 12}, 895 (1995)
[arXiv:gr-qc/9501018].
}

\lref\HoweZM{
P.~S.~Howe, J.~M.~Izquierdo, G.~Papadopoulos and P.~K.~Townsend,
Nucl.\ Phys.\ B {\bf 467}, 183 (1996)
[arXiv:hep-th/9505032].
}

\lref\BrownNW{
J.~D.~Brown and M.~Henneaux,
``Central Charges In The Canonical Realization Of Asymptotic Symmetries: An
Example From Three-Dimensional Gravity,''
Commun.\ Math.\ Phys.\  {\bf 104}, 207 (1986).
}

\lref\StromingerEQ{
A.~Strominger,
``Black hole entropy from near-horizon microstates,''
JHEP {\bf 9802}, 009 (1998)
[arXiv:hep-th/9712251].
}

\lref\AchucarroVZ{
A.~Achucarro and P.~K.~Townsend,
``A Chern-Simons Action For Three-Dimensional Anti-De Sitter Supergravity
Theories,''
Phys.\ Lett.\ B {\bf 180}, 89 (1986);
E.~Witten,
``(2+1)-Dimensional Gravity As An Exactly Soluble System,''
Nucl.\ Phys.\ B {\bf 311}, 46 (1988).
}

\lref\LarsenXM{
F.~Larsen,
``The perturbation spectrum of black holes in N = 8 supergravity,''
Nucl.\ Phys.\ B {\bf 536}, 258 (1998)
[arXiv:hep-th/9805208].
}

\lref\LarsenUK{
F.~Larsen and E.~J.~Martinec,
``U(1) charges and moduli in the D1-D5 system,''
JHEP {\bf 9906}, 019 (1999)
[arXiv:hep-th/9905064].
}

\lref\MaldacenaSS{
J.~M.~Maldacena, G.~W.~Moore and N.~Seiberg,
``D-brane charges in fivebrane backgrounds,''
JHEP {\bf 0110}, 005 (2001)
[arXiv:hep-th/0108152].
}

\lref\CarlipBS{
S.~Carlip and I.~I.~Kogan,
``Three-Dimensional Topological Field Theories And Strings,''
Mod.\ Phys.\ Lett.\ A {\bf 6}, 171 (1991).
}

\lref\LuninIZ{
O.~Lunin, J.~Maldacena and L.~Maoz,
``Gravity solutions for the D1-D5 system with angular momentum,''
arXiv:hep-th/0212210.
}
\lref\MartinecCF{
E.~J.~Martinec and W.~McElgin,
``String theory on AdS orbifolds,''
JHEP {\bf 0204}, 029 (2002)
[arXiv:hep-th/0106171].
}

\lref\KutasovXU{
D.~Kutasov and N.~Seiberg,
``More comments on string theory on AdS(3),''
JHEP {\bf 9904}, 008 (1999)
[arXiv:hep-th/9903219].
}

\lref\CowdallBU{
P.~M.~Cowdall and P.~K.~Townsend,
``Gauged supergravity vacua from intersecting branes,''
Phys.\ Lett.\ B {\bf 429}, 281 (1998)
[Erratum-ibid.\ B {\bf 434}, 458 (1998)]
[arXiv:hep-th/9801165].
}

\lref\BoonstraYU{
H.~J.~Boonstra, B.~Peeters and K.~Skenderis,
``Brane intersections, anti-de Sitter spacetimes and dual superconformal
theories,''
Nucl.\ Phys.\ B {\bf 533}, 127 (1998)
[arXiv:hep-th/9803231].
}

\lref\GauntlettKC{
J.~P.~Gauntlett, R.~C.~Myers and P.~K.~Townsend,
``Supersymmetry of rotating branes,''
Phys.\ Rev.\ D {\bf 59}, 025001 (1999)
[arXiv:hep-th/9809065].
}

\lref\RocekVK{
M.~Rocek, K.~Schoutens and A.~Sevrin,
``Off-shell WZW models in extended superspace,''
Phys.\ Lett.\ B {\bf 265}, 303 (1991);
M.~Rocek, C.~h.~Ahn, K.~Schoutens and A.~Sevrin,
``Superspace WZW models and black holes,''
arXiv:hep-th/9110035.
}

\lref\StromingerSH{
A.~Strominger and C.~Vafa,
``Microscopic Origin of the Bekenstein-Hawking Entropy,''
Phys.\ Lett.\ B {\bf 379}, 99 (1996)
[arXiv:hep-th/9601029].
}

\lref\AharonyTI{
O.~Aharony, S.~S.~Gubser, J.~M.~Maldacena, H.~Ooguri and Y.~Oz,
``Large N field theories, string theory and gravity,''
Phys.\ Rept.\  {\bf 323}, 183 (2000)
[arXiv:hep-th/9905111].
}

\lref\braam{P. Braam and J. Hurtubise,
``Instantons on Hopf surfaces and monopoles on solid tori,''
J. Reine Angew. Math. {\bf 400} (1989), 146--172. }

\lref\MaldacenaPB{
J.~M.~Maldacena and H.~Nastase,
``The supergravity dual of a theory with dynamical supersymmetry  breaking,''
JHEP {\bf 0109}, 024 (2001)
[arXiv:hep-th/0105049].
}

\lref\GauntlettUR{
J.~P.~Gauntlett, N.~W.~Kim, D.~Martelli and D.~Waldram,
``Fivebranes wrapped on SLAG three-cycles and related geometry,''
JHEP {\bf 0111}, 018 (2001)
[arXiv:hep-th/0110034].
}

\lref\AcharyaMU{
B.~S.~Acharya, J.~P.~Gauntlett and N.~Kim,
``Fivebranes wrapped on associative three-cycles,''
Phys.\ Rev.\ D {\bf 63}, 106003 (2001)
[arXiv:hep-th/0011190].
}

\lref\GS{
S.~Gukov and J.~Sparks,
``M-theory on Spin(7) manifolds,''
Nucl.\ Phys.\ B {\bf 625}, 3 (2002)
[arXiv:hep-th/0109025].
}

\lref\SchvellingerIB{
M.~Schvellinger and T.~A.~Tran,
``Supergravity duals of gauge field theories from SU(2) x U(1) gauged
supergravity in five dimensions,''
JHEP {\bf 0106}, 025 (2001)
[arXiv:hep-th/0105019].
}

\lref\WittenSC{
E.~Witten,
``Solutions of four-dimensional field theories via M-theory,''
Nucl.\ Phys.\ B {\bf 500}, 3 (1997)
[arXiv:hep-th/9703166].
}

\lref\AharonyUB{
O.~Aharony, M.~Berkooz, D.~Kutasov and N.~Seiberg,
``Linear dilatons, NS5-branes and holography,''
JHEP {\bf 9810}, 004 (1998)
[arXiv:hep-th/9808149].
}

\lref\MaldacenaBP{
J.~M.~Maldacena, G.~W.~Moore and A.~Strominger,
``Counting BPS black holes in toroidal type II string theory,''
arXiv:hep-th/9903163.
}

\lref\IvanovAI{
S.~Ivanov and G.~Papadopoulos,
``Vanishing theorems and string backgrounds,''
Class.\ Quant.\ Grav.\  {\bf 18}, 1089 (2001)
[arXiv:math.dg/0010038].
} 

\lref\MooreGB{
G.~W.~Moore and E.~Witten,
``Self-duality, Ramond-Ramond fields, and K-theory,''
JHEP {\bf 0005}, 032 (2000)
[arXiv:hep-th/9912279].
}
\lref\MaldacenaSS{
J.~M.~Maldacena, G.~W.~Moore and N.~Seiberg,
``D-brane charges in five-brane backgrounds,''
JHEP {\bf 0110}, 005 (2001)
[arXiv:hep-th/0108152].
}
\lref\SchomerusDC{
V.~Schomerus,
``Lectures on branes in curved backgrounds,''
Class.\ Quant.\ Grav.\  {\bf 19}, 5781 (2002)
[arXiv:hep-th/0209241].
}
\lref\MooreVF{
G.~Moore,
``K-theory from a physical perspective,''
arXiv:hep-th/0304018.
}

\lref\SeibergXZ{
N.~Seiberg and E.~Witten,
``The D1/D5 system and singular CFT,''
JHEP {\bf 9904}, 017 (1999)
[arXiv:hep-th/9903224].
}

\lref\emillectures{E. Martinec, 
Lectures given at the Komaba workshop, November 1999;
notes available at
http://theory.uchicago.edu/$\sim$ejm/japan99.ps}

\lref\indexsummary{S. Gukov, E. Martinec, G. Moore, and A. Strominger,
``An index for $2D$ field theories
with large $N=4$ superconformal symmetry,'' to appear.}

\lref\massivecs{S. Gukov, E. Martinec, G. Moore, and A. Strominger,
``Massive Chern-Simons gauge theory 
and singletons in the $AdS_3/CFT_2$ correspondence,'' to appear.}


\lref\MBerg{M.~Berg and H.~Samtleben,
``An exact holographic RG flow between 2d conformal fixed points,''
JHEP {\bf 0205} (2002) 006, hep-th/0112154;
``Holographic correlators in a flow to a fixed point,''
JHEP {\bf 0212} (2002) 070, hep-th/0209191.}

\lref\LuPoritz{
H.~Lu and J.~F.~Vazquez-Poritz,
``Penrose limits of non-standard brane intersections,''
Class.\ Quant.\ Grav.\  {\bf 19} (2002) 4059, hep-th/0204001.}

\lref\GiveonP{A.~Giveon and A.~Pakman,
``More on superstrings in AdS(3) x N,''
JHEP {\bf 0303} (2003) 056, hep-th/0302217.}

\lref\Ivanov{E.A. Ivanov, S.O. Krivonos, ``N=4 super-Liouville equation'',
J.Phys. A: Math. Gen., 17 (1984) L671; ``N=4 superextension of the
Liouville equation with quaternionic structure'', Teor. Mat. Fiz.
63 (1985) 230 [Theor. Math. Phys. 63 (1985) 477];
E.A. Ivanov, S.O. Krivonos, V.M. Leviant, ``A new class of
superconformal sigma models with the Wess-Zumino action'',
Nucl. Phys. B304 (1988) 601;
E.A. Ivanov, S.O. Krivonos, V.M. Leviant, ``Quantum N=3, N=4
superconformal WZW sigma models'', Phys. Lett. B215 (1988) 689; B221
(1989) 432E.}

\lref\Blauetal{
M.~Blau, J.~Figueroa-O'Farrill, C.~Hull and G.~Papadopoulos,
``A new maximally supersymmetric background of IIB superstring theory,''
JHEP {\bf 0201} 047 (2002), hep-th/0110242;
``Penrose limits and maximal supersymmetry,''
Class.\ Quant.\ Grav.\  {\bf 19} L87 (2002), hep-th/0201081.}


\def\modspace{{1}}
\def\branefig{{2}}


%
\Title{\vbox{\baselineskip12pt
\hbox{hep-th/0403090}
\hbox{HUTP-04/A003}
\hbox{RUNHETC-2004-03}
\hbox{EFI-04-04}
}}
{\vbox{\centerline{
The Search for a Holographic Dual
to $AdS_3 \!\times \!S^3\!\times \!S^3\!\times \!S^1$
}}}
\centerline{
Sergei Gukov\footnote{$^*$}{\it Jefferson Physical Laboratory,
Harvard University, Cambridge, MA 02138},
Emil Martinec\footnote{$^{**}$}{\it
Enrico Fermi Institute and Department of Physics, University of Chicago,
Chicago, IL 60637},
Gregory Moore\footnote{$^\dagger$}{\it Department of Physics and
Astronomy, Rutgers University, Piscataway, NJ 08854-8019}
and Andrew Strominger$^*$}
\vskip.1in
\vskip.1in \centerline{\bf Abstract} The problem of
finding a holographic CFT dual to string theory on $AdS_3 \times
\S^3 \times \S^3 \times \S^1$ is examined in depth.  This background
supports a large $\CN=4$ superconformal symmetry. While in some
respects similar to the familiar small $\CN=4$ systems on
$AdS_3\times \S^3\times K3$ and $AdS_3\times \S^3\times T^4$, there
are important qualitative differences. 
Using an analog of the elliptic genus for large $\CN=4$ theories we rule out 
all extant proposals~-- in their simplest form~--
for a holographic duality to
supergravity at generic values of the background fluxes.
Modifications of these extant proposals and other possible duals
are discussed.

\Date{March 8, 2004}

\listtoc\writetoc


\newsec{Introduction and Summary}

The AdS/CFT correspondence has been a powerful tool in
understanding nonperturbative string theory (for a review, see
\AharonyTI). This is especially true in two dimensions, due to the
infinite dimensional structure of the conformal group. The
examples most studied are the conformal field theories dual to
type II string theory on geometries of the form
$AdS_3\times\S^3\times\CM$, with $\CM=K3$ or $T^4$. These
geometries arise from the near-horizon limit of $Q_1$ onebranes
coincident with $Q_5$ fivebranes, with the fivebranes wrapping
$\CM$ and the onebranes transverse to $\CM$. The dual CFT's
obtained in this way are sigma models on the moduli space of $Q_1$
instantons in $U(Q_5)$ gauge theory on $\CM$.  They possess {\it
small} $\CN=(4,4)$ superconformal symmetry, in which the four
(anti)holomorphic supercurrents are charged under a single $SU(2)$
$R$-symmetry current.  They also form a doublet under a global,
custodial $SU(2)$ $R$-symmetry. U-duality implies
\refs{\SeibergXZ,\LarsenUK} that the CFT's for different $Q_1$, $Q_5$
having the same product $N=Q_1Q_5$, are different descriptions of
the same theory appropriate to different asymptotic regimes of its
moduli space.  These CFT's are all deformations of the
much-studied symmetric product orbifold $\sym^N(\CM)$
\StromingerSH.

Type II string theory also has a solution with the geometry $AdS_3
\times \S^3_+\times \S^3_-\times \S^1$, where the three-spheres
$\S^3_\pm$ are threaded by integral fivebrane flux $Q_5^\pm$, and
there is also a onebrane charge $Q_1$
\refs{\CowdallBU\BoonstraYU\GauntlettKC\ElitzurMM\GiveonP-\deBoerRH}. This
solution is distinguished in having 16 Killing spinors and a
corresponding {\it large} $\CN=(4,4)$ superconformal symmetry.
Large $\CN=4$ supersymmetry is distinguished from its small
counterpart in that both $SU(2)$ $R$-symmetries under which the
supercharges transform give rise to current algebras (at levels
$k^\pm$ related to the background fluxes). Despite this enhanced
symmetry, this example is much less well understood than that of
its $AdS_3 \times \S^3\times K3$ or $AdS_3 \times \S^3\times T^4$
cousins. In particular, the holographic dual has not been
established.

For the special case $Q_5^+=Q_5^-\equiv Q_5$, a seemingly obvious
candidate dual is obtained by replacing $K3$ or $T^4$ with
$\S^3\times \S^1$ in the symmetric product CFT. This was first
suggested in \ElitzurMM, and further studied and elaborated in
\deBoerRH. More specifically one takes (deformations of) the
symmetric product $\sym^{Q_1Q_5} (\CS )$, where $\CS \sim \S^3
\times \S^1$ is the supersymmetric  $U(2)$ WZW model with central
charge $c=3$. $\CS$ can be described by a free boson and four free
fermions and is the smallest large $\CN=4$ CFT. Many aspects of
this construction appear promising. First, it carries the large
$\CN=(4,4)$ superconformal symmetry and has a central charge
$c=6Q_1Q_5$ which agrees with the Brown-Henneaux formula \BrownNW\
as applied to $AdS_3\times \S^3 \times \S^3 \times \S^1$ . It has
the small RR-sector gap (of order ${ 1 \over Q_1Q_5}$) required
for agreement with black hole thermodynamics \preskill, and the
low-lying states of the Hilbert space have the structure of a Fock
space, much like supergravity/string theory quanta in the $AdS$
background. Indeed it is hard to see how one could satisfy these
requirements in any way other than with a $Q_1Q_5$-fold symmetric
product. Given the assumption of a $Q_1Q_5$-fold symmetric
product, $\CS$ is the only game in town with the required central
charge $c=3$. On top of this, we match the CFT and supergravity
moduli as well as the indices (as far as they can be compared) in
the sector of the theory with zero $\S^1$ charge.

Despite these promising features, this proposed duality has a
fatal flaw (for generic $Q_5$)  in its simplest form. The basic
problem is that $\sym^{Q_1Q_5} (\CS )$ depends only on the product
$Q_1Q_5$, while the natural formulation of string theory on
$AdS_3\times \S^3 \times \S^3 \times \S^1$ depends on $Q_1$, $Q_5$
separately,
as we will deduce from by comparing a certain index of the 
conformal field theory with a partition function of the 
supergravity theory.  
\foot{More precisely, the index of $\sym^N(\CS)$ depends on all
the prime factors of $N$ ``democratically'' but the supergravity 
depends on the particular factorization $N=Q_1 Q_5$.} 
The $K3$ and $T^4$ cases are rescued from
such a contradiction by a large $U$-duality group which relates
all theories with the same value for the product $Q_1Q_5$. In
striking contrast, we find in section two below that the U-duality
group is extremely limited for $\S^3\times \S^1$ and does not relate
theories of the same central charge and different $Q_1$, $Q_5$. It
is possible that this difficulty may be overcome by some kind of
modification or twisting of the symmetric product but we do not
have a concrete suggestion.

The considerations of the preceding paragraph do not rule out the
possibility of a 'duality' to $\sym^{Q_1Q_5} (\CS )$ when $Q_5=1$.%
\foot{ Such a duality may well ultimately make sense, but at
present it is not so well-defined because supergravity is strongly
coupled when $Q_5=1$.  There may be  a duality to a bulk string
theory when $Q_5=1$, but at our current level of string technology
this is not well-understood -- even in the NS case there are
singularities \SeibergXZ. Nevertheless in this paper we shall continue to
speak of a $Q_5=1$ duality with the idea that the difficulties on
the bulk side may eventually be overcome. } One can generalize
this proposal to the case where only one of $Q_5^\pm$ equals one;
then the symmetric product $\sym^{Q_1} (\CS)$ is still a viable
candidate (one of the $SU(2)$ $R$-symmetries of the component
$\S^3\times\S^1$ CFT is then a current algebra of level $Q_5'>1$).
For general $Q_5^+\neq Q_5^-$ there is  not even a full conjecture
for a dual. (An interesting and tentative partial proposal was
made in \deBoerRH.)

 For general values of $Q_5$ alternatives should be considered.
One possibility is the low-energy dynamics of fivebranes wrapped
on $\S^3\times \S^1$. The gauge theory and related supergravity
solutions for $Q_5$ fivebranes wrapped on a special Lagrangian
$\S^3$ threaded by $Q'_5$ units of three-form flux, were
considered in \refs{\AcharyaMU \SchvellingerIB
\MaldacenaPB-\GauntlettUR}. The worldvolume of the fivebranes is
$\IR^{1,2}\times\S^3$ with a warp factor for the $\S^3$. The
solution in section 3.1.1 of \AcharyaMU\ has 1/16 supersymmetry,
and $SU(2)^3$ symmetry. We conjecture that, with $\IR^{1,2}$
compactified to $\IR^{1,1}\times \S^1$ and $Q_1$ instantons on
$\S^3\times \S^1$, the theory will flow in the IR to a sigma model
with large $\CN=(4,4)$ superconformal symmetry, \ie\ 1/2
supersymmetry and its associated $SU(2)^4$ global symmetry. This
sigma model should be closely related to the sigma model on the
moduli space of $Q_1$ instantons in $U(Q_5)$ gauge theory on
$\S^3\times\S^1$. This sigma model has not been studied (some
relevant mathematical results can be found in \braam); indeed, it
is not known if this model has large $\CN=4$ supersymmetry.

The difficulties in establishing a holographic duality might seem
surprising.  One might have expected that the enhanced large
$\CN=4$ supersymmetry would give greater control for this case.
While that may ultimately prove correct, there are substantial
qualitative differences between large and small $\CN=4$ which
prevent us from drawing on the familiar bag of tricks. To name a few:
\item{1.}
The BPS bound is nonlinear in the charges and implies that
some BPS states must get mass corrections at every order in perturbation
theory.
\item{2.} The large $\CN=4$ algebra has a finite dimensional $\CN=4$
superconformal subalgebra $\dto$.  However, BPS states
of the global $\dto$ subalgebra  are not in general BPS
states of the large $\CN=4$ super Virasoro algebra.
\item{3.}
There can be any number -- odd or even -- of moduli,
and there are few known constraints on the moduli space geometry.%
\foot{We will demonstrate one constraint in section 4.4 -- that the
moduli space is a real slice of a self-mirror $\CN=2$ theory,
which is also fixed under the mirror map.}

\noindent
Even so, we will report on progress in understanding
both sides of the correspondence.

On the supergravity side, we revisit in section 2 the solution of
the supergravity equations of motion on this background, for both
NS and R background fluxes.  We determine the massless moduli,
which can be parametrized by the string coupling $g_s$ and (in the
IIB theory)  a linear combination of RR axion $C_0$ and four form
$C_4$ . The radius of the $\S^1$ is determined in terms of $g_s$
and the charges.\foot{This formula differs from the one in
\deBoerRH.} We discuss the global structure of the moduli space,
the low-energy descriptions appropriate to various regimes, and
the locus in moduli space where the CFT becomes singular. In
section 3 we discuss the relation of the solution to the
near-horizon geometry of intersecting branes.

On the CFT side, we review in section 4 (following
\refs{\STVP\SpindelSR\GPTVP\PT\PTtwo-\OPT}) the large $\CN=4$
superconformal algebra and its representation theory.  We
demonstrate the fact mentioned above, that the BPS bound of large
$\CN=4$ superconformal symmetry in general differs from that of
its global subalgebra $D(2,1)$ which comprises the
super-isometries of $AdS_3\times \S_+^3\times\S_-^3$. We exhibit
the general structure of marginal deformations, and we examine the
question of whether  an $h=1/2$ chiral primary field   generates a
modulus that preserves large $\CN=4$. In yet another surprise,
Dixon's proof of this fact for $\CN=(2,2)$ \DixonBG\ does not
immediately apply to the case of large $\CN=(4,4)$ supersymmetry.
We will nevertheless find an appropriate generalization of Dixon's
proof which does apply to large $\CN=4$.

We also introduce an index for theories with large $\CN=4$
supersymmetry, with rather remarkable properties: The index is not a number,
rather it is a nontrivial modular form.  Consequently,
the analogue of the elliptic genus is not holomorphic.
 We introduce the index and the analogue of the elliptic
genus in section 4 and evaluate the index in section 6
for the symmetric product $\sym^N(\S^3\times\S^1)$.
Detailed derivations and further discussion of these indices
will be the subject of a companion paper \indexsummary.
The related BPS spectrum and the moduli of the symmetric product
are exhibited in section 5.

Sections 7 and 8 analyze the BPS and near-BPS spectra of supergravity
and compare them to the symmetric product.
We find that the BPS spectra do not match, in that
the one-particle states of the classical supergravity limit
with different spins $\ell^+\ne\ell^-$ on $\S^3_\pm$
do not have a BPS counterpart in the symmetric product
(this was already noted in \ElitzurMM\ for a special case).
This might indicate that such states are not protected
by large $\CN=4$ supersymmetry (assuming that the correct
dual has been identified).  Indeed, as mentioned above,
the BPS bound already requires that the masses receive
perturbative corrections; furthermore, we show
that the (BPS) short multiplets of supergravity occur
in combinations that can naturally pair up into
(non-BPS) long multiplets,
so there is no reason {\it a priori} that they
should survive across moduli space.
The near-BPS spectrum is of course also not protected,
but in recent studies \BMN\ has been seen to be remarkably robust.
In our case, the spectrum provides an indication
that the symmetric product orbifolds indeed only
describe the situation where one of the fivebrane charges is one.

Finally, in section 9 we discuss aspects of the
$U(1)\times U(1)$ Chern-Simons gauge theory which
appears in low-energy supergravity.
A study of the associated topological field theory
yields further constraints on the structure of the
holographic dual, and provides further strong evidence
that the symmetric product has $Q_5^+=1$ or $Q_5^-=1$.
Again, details and generalizations are deferred
to another companion paper \massivecs.

While our results should help guide the search for
holographic duals for supergravity backgrounds with
large $\CN=4$ supersymmetry, many open questions remain.
To list a few:
\item{{\it i.}}
What are the geometrical conditions on a sigma model target
space in order that it admit large $\CN=4$ supersymmetry?
The examples discussed to date are based on current algebra
cosets \SpindelSR.
Are all models with large $\CN=4$ automatically conformally invariant,
as is the case for small $\CN=4$?
\item{{\it ii.}}
%
%
What is the geometrical interpretation of the large $\CN=4$ index?
\item{{\it iii.}}
Does the sigma model on the moduli space of $Q_1$ instantons
in $U(Q_5)$ gauge theory on $\S^3\times \S^1$
have large $\CN=4$ supersymmetry?
Is it a viable dual for $Q_5^+=Q_5^-$?
\item{{\it iv.}}
Are there possible alternatives to the naive
orbifold  $Sym^N(\S^3\times\S^1)$
(Making use, for example, of discrete torsion, extensions of the
orbifold group $S_N$,
asymmetric shifts on the $\R$ factor, \etc.),
which could serve as candidate duals?
For $Q_5^+=Q_5^-$, are such orbifold theories on the moduli space
of the sigma model proposed in {\it iii.}?
\item{{\it v.}}
What can we say about the (Zamolodchikov) metric
and the corresponding geometry of
moduli space as a consequence of large $\CN=4$ superconformal symmetry?
\item{{\it vi.}}
The new large $\CN=4$ index predicts ``long string'' BPS states.
What are the corresponding geometries/bulk states?
(A natural conjecture is that they are generalizations
of the supertube solutions found in \LuninIZ.)
\item{{\it vii.}}
The intersecting D-brane configurations that naively give rise
to large $\CN=4$ supersymmetry have chiral fermions bound to
the intersection.  What is their role, and does their presence
imply any constraint on the CFT dual?  Do they decouple,
as we will assume below? (See the discussion near equation (3.5)
below.)

\noindent
These and many other questions remain for future research.


\newsec{Supergravity solutions}

\subsec{Type II conventions}

The  IIB Lagrangian is\foot{We set $\alpha^\prime ={1 \over (2\pi)^2}$.
In the notation of \PolchinskiRQ, we have $\kappa_{10}^2={1 \over 4 \pi}$,
$\tilde F_k=R_k$ and $\mu^p=2\pi$.} 
\eqn\iiblag{ \eqalign{
& {2\pi \over g_B^2 } \int \sqrt{- g}
    e^{-2\phi} \bigl( \CR + 4 (\nabla \phi)^2\bigr)
- {\pi \over g_B^2 } \int  e^{-2\phi} H\wedge * H\cr
&\hskip 1cm
    -{\pi }\int R_1\wedge * R_1 - {\pi } \int R_3 \wedge * R_3
    - \half \pi \int R_5 \wedge * R_5
    + \pi\int C_4\wedge H\wedge F_3 \quad .} }
Here $R_1$ has integral periods; locally $R_1= dC_0$.
$R_3$ satisfies the Bianchi identity
\eqn\bianiiib{
dR_3 + R_1 \wedge H = 0\ .
}
When $R_1= dC_0$ can be trivialized then
\eqn\fthree{
F_3 = R_3+ C_0 H
}
is closed
and has integral periods.
$R_5$
has integral periods when $H=0$, is self-dual and obeys
\eqn\tyk{dR_5=H\wedge F_3.}
The IIA Lagrangian is similarly 
\eqn\iialag{ \eqalign{
& {2\pi \over g_A^2 } \int \sqrt{-g}\, e^{-2\phi}
    \bigl( \CR + 4 (\nabla \phi)^2\bigr)
    - {\pi \over g_{\!A}^2 } \int  e^{-2\phi} H\wedge * H\cr
& \hskip 1.5cm
    -{\pi }\int R_2\wedge * R_2 - {\pi } \int R_4
    \wedge * R_4 +  \pi  \int C_3 \wedge d C_3\wedge H\cr} }
where $R_4=dC_3 - H\wedge C_1$.

We now look for $AdS$ solutions
to the equations of motion following from
\iiblag, \iialag\ on $AdS_3 \times \S^3 \times \S^3 \times \S^1$ with either
NS or RR three-form background fluxes.

\subsec{Pure NS solutions}

We take $R_5=0,F_3=0$, $\phi=0$ and
\eqn\hthree{ H = \lambda_0  \omega_0 + \lambda_+ \omega_+ +
\lambda_- \omega_- }
where the volume forms
\eqn\volforms{\eqalign{
\omega_0 & = \vol(AdS_3) = (\ell/x_2)^3 dt \wedge dx^1 \wedge dx^2 \cr
\omega_\pm & = \vol(\S^3_\pm)
}}
are normalized so that $\int_{\S^3_{\pm}} \omega_{\pm} = 2\pi^2 R_{\pm}^3$.
We take the metric
\eqn\iibmet{
ds^2 = \frac{\l^2}{x_2^{~2}}\Bigl(-dt^2 + (dx_1)^2 + (dx_2)^2\Bigr)
    + R_+^2 ds^2(\S^3_+) + R_-^2 ds^2(\S^3_-) + L^2 (d\theta)^2
}
with $\theta \sim \theta + 1$.
The curvatures are
\eqn\adscurv{
\eqalign{
R_{\mu\nu\lambda\rho} & = - \ell^{-2}(g_{\mu\lambda} g_{\nu \rho}
- g_{\mu \rho} g_{\nu \lambda}) \cr
\CR_{\mu\nu} & = -2 \l^{-2} g_{\mu\nu} \cr
\CR & = - 6 \l^{-2} \quad.}
}

Similarly, $ds^2(\S^3)$ is the
round metric of $\S^3$ normalized as in the unit sphere in Euclidean $\R^4$.
With this normalization we have curvatures:
\eqn\sthreecurv{
\eqalign{
R_{\mu\nu\lambda\rho} & = R^{-2}(g_{\mu\lambda} g_{\nu \rho}
- g_{\mu \rho} g_{\nu \lambda}) \cr
\CR_{\mu\nu} & = 2 R^{-2} g_{\mu\nu} \cr
\CR & =  6 R^{-2} \quad .}
}
We look for solutions with constant dilaton.
The $\phi$ equation of motion then
forces $H_{MNP}H^{MNP}=0$,%
\foot{We absorb the constant mode of the
dilaton in $g_s$, and set $\phi=0$ at infinity.}
\eqn\taux{ {1 \over 6} H_{MNP}H^{MNP} = - \lambda_0^2 + \lambda_+^2 +
\lambda_-^2=0 \ .}
The stress-energy simplifies and $\CR=0$.
The Einstein equations then give
\eqn\einst{
\eqalign{
\ell^{-2}   & = {1\over 4}   \lambda_0^2\cr
R_+^{-2} & = {1\over 4}   \lambda_+^2 \cr
R_-^{-2} & = {1\over 4}   \lambda_-^2 \quad.}
}
The fivebrane charges on the two $\S^3$s are
\eqn\normflx{ \int_{\S^3_\pm} H = Q_5^\pm = 4\pi^2 R_\pm^2 }
with $Q_5^\pm$ integers. The fundamental string charge -- also an
integer -- is 
\eqn\exlf{
  Q_1 = {1\over g_B^2} \int  * H
    = {8\pi^4   R_+^3 R_-^3 L \over \ell g_B^{2} }\quad .
}

In summary we have 
\eqn\lengths{\eqalign{
\ell & = { 1\over 2\pi }  \sqrt{ Q_5^+Q_5^-\over Q_5^++ Q_5^-} \cr
R_\pm & = { 1\over 2\pi }\sqrt{Q_5^\pm} \cr
L & = {4 \pi  g_B^2 Q_1 \over Q_5^+Q_5^- \sqrt{Q_5^++ Q_5^-}}
\quad.}}
Note that the radius $L$ and the string coupling $g_B^2$
are not separate moduli, rather their
ratio is fixed by this relation in terms of the charge quanta.

The pure NS solution considered here can be constructed as an
exact worldsheet conformal field theory, using products of $SU(2)$
level $Q_5^\pm$, $SL(2,R)$ level $\frac{Q_5^+ Q_5^-}{Q_5^++Q_5^-}$
and $U(1)$ WZW models \refs{\ElitzurMM,\GiveonP}.
This conformal field theory provides solutions
of all the $d=10$ superstring theories.

For the case of the IIA string almost the same equations
apply. The result is exactly  \lengths\  with $g_B \to g_A$. Note that this makes good sense
since under $T$-duality
\eqn\smsys{
L_B/g_B^2 = L_A/g_A^2
}
and this is the quantity which is fixed when we have purely NS
sector fluxes.


\subsec{Pure RR solutions}

The case of purely RR fluxes, which is related to
the near-horizon geometry of the intersecting D1-D5-D5' system
in the next section, is also of interest.
The spacetime solution is easily obtained using the S-duality of the
supergravity equations of motion, under which $g_B \to {1 /g_B}$,
lengths are rescaled%
\foot{Since they are referred to the string tension.
The fundamental and D-string tensions differ
by a factor of $g_B$, so a factor of $\sqrt{g_B}$
takes into account the change in conventions.}
by a factor of $\sqrt{g_B^{~}}$
and the integer NS
charges $(Q_1, Q_5^+,Q_5^-)$ become integer RR charges which we
continue to denote $(Q_1, Q_5^+,Q_5^-)$. The relations \lengths\
become 
\eqn\lngths{ \eqalign{ \ell & = {1 \over 2\pi }  \sqrt{
g_B Q_5^+g_B Q_5^-\over g_B Q_5^+ + g_B Q_5^-} \cr
R_\pm & = {1\over 2\pi}
\sqrt{g_B Q_5^{\pm}} \cr L
& = {4 \pi g_B Q_1 \over g_B Q_5^+g_B Q_5^-
\sqrt{g_B Q_5^{+} + g_B Q_5^{-}}} \quad.} }
We have written the expression in a manner which emphasizes the
fact that $R_\pm$, $L$ and $\ell$ are finite in the $Q\to \infty$
limit with $g_B Q$ held fixed.


\subsec{Chern-Simons terms and central charges}

The central charge of the spacetime conformal field theory
can be computed from an analysis of the algebra of diffeomorphisms
near the conformal boundary of $AdS_3$ \BrownNW; the result is
\eqn\bhcf{
c = {3\l \over 2 G_{N}^{(3)}} \quad .
}
By dimensional reduction (in the NS background) we have 
\eqn\thrdee{ {1\over 16 \pi G_{N}^{(3)}}
= {8 \pi^5 R_+^3 R_-^3 L \over g_B^2 }
= {1 \over 2} Q_1 \sqrt{ { Q_5^+Q_5^-\over Q_5^++ Q_5^-} \ .}
}
and so
\eqn\cexpl{ c = 6 Q_1 { Q_5^+Q_5^-\over Q_5^++ Q_5^-}\quad . }
Similarly, the left and right
$SU(2)\times SU(2)\times U(1)$ isometries of
$\S^3\times\S^3\times\S^1$
yields a set of corresponding gauge fields from
the Kaluza-Klein reduction of the metric and NS $B$-field.
The action for these fields on $AdS_3$ contains Chern-Simons terms. 
(The abelian Chern-Simons term is discussed in more depth in  in section 9).
The result is \refs{\DKSS,\LPS,\LPSb,\APT,\Mathur,\NS}:
\eqn\csresult{
\eqalign{ S &= {1 \over 16 \pi G_N^{(3)}} \int\! d^{3}\!x
    \sqrt{-g} \Bigl( \CR^{(3)} + {2 \over \ell^2} \Bigr)
    + \Bigl[{Q_1 Q^+_5 \over 8 \pi} \int\Tr\Bigl(\CA^+_L d\CA^+_L
        + {\coeff 23}{\CA^+_L}^3\Bigr) +\cr &~~~{Q_1 Q^-_5 \over 8 \pi}
\int \Tr\Bigl( \CA^-_L d \CA^-_L+{\coeff 23}{\CA^-_L}^3\Bigr)
+\frac{Q_1}{8\pi} \int\!A_LdA_L\Bigr]-(L \leftrightarrow R)
 }}
where $\CA^\pm_{L,R}$ are the gauge fields in $AdS_3$
that transform under left- and right-handed $SU(2)$
isometries of $\S^3_\pm$,
and $A_{L,R}$ are the corresponding $U(1)$ gauge fields for $\S^1$.
We have also included the 3d Einstein term,
which can be written as a Chern-Simons form \AchucarroVZ.
These Chern-Simons terms enforce the integer quantization
of the background charges $Q_1$, $Q_5^\pm$.

While we have isolated this apparently three-dimensional
action for the bosonic modes of an $AdS_3$ supergravity,
it is important to note that the
radii of $\S^3_\pm$ are typically of the same order
as the curvature radius of $AdS_3$, and set the
scale of the masses of KK modes.  There is no sense in
which the bulk theory is effectively 2+1 dimensional;
the reason for exhibiting the Chern-Simons forms \csresult\
is to manifest the central extensions
of the various current algebras in the spacetime CFT.
Indeed, the dual CFT
contains left and right $SU(2)\times SU(2)\times U(1)$
current algebras; the SU(2) current algebras are at levels
$k_\pm=Q_1 Q_5^\pm$, possibly up to $O(1)$ corrections
that are invisible in the classical supergravity limit.%
\foot{The D-brane analysis of the next section
provides evidence that there are no such $O(1)$ corrections.}
In the gauge field equations of motion
(see for example \deBoerIP), the Chern-Simons term
gives mass to half of the components, such that
their lowest modes have conformal weight $(h_L,h_R)=(1,2)$
or $(2,1)$; the lowest modes of the other components
are the `singleton' modes of weight $(1,0)$ or $(0,1)$,
dual to the respective $(0,1)$ and $(1,0)$
$SU(2)\times SU(2)\times U(1)$ currents $j_{L,R}$
of the dual CFT via the usual boundary coupling
\eqn\bcoup{
  \int_{\partial AdS_3}\Bigl( \CA_L j_R+\CA_R j_L\Bigr)\ .
}

The spacetime supersymmetry of the background requires
a supersymmetric completion of this
$SU(2)\times SU(2)\times U(1)$ current algebra,
and an action of two-dimensional conformal symmetry.
The supersymmetry currents must transform as $(\hf,\hf)$
under $SU(2)\times SU(2)$.
The only known algebra with these properties is the large $\CN=4$
superconformal algebra of \STVP\
with generators
\eqn\lnfourgens{
  T\ ;\quad G^a\ ;\quad A_+^i\ ,\quad A_-^i\ ,\quad U\ ;\quad Q^a
}
where $a=0,1,2,3$ and $i=1,2,3$.  The currents $A_+$, $A_-$, and $U$
are dual to the gauge fields $\CA^+$, $\CA^-$, and $A$,
respectively.  The supersymmetry generated by $G^a$ relates
the $U(1)$ current $U$ to a set of four free fermions $Q^a$,
which are thus required for completion of the algebra.

We will describe this large $\CN=4$ algebra
in more detail below in section 4.
However, at this point we wish to point out a surprise:
In contrast to other supergravity backgrounds
based on $AdS_3$, the large $\CN=4$ superalgebra
does {\it not} have a realization as an Chern-Simons-type $AdS_3$
supergravity --- at least not an obvious one.%
\foot{Thus providing another reason why \csresult\
is not the whole story when we wish to compare
the spacetime CFT with the effective supergravity theory.}
Extended $AdS_3$ supergravities can be written as Chern-Simons
theories \AchucarroGM\ with gauged supergroup containing 
$SL(2,R)_{L,R}$ factors
for the isometries of $AdS_3$, as well as
factors for the gauged $R$-symmetry
(in this case
$(SU(2)_+\times SU(2)_-)_{L,R}$).
The unique supergroup with this bosonic subalgebra
and fermionic generators transforming as $(\hf,\hf)$
is the supergroup $D(2,1|\alpha)$.
In $AdS_3$ supergravities with $\CN=0,1,2,3$ or
small $\CN=4$ supersymmetry, there is a Chern-Simons
action using the super-isometry group $\CI$;
the superconformal algebra of the spacetime CFT
is a Hamiltonian reduction of the affinization $\hat\CI$
imprinted on the boundary by Chern-Simons
gauge transformations.
However, the symmetry generators \lnfourgens\
are {\it not} a Hamiltonian reduction of the currents of
affine $D(2,1|\alpha)$
(although \lnfourgens\ contains $D(2,1|\alpha)$ as a subalgebra).
\foot{We should note, however, that ref.   \IMadsen\ shows
 that the Hamiltonian reduction of the affinization 
of $D(2,1|\alpha)$ leads to $\tilde \CA_{\gamma}$.}

In fact, it is easy to see that there is no finite
dimensional superalgebra that could serve as
the basis for an $AdS_3$ supergravity corresponding
to the large $\CN=4$ algebra.  Such an algebra would
have to contain the $D(2,1|\alpha)$ subalgebra
generated by $L_{\pm1}$, $L_0$, $G_{\pm1/2}^a$,
and $A_{0}^{\pm,i}$.  Adding the zero mode of
the $U(1)$ current, $U_0$, then requires us
to add the fermion modes $Q_{-1/2}^a$ by supersymmetry.
But then the anticommutator $\{Q_{-1/2}^a,G_{-1/2}^b\}$
includes $A_{-1}^{\pm,i}$, and so on --
we end up generating the entire large $\CN=4$ algebra.

\subsec{Moduli}

In this subsection we analyze the moduli of the solution. It turns
out to be simplest to analyze the pure NS form of the solution.
Since the action is even in RR fields, the linearized equations of
motion which determine the number of massless moduli do not mix RR
and NS fluctuations. Hence the two possible types of moduli can be
analyzed separately.

We begin with the NS fluctuations.

\item{1.} {\it The metric}. The only possible scalar fluctuations
of the metric are parameterized as
\eqn\diibmet{ ds^2 = \frac{\l^2}{x_2^{~2}}
\Bigl(-dt^2 + (dx_1)^2 + (dx_2)^2\Bigr)
    + {Q_5^+\over 4 \pi^2} ds^2(\S^3_+)
    + {Q_5^-\over 4 \pi^2} ds^2(\S^3_-)
    + (L+ \delta L(x))^2 (d\theta)^2 }
where $\delta L(x)$ is a scalar depending only on the $AdS_3$
coordinates. It is clear from the construction of the solution as
a worldsheet CFT that the $\S^3$ and $AdS_3$ radii cannot be moduli
because they appear as levels of WZW models.
%
\item{2.} {\it The dilaton}. We also get a scalar   $\phi(x)$,
whose zero mode we have absorbed into the string coupling $g_B$.
However we have already seen in equation \lengths\ or \lngths\
that this is not a separate modulus, but rather is fixed in terms
of the $\S^1$ radius and the charges.  A direct Kaluza-Klein
reduction reveals a mass for fluctuations which change
the sizes of the $\S^1$ and $\S^3$ radii
\lengths\ or \lngths.
\item{3.} {\it The NS $B$-field}. Again
there are no possible moduli here as the $H$ fluxes are quantized.

\bigskip
\bigskip 
Now we consider possible RR moduli in the IIA context
in order to avoid subtleties related to the self-duality
of the RR four-form $C_4$ in the IIB description.
The RR equations of motion following from \iialag\ are
\eqn\iiaeomi{ d*dC_1 + H \wedge * (dC_3- H \wedge C_1) =0 }
\eqn\iiaeomii{ d*(dC_3- H \wedge C_1) - dC_3 \wedge H  =0\ . }
Our ansatz is
\eqn\ceeone{ C_1 = c_1 + \sigma d\theta }
where $c_1$ is a 1-form on $AdS_3$ and $\sigma$ is a scalar on
$AdS_3$. Even though it is a gauge field and not a scalar we include
$c_1$ at this point because it eats one of the scalars. We also take
\eqn\ceeetrhee{ C_3 = \alpha_+(x) \omega_+ + \alpha_-(x) \omega_-
 }
where $\alpha_\pm$ are scalars. Because of large $C_3$ field gauge
transformations, they are periodic scalars (more on this below).
%
%
%
%

Choosing the orientation to be
$\omega_0 \wedge \omega_+ \wedge \omega_- \wedge d\theta$ we obtain
\eqn\iiaeomia{ \eqalign{ &\nabla^2 \sigma  = 0 \cr & d*_3 d c_1
+  \lambda_+ *_3 (d\alpha_+ + \lambda_+ c_1) + \lambda_- *_3
(d\alpha_- + \lambda_- c_1) =0 } }
from \iiaeomi. Here $\nabla^2 \sigma= *d*d\sigma/\omega_0$. We
also get
\eqn\iiaeocmiia{ \eqalign{ & d*_3(d\alpha_+ + \lambda_+ c_1) =0 \cr
& d*_3(d\alpha_- + \lambda_- c_1)  =0 \cr
&d(\lambda_0 \sigma)
+L (\lambda_- d\alpha_+ - \lambda_+ d\alpha_-) =0 \cr} }
from \iiaeomii. The third equation of \iiaeocmiia\  freezes one
linear combination of $\alpha_\pm$ to equal $\sigma$.  We also
recognize the other linear combination as the Goldstone boson
eaten by $c_1$. The remaining scalar $\sigma$ has mass
\eqn\plms{ m^2_\sigma = 0\ . }
Hence there is one massless modulus in the RR sector.


\subsec{Effects of the second modulus}

Now, let us consider effects associated with the second modulus.
In type IIB setup with NS background,
it corresponds to a combination of the axion field $C_0$
and the 4-form field.
Expanding the 4-form field $C_4$ in terms of \volforms,
\eqn\cfourexp{
  C_4 = (\alpha_+\omega_+ + \alpha_-\omega_-)\wedge d\theta\ ,
}
the charge quantization conditions in the RR sector
\eqn\rrcharges{\eqalign{
& \int_{\S^3_+ \times \S^3_- \times \S^1} \bigl[
* \left( C_0 H - F_3 \right) - H \wedge C_4\bigr] = 0 \cr
& \int_{\S^3_{\pm}} F_3 = 0
}}
are solved by
\eqn\cfldmod{
  \lambda_+\alpha_- - \lambda_-\alpha_+ = \lambda_0 L C_0\ .
}
{}From \iiaeomia, we learn that the linear
combination
$\lambda_+ \alpha_+ +  \lambda_- \alpha_-$
is proportional to a Goldstone mode for the
gauge field $c_1$ for the IIA theory,  (and $\int_{S^1} C_2$
for the IIB theory).
The orthogonal combination $-\lambda_- \alpha_+ + \lambda_+ \alpha_-$
is a modulus. From \cfldmod\ when  $C_0$ is turned on we must also turn on the
RR potential $C_4$. If we set  the Goldstone mode to zero
then we may write
\eqn\cfldmodnew{
C_4 = {C_0 L \over \lambda_0} \left( - \lambda_- \omega_+ \wedge d\theta
+ \lambda_+ \omega_- \wedge d\theta \right)
}

The only equation of motion that gets modified
in the background \cfldmod\ is the Einstein
equation. Now, instead of \einst, it gives: 
\eqn\einstnew{\eqalign{
\ell^{-2}& = {1\over 4}   \lambda_0^2 \left( 1 + (g_B C_0)^2 \right) \cr
R_{\pm}^{-2} & = {1\over 4}
\lambda_{\pm}^2 \left( 1 + (g_B C_0)^2 \right)
\ .}}
Evaluating the NS5-brane charges, {\it cf.} \normflx,
\eqn\nsfchrg{\int_{\S^3_{\pm}} H = Q_5^{\pm} }
we find a relation between $Q_5^{\pm}$ and $R_{\pm}$
$$
Q_5^{\pm} = 2\pi^2 \la_{\pm} R_{\pm}^3
$$
which together with \einstnew\ yields 
\eqn\rrpm{
R_{\pm} = {1 \over 2\pi} \sqrt{Q_5^{\pm}}
\left( 1 + (g_B C_0)^2 \right)^{1/4}\ .
}
Since the background $H$-flux \hthree\ is still defined
so that \taux\ holds, we can use this equation to find the AdS radius, 
\eqn\adssize{
\ell = {1 \over 2\pi} \sqrt{ {Q_5^+ Q_5^- \over Q_5^+ + Q_5^-} }
\left( 1 + (g_B C_0)^2 \right)^{1/4}\ .
}

Finally, from the fundamental string charge quantization condition
%
\eqn\fundq{
 \int * (\vert \tau_B \vert^2 H - C_0 F_3) = Q_1 \in \IZ
}
we find
\eqn\eelesnew{ 4 \pi^4 R_+^3 R_-^3 L \la_0 |\tau_B|^2 = Q_1 .}
Here $\tau_B$ is the complexified type IIB coupling 
\eqn\iibtau{
\tau_B =C_0 + {ie^{-\phi}\over g_B}\ }
evaluated at $\phi=0$.
This relation can be used to solve for the size of the $\S^1$.
Thus, substituting \einstnew, \rrpm, and \adssize, we find 
\eqn\newl{
L = Q_1 {4 \pi g_B^2 \over Q_5^+ Q_5^- \sqrt{Q_5^+ + Q_5^-} }
\left( 1 + (g_B C_0)^2 \right)^{-7/4}
}
which is similar to the previous expression, except for the last factor.

To summarize, turning on the second (axion) modulus in our NS
background modifies the expressions \lengths\ for the radii in the
following way 
\eqn\newlengths{\eqalign{
\ell & = { 1\over 2\pi }
\sqrt{Q_5^+Q_5^-\over Q_5^+ + Q_5^-}
    \left( 1 + (g_B C_0)^2 \right)^{1/4} \cr
R_\pm & = { 1\over 2\pi }\sqrt{Q_5^\pm}
    \left( 1 + (g_B C_0)^2 \right)^{1/4} \cr
L & = {4 \pi  g_B^2 Q_1 \over Q_5^+ Q_5^- \sqrt{Q_5^+ + Q_5^-}}
    \left( 1 + (g_B C_0)^2 \right)^{-7/4}\ .
}}
These modifications leave the Brown-Henneaux central
charge \bhcf\ unchanged.
It will be useful for later purposes to note
that
\eqn\strfct{
 1 + (g_B C_0)^2 = \left( {\vert \tau_B \vert \over \Im \tau_B } \right)^2.
}
%


\subsec{Moduli space metric}

The metric on the moduli space is most easily computed for the
case of RR charges, which may be obtained from \newlengths\ by
S-duality. Quite generally, under $SL(2,\IZ)$ transformations we
have
\eqn\trmsnrl{
\eqalign{
\ell' & = \ell \left( {\Im \tau_B \over \Im \tau_B' } \right)^{1/4}\cr
R_\pm' & = R_\pm \left( {\Im \tau_B \over \Im \tau_B' } \right)^{1/4} \cr
L' & = L \left( {\Im \tau_B \over \Im \tau_B' } \right)^{1/4}
}}
It is convenient to write the answer in terms of $g_B Q$
which is held fixed in the classical limit
of the RR background. One finds the simple
$C_0$-independent expressions 
\eqn\newlgths{\eqalign{ \ell & = { 1\over 2\pi }
\sqrt{g_B Q_5^+ g_B Q_5^-\over g_B Q_5^+ + g_B Q_5^-}  \cr
R_\pm & = { 1\over 2\pi }\sqrt{g_B Q_5^\pm}  \cr
L & = {4\pi g_B Q_1 \over  g_B Q_5^+ g_B Q_5^-
\sqrt{g_B Q_5^+ + g_B Q_5^-}} \ .}}

The moduli space metric follows from the kinetic terms of the
three-dimensional low-energy effective action.
These in turn descend from the ten-dimensional kinetic terms
in \iiblag\foot{In the ten-dimensional Einstein frame,
the first three terms in this action can be written as
${2\pi \over g_B^2} \int d^{10}x\sqrt{-g}
\bigl(\CR-{d\tau_B d\bar\tau_B \over 2(\Im \tau_B)^2}\bigr)$.
Also, the contribution of the 4-form field $C_4$ is best described
in the T-dual type IIA theory, which automatically
avoids subtleties related to self-duality.} 
\eqn\tensd{\int d^{10}x\sqrt{-g}\Bigl({2\pi e^{-2\phi} \over g_B^2}
\bigl(\CR + 4 (\nabla \phi)^2\bigr) -\pi(d C_0)^2
-{\pi\over 5!}(d C_4)^2\Bigr)\ ,}
with the metric ansatz
\eqn\iibmt{ ds^2 =e^{\phi(x)}g^{(3)}_{\mu\nu}dx^\mu dx^\nu
+e^{\phi(x)} R_+^2 ds^2(\S^3_+)
+e^{\phi(x)} R_-^2 ds^2(\S^3_-) + e^{-3\phi(x)}L^2
(d\theta)^2 }
corresponding to the modulus generated by taking the coupling
$g_B \to g_B e^{\phi(x)}$ in \newlgths\
(together with a Weyl rescaling of $g^{(3)}$). 
Using the S-dual of relation \cfldmod\ 
to express $C_4$ in terms of $C_0$ one finds, 
after some computation, the three-dimensional effective action
\eqn\thnsd{{2\pi V \over g_B^2}\int
d^{3}x\sqrt{-g^{(3)}}\bigl(\CR^{3}-4(\nabla \phi)^2
- g_B^2 e^{2 \phi}(\nabla C_0)^2\bigr)}
where $V=4\pi^4 R_+^3 R_-^3 L$ is the internal volume.
{}From this we can read off the moduli space metric 
\eqn\dxc{ds^2={d \tau d\bar{\tau}\over
(\Im \tau)^2}, ~~~~\tau=C_0 + {2ie^{-\phi} \over g_B} }
which is the hyperbolic metric on the upper half plane.
Note that the $\tau$ in \dxc\ is not the same as the ten-dimensional
coupling $\tau_B$.


\subsec{Speculations on the global structure of the moduli space}

We now consider the RR gauge transformations which preserve the
IIB solution with NS-sector fluxes described in section 2.6,
in equations \cfourexp - \strfct. The unbroken gauge group is
generated by three types of transformations. First, there
are $SL(2,\IZ)$ transformations
\eqn\essdul{
\tau_B' =  {a \tau_B + b \over c \tau_B + d}
\qquad\qquad \pmatrix{F_3'\cr H_3'\cr} =
\pmatrix{a & b \cr c & d \cr} \pmatrix{F_3\cr H_3\cr}
}
leaving $C_4$ invariant. The background 3-fluxes break the S-duality group down to the group
of transformations:
\eqn\ubrnkd{
\pmatrix{1 & 0 \cr m & 1 \cr}
}
with $m\in \IZ$.

Next there are small $RR$ gauge transformations.
Of these the only significant ones are the shifts of $C_2$ by exact forms.
Only the combination $R_5 = dC_4 - C_2 H$ is gauge invariant, so in the
presence of $H$-flux small $C_2$ gauge transformations acting by
\eqn\smgt{
C_2 \to C_2 + d (\chi d \theta)
}
must be accompanied by
\eqn\smallgt{
\eqalign{
\alpha_+ & \to \alpha_+ - \lambda_+ \chi \cr
\alpha_- & \to \alpha_- - \lambda_- \chi \cr}
}
Note that $\chi\in \IR$ is an arbitrary {\it real number}.
We define the Goldstone mode to be
\eqn\goldstone{
\phi_{GB} = {\lambda_+ \alpha_+ + \la_- \a_- \over \la_0^2}
}
Then \smallgt\  shifts $\phi_{GB} \to \phi_{GB} - \chi$, but leaves
$\lambda_+ \alpha_- - \lambda_- \alpha_+$ and $C_0$ invariant.
Finally there are large $C$ field gauge transformations.
These act by
\eqn\alphper{
\eqalign{
\alpha_+ & \to \alpha_+  + {\lambda_+ \over Q_5^+} n_+ \cr
\alpha_- & \to \alpha_-  + {\lambda_- \over Q_5^-} n_- \cr
C_0 & \to C_0 \cr}
}
where $n_\pm \in \IZ$ are independent integers.
Defining
\eqn\cfour{
C_4^\pm :=
\int_{S_\pm^3\times S^1}  C_4
}
they act by
\eqn\shftper{
 C_4^\pm \to  C_4^\pm + n_\pm
}
Note that these transformations are not all independent. For example,
some transformations of the type \alphper\ are in fact of the form \smallgt.

The transformations \ubrnkd\smallgt\alphper\ generate a commutative group
of unbroken gauge transformations. However, we must consider the subgroup
of transformations which preserve the condition \cfldmod.
The transformation of lengths under $SL(2,\IZ)$ \trmsnrl\ shows that
\eqn\trmsnrlp{
\eqalign{
\lambda_\pm'
& = \lambda_\pm  \left( {\Im \tau_B'\over \Im \tau_B} \right)^{3/4}\cr
\a_\pm'  & = \a_\pm \left( {\Im \tau_B'\over \Im \tau_B} \right)^{3/4} \cr}
}
and hence the subgroup of transformations preserving \cfldmod\ is
determined from
\eqn\subpp{
\la_+ ( \alpha_-  + {\lambda_- \over Q_5^-} n_-) -
\la_- ( \alpha_+   + {\lambda_+ \over Q_5^+} n_+)
= \lambda_0 L \left( {\Im \tau_B \over \Im \tau_B'} \right) C_0'
}
A little bit of algebra reveals that this is true iff
\eqn\finsub{
Q_5^+ n_- - Q_5^- n_+ = m Q_1 \ . }
Let us now introduce  $d:= gcd(Q_5^+, Q_5^-)$
and $Q_5^\pm := d \hat Q_5^\pm$. Moreover, we
make the 1-1 invertible change of variables:
\eqn\newenn{
\pmatrix{n_+ \cr n_-\cr} = \pmatrix{S_+ & \hat Q_5^+ \cr S_-& \hat Q_5^-\cr}
\pmatrix{\tilde n_+ \cr \tilde n_-\cr}
}
where $S_\pm$ are integers with  $S_- Q_5^+ - S_+ Q_5^- = d$.
The parameter $\tilde n^-$ is equivalent to a small gauge transformation
\smallgt\ and hence can be dropped. If we fix the gauge by
setting $\phi_{GB}=0$ then the unbroken symmetry group is $\IZ$,
generated by $(\tilde n_+=1, \tilde n_- =0)$ (which must be
accompanied by a small gauge transformation $\chi$ to preserve
the gauge condition $\phi_{GB}=0$).

The resulting unbroken gauge transformations are {\it much} more simply
expressed in the $S$-dual background related by $\tau_B \to -1/\tau_B$.
In this background \ubrnkd\ is mapped to the usual RR shift symmetry
$C_0 \to C_0 -m$. Henceforth we shall work in this $S$-dual picture. 
In this picture the unbroken gauge group has a generator acting by
\eqn\rrshift{
\eqalign{
C_0 &\to C_0 - d \cr
C_4^+ &\to C_4^+ + S_+ Q_1 \cr
C_4^- &\to C_4^- + S_- Q_1 \ ,
}}

The equivalence relation \rrshift\ is very much analogous
to an identification of
the moduli space in the D1-D5 system
\LarsenUK.
Since we parametrize the moduli space by $\tau$, equation \dxc,
we say that the moduli space is identified under
shifts $\Re \tau \to \Re \tau + d$.  Note that $d$ depends on
the arithmetic of $Q_5^\pm$. If $Q_5^+, Q_5^-$ are
relatively prime, or have small common divisors, then the
identification is by a distance of order $g_B$. On
the other hand,   if $Q_5^+ = Q_5^-$ then
$d= Q_5^+$. In the scaling required for the supergravity
limit this shift is very large, of order $1/g_B$, and the 
distance on moduli space is order $1$. 
Such shifts mix up all orders of string perturbation theory,
and our supergravity analysis cannot reliably conclude that
\rrshift\ is a symmetry of the exact theory.%
\foot{This is why the title of our section contains the word ``speculations.''}
In principle
it could be spoiled  by D3 instantons, for example.
Nevertheless we proceed in the rest of this subsection
under the assumption that supergravity
is indeed a reliable guide in this case.

Apart from this RR shift symmetry, the U-duality
group is generated by various `inversion' transformations:
\item
1) T-duality, which sends $L \to 1/L$, $g \to g/L$, and interchanges IIA/B;
\item
2) S-duality in IIB, which sends $g \to 1/g$, $L \to L/\sqrt g$,
and interchanges NS and RR backgrounds;
\item
3) `9/11 flip', which sends
$L \to \sqrt{L g}$, $g \to L^{3/2}/g^{1/2}$ in type IIA.

\noindent
One easily checks that $TFT=S$, so essentially there is just T-duality and
S-duality -- the flip is just the image of S-duality in type IIA.
Clearly T-duality interchanges
momentum and fundamental string winding on the circle
in the NS background, but this interchanges IIA/B.
In the IIB D-brane background,
the equivalent operation is STF, which again interchanges IIA/B.
Hence there is no inversion automorphism of the theory leaving the
background charges fixed -- the only candidate is
S-duality, and that interchanges the RR and NS descriptions
of the background.

The only identification of the moduli space is thus
the RR shift symmetry \rrshift.  If we parametrize
the modulus by $\tau$, equation \dxc,
we find the fundamental domain of the moduli space is
the strip in the upper half plane with
$\Re\,\tau\in(-\hf d ,\hf d )$.

The restricted scope of the U-duality group is quite
different from the situation in the D1-D5 system
on $\MM=K3$ or $T^4$.  There it was found that all supergravities
with the same value of the product $N=Q_1Q_5$ of
background charges were located in different
cusps of the moduli space \refs{\SeibergXZ,\LarsenUK}.
In particular, this allowed the symmetric product
orbifold $\sym^N(\MM)$, which was naturally associated
to the background with $Q_5=1$, to be continuously
connected to all other backgrounds with the same central charge.
In the present case, all distinct sets of charges
$(Q_1,Q_5^+,Q_5^-)$ lead to distinct, disconnected
moduli spaces of theories.  This leads to the
possibility that symmetric product orbifolds
will only lie on a subset of these moduli spaces
of theories; for instance, they might not describe
both $(pQ_1,Q_5,Q'_5)$ and $(Q_1,pQ_5,pQ'_5)$,
which are backgrounds having the same central charge
in the spacetime CFT but lying on disconnected
moduli spaces.


\subsec{Regions of the moduli space}

We now turn to a discussion of the various regions of the moduli space.
Different low-energy descriptions are appropriate
in different regions.
To fix notation, let us refer all quantities
to the IIB RR background, via the appropriate dualities.
In that frame, the moduli space is the strip in the UHP $|\Re\tau|<\hf d$.
Weakly coupled IIB string theory is appropriate as
we move up into the cusp of the moduli space at large $\Im\tau$.
The cycle sizes and curvature radii are not too small provided
$\ell,R_\pm,L>1$ in string units;
for instance we want $g_B Q_5^\pm>1$.
If this is not true, then
we are in a regime of weak coupling of the dual CFT
(just like $g_BQ_3 < 1$ is weak coupling for $\CN=4$
super Yang-Mills),
and the geometrical interpretation of the target breaks down.
Thus, the region far up in the cusp
is the perturbative regime of the spacetime CFT.

{}From equations \newlgths\ and their various duals,
\eg\ \newlengths,
we have the following criteria to impose:
\item{1)}
If $g_B>1$ we should S-dualize to the NSB description.
\item{2)}
If $L<1$ we should T-dualize the $\S^1$ to type IIA.
This will be the F1-NS5A-NS5A' background
if we arrive from the NSB description,
otherwise we arrive from RRB and get D2-D4-D4'.
Referred to the RRB background, the condition to T-dualize is
\eqn\tdualcond{
g_B^{3/2} > {Q_1 \over Q_5^+ Q_5^-\sqrt{Q_5^+ + Q_5^-}}
}
\item{3)}
If the IIA coupling becomes strong, we go to M-theory
with the charges M2-M5-M5'.
Referred back to the RRB frame, the condition is
\eqn\mcond{
g_B > {Q_1 \over Q_5^+ Q_5^-\sqrt{Q_5^+ + Q_5^-}}
}
(note that the RHS is the same as in \tdualcond).

\noindent
Note that the natural boundary at $\Im\tau=0$ is arrived at from
an effective M-theory description.
In the M-theory description, the RR axion has
transformed into the shear of the $T^2$ comprised of the
$\S^1$ and the M-theory circle.

To summarize: The cusp region is the weakly coupled dual CFT.
Coming down from the cusp, we encounter RRB sugra.
Then, depending on whether (1) or (2) is satisfied first,
we go to either
(a) NSB supergravity, then NSA after T-duality, then M-theory; or
(b) RRA supergravity by T-duality, then M-theory.
A sketch of the first possibility is given in figure 1.

\bigskip
{\vbox{{\epsfxsize=2.5in
        \nobreak
    \centerline{\epsfbox{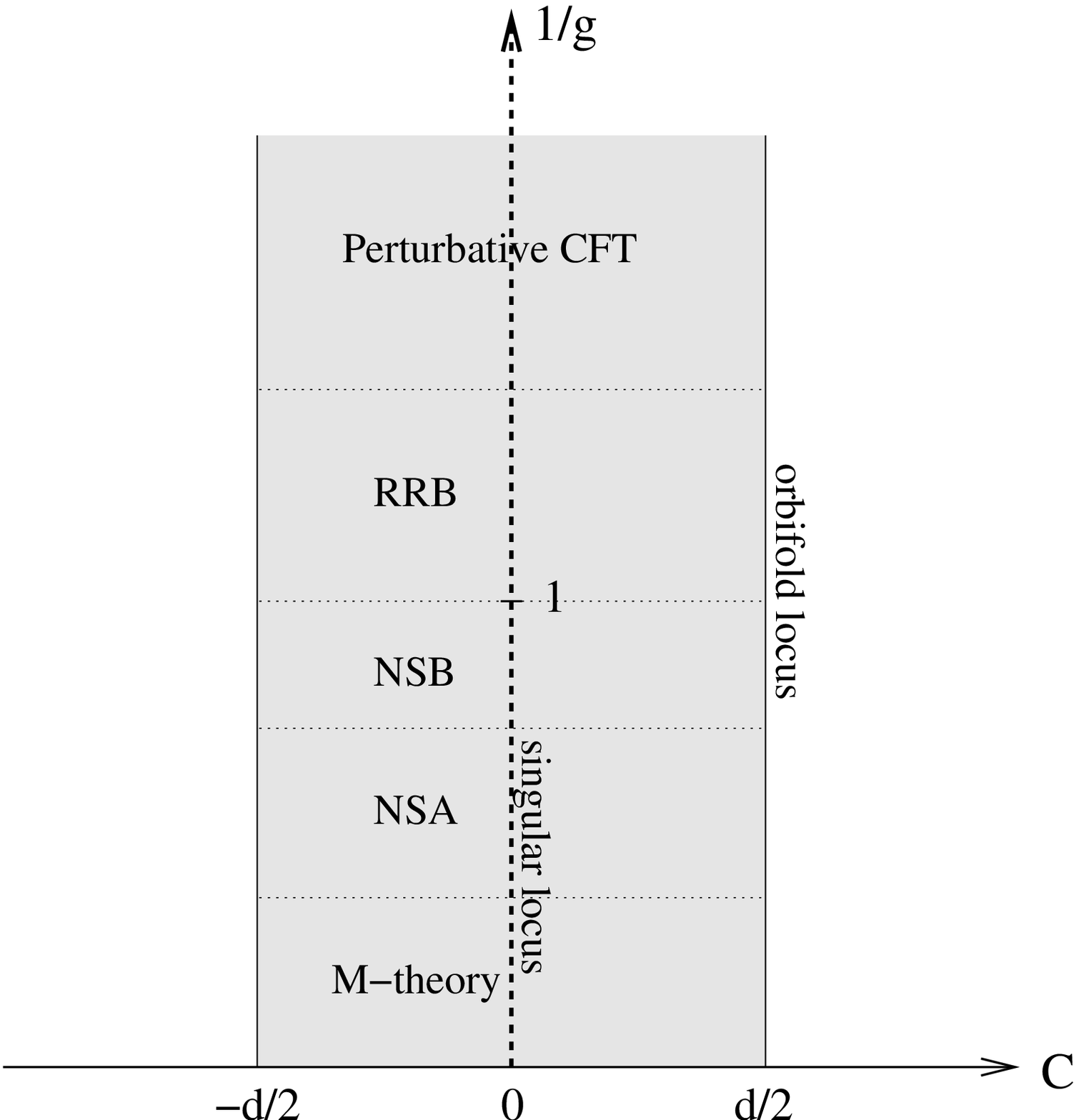}}
        \nobreak\bigskip
    {\raggedright\it \vbox{
{\bf Figure \modspace.}
{\it
Sketch of the moduli space, and the regimes in
which various low energy descriptions are valid.
The dashed line is the singular locus, where a long string continuum appears
(see section 2.10).
The line $C=d/2$ is 
argued to be the location of the  symmetric product orbifolds.
} }}}}
\bigskip}

The region of RRB supergravity can be vanishingly small if \eg\
$Q_5^\pm$ are small.  However, there will be a regime
described by the perturbative string formalism
of \refs{\GiveonNS,\ElitzurMM,\GiveonP} for $g_B>1$
(recall we are referring everything to the RR duality frame),
although it may only involve type IIA.

The dual CFT is typically perturbative up in the cusp.
In the D1-D5 and related systems, the dual CFT had
a description as a symmetric orbifold along a line
$\Re\tau=\hf$ in a duality frame where the background
charges were RR and $Q_5=1$ \LarsenUK.
Below we will argue that similarly, there is a
symmetric product orbifold when one of the
fivebrane charges is one, say $Q_5^+=1$;
and then the orbifold line is $\Re\tau=\hf$.

The perturbative string description of \GiveonNS\
is only equipped to handle backgrounds with
vanishing RR potentials, \ie\ $\Re\tau=0$.
The spacetime CFT is actually singular on this
subspace of the moduli space.
We now turn to a discussion of this phenomenon.


\subsec{Long strings and singular CFT's}

A common feature of conformal field theories dual to $AdS_3$
string backgrounds is that, in certain regions of the moduli
space, they exhibit a continuum of states above a gap $\Delta_0$.
The continuum is associated to the appearance
of a new branch of the configuration space
where the onebrane-fivebrane ensemble can
fragment into separate pieces \refs{\fragm,\GiveonNS,\SeibergXZ}.
The new branches of the configuration space
describe separating clusters of onebranes and fivebranes,
often called `long strings', since they are codimension one
objects in $AdS_3$ whose proper length grows to infinity
as they approach the $AdS_3$ boundary.

Typically one thinks of the spacetime CFT dual to
$AdS_3$ as the Higgs branch of the onebrane-fivebrane system,
where the onebranes are dissolved in the fivebranes
as finite-size instantons.  For instance, in the D1-D5 system,
the CFT is the sigma model on the moduli space of
instantons on $T^4$ or $K3$ (see \eg\
\refs{\DijkgraafGF,\SeibergXZ,\LarsenUK,\LarsenDH,\emillectures} 
 for  reviews and further references).
In this dual CFT, the new branches of the configuration space
are Coulomb branches, where some number of instantons
shrink to zero size \SeibergXZ.
Whereas the instanton of
non-zero size gives a potential for the coordinates
of the dissolved string in the directions transverse
to the fivebrane, the zero-size instanton string
allows these fields to turn on, so that the string moves
away from the fivebrane background (out to the boundary
of $AdS_3$).

In either description, long strings or zero-size instantons,
the appearance of the continuum results in singularities in
correlation functions.  The singularity arises only
when strings can become infinitely long at finite
energy cost (or correspondingly instantons can shrink
to exactly zero size).  This is not allowed at generic
points on the moduli space, typically only when the RR potentials
vanish.

In the NS duality frame, the tension of a long string
receives compensating contributions from the tension
of a fundmental string (determined \eg\ from the
Nambu-Goto action) and from the background $B$ field
(the Wess-Zumino term in the $AdS_3=SL(2,R)$ sigma model).
We choose global coordinates for $AdS_3$:
\eqn\adsgl{
ds^2 = \ell^2 (-\cosh^2\!\rho \,dt^2 + \sinh^2\!\rho \,d\phi^2 + d \rho^2)
}
and we choose the duality frame with NS fluxes turned on so that
\eqn\hef{
H = \lambda_0 \omega_0 = \half \lambda_0 \ell^3 \sinh(2\rho)
    dt \wedge d\phi \wedge d \rho\ .
}

Consider a string at fixed $\rho$, with a worldsheet
that spans a time $\Delta t$.
The action consists of two pieces
\eqn\action{
S = S_{\rm NG} - S_{W\rm Z} =
    2 \pi \int \!\sqrt{-h} - 2\pi \int\! B\ ;
}
we define $q$ (following \fragm)
to be the ratio $S_{\rm WZ}/S_{\rm NG}$ as $\rho \to \infty$.
Thus, $(1-q)$ measures the coefficient of a ``cosmological term''
$\sim e^{2\rho}$ giving the energy cost per unit proper length
of the string.   When $q\ne 1$, it costs infinite energy
to take the string to the boundary of $AdS_3$
and so it is effectively bound to the system.
There is no continuum in the spectrum.

In our conventions,
\eqn\sng{
S_{\rm NG} = 2 \pi^2 \ell^2 \,\sinh( 2\rho) \;\Delta t \ .
}
Next we choose a gauge $B= {1\over 4} \lambda_0 \ell^3 \cosh 2\rho dt d\phi$,
so that
\eqn\swz{
S_{\rm WZ} = {1\over 4} \lambda_0 \ell^3 (2\pi)^2 \cosh(2\rho) \;\Delta t \ .
}
Using the Einstein equation \einstnew\ we compute 
\eqn\qufs{
q = \lim {S_{\rm WZ} \over S_{\rm NG} } =
\half \lambda_0 \ell = {1\over g_B \vert \tau_B \vert}
= {1\over \sqrt{ 1+ (g_B C_0)^2 } }
}
We thus conclude that the singular locus on moduli space is at $C_0=0$,
the positive imaginary axis for $\tau$.

The energy cost of a long string
is related to the change of the central charge \cexpl\
resulting from pulling it completely out
of the background.
The difference in the ground state (Casimir) energies is given
by the change in the central charge as $\delta h = -\delta c/24$.
One finds the gap $\Delta_0$ to the continuum of long string states
\eqn\fstrgap{
  \Delta_0=-\delta c/24=\frac{Q_5^+ Q_5^-}{4(Q_5^++Q_5^-)}
}
associated to pulling out a onebrane.

Long strings can also carry non-zero fivebrane
charge -- strings in $AdS_3$ can be obtained by wrapping
fivebranes over $\S_+^3 \times \S^1$ or  $\S_-^3 \times \S^1$.
We have computed $q$ for these strings. This is a more difficult
computation, but it does indicate that sometimes these strings   can
produce singularities, again on the locus of vanishing $C_0$.
As the details would take us somewhat far afield 
we do not include them here.


\newsec{Relation to intersecting D-branes}

Anti-de~Sitter backgrounds are often realized as near-horizon
limits of the geometry surrounding intersecting brane sources.
In this section we study the geometries surrounding
intersecting brane configurations that are expected to
exhibit large $\CN=4$ supersymmetry in their infrared dynamics.
While we have not found a brane configuration with
all the desired properties, we discuss three different ones
which exhibit different aspects of the dynamics:
\item{1.}
The collection of branes in $\IR^{1,9}$
\eqn\bkgdconfig{\matrix{
 Q_1 & {\rm ~~D1~branes~along~~} &\quad x^0, ~x^5 \hfill\cr
 Q_5^+& {\rm ~~D5~branes~along~~} & \quad x^0,~x^5,~x^6, \ldots,~x^9 \hfill\cr
 Q_5^-& {\rm ~~D5'~branes~along~~} & \quad x^0,~x^5,~x^1, \ldots, ~x^4
\hfill
}}
preserves 1/8 supersymmetry, and has near-horizon geometry
\eqn\adsssr{
AdS_3 \times \S^3 \times \S^3 \times \R\ .
}
There are thus strong reasons to believe that the IR theory has
large $\CN=4$ supersymmetry.  However, what sort of dynamics
describes the intersecting fivebranes is not understood,
and it is not clear how to implement a
compactification of $\R$ to $\S^1$.
\item{2.}
For $Q_5^+=Q_5^-$, the locus of fivebrane sources above can be deformed
to a special Lagrangian 4-manifold $\CM\subset\R^8$;
the two sets of intersecting branes deforms to a single
set of branes, much as in \WittenSC.
The $SO(4)\times SO(4)$ symmetry of the branes in (1)
is broken to the diagonal $SO(4)$.  The near-horizon geometry
is still \adsssr, but the fivebrane dynamics in the IR
appears to be a more conventional $U(Q_5)$ gauge theory;
the addition of onebranes dissolved in the fivebranes
should be described as the Higgs branch
of the corresponding D1-D5 system.
\item{3.}
While this second configuration points toward the
appropriate IR dynamics (for $Q_5^+=Q_5^-$),
in the near-horizon geometry
the fivebranes are wrapping $\S^3\times \R$ and not $\S^3\times\S^1$.
To find the latter, we can change the setup somewhat
and consider $Q_5^+$ fivebranes wrapping a special Lagrangian $\S^3$
supported by $Q_5^-$ units of three-form flux.  The remaining
directions on these branes can be taken to be $\R^{1,1}\times\S^1$.
As we will see below, there are good reasons to expect that
with onebranes along $\R^{1,1}$ included, the geometry in the infrared
flows to a theory with large $\CN=4$ supersymmetry;
and at the same time, the dynamics is that of the onebranes
dissolved in the fivebranes -- a sigma model on the moduli
space of instantons on $\S^3\times\S^1$.

\noindent
We will now describe each of these brane configurations in more detail.

First consider the configuration \bkgdconfig\ of flat branes intersecting
in flat spacetime.
Each of these D-branes is invariant under half of the supersymmetries,
and altogether the D-branes preserve only ${1 \over 8} \times 32 = 4$
supersymmetries. Hence, the two-dimensional field theory
on the D1-branes has $\CN=(0,4)$ supersymmetry.
The D-brane configuration \bkgdconfig\ breaks the Lorentz
group $SO(1,9)$ to the subgroup
\eqn\lorentz{
SO(1,1)_{05} \times \left[ SU(2)_L \times SU(2)_R \right]_{1234}
\times \left[ SU(2)_L \times SU(2)_R \right]_{6789}\ .
}
The $SU(2)$ factors in this symmetry group play the role of
the $R$-symmetry in the effective two-dimensional field theory
on the intersection.

The $05$ field theory on the world-volume of intersecting D-branes
is composed of $11$, $15$, $15'$ and $55'$ string states,
which form complete representations under the unbroken symmetry group \lorentz.
Among various states, the $15$ and $15'$ strings are in a chiral
representation of the rotational $R$-symmetry
and contribute to an $R$-charge anomaly.
Since the $15$ ($15'$) fermions are invariant under
$[SU(2)_L \times SU(2)_R]_{1234}$
(respectively $[SU(2)_L \times SU(2)_R]_{6789}$),
the computation of this contribution to the anomaly
is exactly as in the standard D1-D5 system.
Specifically, one has
\eqn\fcv{
  k^+_L =Q_1Q_5^+\ ,\quad k^+_R =-Q_1Q_5^+\ ,\quad
  k^-_L =Q_1Q_5^-\ ,\quad k^-_R =-Q_1Q_5^-\ .}
At the IR fixed point,
the theory must have $\CN=(0,4)$ supersymmetry;
$(0,4)$ supersymmetry implies at least one of the
$SU(2)$ $R$-symmetries must become an
$SU(2)$ current algebra.  However both right-moving
$SU(2)$ $R$-symmetries are on the same footing.
In other words, we have at
least the large $\CN=4$ supersymmetry algebra on the right.
The large $\CN=4$ supersymmetry algebra
has the four supercharges transforming as $(\hf,\hf)$ under
two $SU(2)$ $R$-symmetry currents.  We will describe
this algebra in the next section.
Here we simply note that the $SU(2)\times SU(2)$ currents have
central extensions $k^+_R = -Q_1Q_5^+$
and $k^-_R = -Q_1Q_5^-$, and the large $\CN=4$
superalgebra with these two $R$-currents indeed
has conformal central charge \cexpl.%
\foot{In particular, we see that there are no $O(\hbar)$ corrections
to the central extensions.}

In the $55'$ spectrum, since there are 8 DN
directions the ground state energy in the NS sector is $+\coeff12$
and there are no massless bosons.  In the R sector, the fermions
are periodic in the two NN directions so there are fermion zero
modes $\psi^0$ and $\psi^1$. The ground state is then in the
$(-\coeff12;0,0;0,0)$ representation of these zero modes, which is a
trivial (left-moving) representation of the right moving
superalgebra.  There is a non-trivial contribution
to the central charge
\eqn\faz{
  c_L-c_R=\coeff12 Q_5^+Q_5^-\ .
}
On the other hand, the supergravity background seems to
respect $\CN=4$ supersymmetry of both chiralities on the
$AdS_3$ boundary.  One way to accomodate these facts
is to suppose that
these $R$-invariant $55'$ fields decouple, becoming free
fermions in the IR,
and that the remaining theory has its symmetry
enhanced to a large $\CN=(4,4)$ superconformal algebra
(the $55'$ fields cannot be fit
into a representation of the large $\CN=(4,4)$ algebra).
It would certainly be helpful to understand this issue better, but
for now we are going to ignore these fermionic `singleton' modes,
and assume that the infrared theory has large $\CN=(4,4)$
superconformal symmetry.

To describe the supergravity solution corresponding to this configuration
of branes, we begin with the geometry of fivebranes intersecting
over a string \GGPT:
\eqn\ffmetric{\eqalign{
ds^2 & = \left( \det U \right)^{-1/2}
\Big[ (-dt^2 + dx_5^2) + U_{ij} d \vec x_i \cdot d \vec x_j \Big] \cr
F_3 & = *_x dU_{11} + *_y dU_{22} \cr
e^{\phi} & = g_B \left( \det U \right)^{-1/2}\ .
}}
Here $U$ is a $2 \times 2$ symmetric matrix,
whose entries are harmonic functions of
the coordinates $\vec x_i = (\vec x, \vec y)$
on the $\R^8 = \R^4 \times \R^4$,
and $*_x$ denotes the Hodge dual on $\R^4$
parametrized by the 4-vectors $\vec x = (x^1, x^2, x^3, x^4)$
(similarly, $*_y$ is the Hodge dual in $\vec y = (x^6, x^7, x^8, x^9)$).

We will in fact consider a slight generalization the
geometry, in which the branes intersect at angles;
this will be useful below when we describe the deformation
to fivebranes wrapping a special Lagrangian submanifold.
Thus, we rotate the D5'-branes by an angle
$\vartheta$ in every two-plane $x^k - x^{k+5}$:
\eqn\rotatedbranes{\matrix{
 Q_5^+& {\rm ~~D5~branes~:} & \quad  \quad 056789 \hfill \cr
 Q_5^-& {\rm ~~D5'~branes~:} & \quad  \quad
05[16]_{\vartheta} [27]_{\vartheta} [38]_{\vartheta}
[49]_{\vartheta} \quad .
\hfill
}}
%
This configuration of intersecting fivebranes
preserves $3/16$ of the original supersymmetry and,
as will be shown below, leads to the same near-horizon
geometry \adsssr\ for {\it any} value of $\vartheta$.
Therefore, all such theories are expected to have
the same IR physics, described by a CFT
with large $\CN=(4,4)$ superconformal symmetry.%
\foot{One can check, using the representation theory
that we will introduce in the next section,
that the symmetries preserved by the deformation
guarantee that it represents an
irrelevant deformation of the infrared physics.}

For fivebranes at the special angles
\rotatedbranes,
the IIB supergravity solution has the form
\eqn\rotatedu{
U = U^{(\infty)}
+ \pmatrix{ {g_B Q_5^+ \over x^2} & 0 \cr
0 & {g_B Q_5^- \over y^2}}\quad .
}
Since constant terms are omitted in the near-horizon limit,
this is the first indication that the near-horizon geometry
of this intersecting D-brane configuration is the same for
any value of $\vartheta \ne 0$.
Explicitly, the rotation angle $\vartheta$ is given
in terms of $U^{(\infty)}_{ij}$ by
\eqn\thetaviau{
\cos\vartheta = - {U^{(\infty)}_{12} \over
\sqrt{ U^{(\infty)}_{11} U^{(\infty)}_{22} }}\ .
}

We may restrict $U^{(\infty)}$ to be such that $\det U^{(\infty)} = 1$.
Specifically, we choose $U^{(\infty)}$ to be
\eqn\uinftychoice{
U^{(\infty)} = \pmatrix{ \cosh \a & \sinh \a \cr \sinh \a & \cosh \a }\ .
}
Then, from \thetaviau\ one finds a relation between $\a$
and the rotation angle $\vartheta$:
\eqn\thviaa{ \cos \vartheta = - \tanh \a \ .}
%

Now let us include D1-branes smeared in the directions
$x^{1,2,3,4}$ and $x^{6,7,8,9}$.
This will further break the supersymmetry from 3/16 to 1/16,
unless $\vartheta=0$ where we preserve 1/8.
The complete supergravity solution looks like (in string frame):
\eqn\rotatedmetric{\eqalign{
ds^2
= & \left( H_1^{(+)} H_1^{(-)} \det U \right)^{-1/2} (-dt^2 + dx_5^2)
+ \sqrt{ H_1^{(+)} H_1^{(-)} }
{U_{11} \over \sqrt{\det U}} (d \vec x)^2 + \cr
& + \sqrt{ H_1^{(+)} H_1^{(-)} }
{U_{22} \over \sqrt{\det U}} (d \vec y)^2
+ {2 U_{12} \over \sqrt{\det U}} d \vec x \cdot d \vec y \cr
F_3 = & dt \wedge dx_5 \wedge d \left( H_1^{(+)} H_1^{(-)} \right)^{-1}
+ *_x dU_{11} + *_y dU_{22} \cr
e^{-2\phi} = & {1 \over g_B^2} {\det U \over H_1^{(+)} H_1^{(-)}}
}}
where
\eqn\hdone{
H_1^{(+)} = 1 + {g_B q_1 \over  x^2}
\quad , \quad
H_1^{(-)} = 1 + {g_B q_1 \over  y^2}
}
Notice, that since D1-branes are smeared along four spatial
directions, these harmonic functions exhibit the same radial
dependence as the fivebrane harmonic functions \rotatedu.
The parameter $q_1$ is the density of onebrane charge
along the fivebrane.
%


Now we are in a position to take the near-horizon limit of
the solution \rotatedmetric\ with the matrix $U$ given by
\rotatedu, \uinftychoice.
Omitting constant terms in the harmonic functions,
we find the near-horizon limit of the metric \rotatedmetric:
\eqn\metriclimita{
ds^2 = { x^2 y^2 \over g_B^2 q_1 \sqrt{Q_5^+ Q_5^-}} (-dt^2 + dx_5^2)
+ g_B q_1 \sqrt{{Q_5^+ \over Q_5^-}}
\Big( {dx^2 \over x^2} + \Omega_+^2 \Big)
+ g_B q_1 \sqrt{{Q_5^- \over Q_5^+}}
\Big( {dy^2 \over y^2} + \Omega_-^2 \Big)
}
By a change of variables,
\eqn\uxtendef{\eqalign{
u & =
xy \left( q_1^2 g_B^3 {Q_5^+ Q_5^- \over Q_5^+ + Q_5^-} \right)^{-1/2} \cr
\hat \th & =
{(Q_5^+ Q_5^-)^{-1/4}\over \sqrt{Q_5^+ + Q_5^-} }
\Biggl[ - Q_5^+ \log x + Q_5^- \log y\Biggr]\cr
}}
we can write the near-horizon metric in the form \iibmet:
%
\eqn\metriclimitc{
ds^2 = \ell^2 ds^2 (AdS_3)
+ R_+^2 d s^2(\S^3_+) + R_-^2 ds^2(\S^3_-) + \hat L^2 (d\hat \theta)^2
}
where
\eqn\rella{\eqalign{
\ell^2 & = g_B
q_1 \Big( {\sqrt{Q_5^+ Q_5^-} \over Q_5^+ + Q_5^-} \Big) \cr
R_+^2 & = g_B q_1 \sqrt{{Q_5^+ \over Q_5^-}} \cr
R_-^2 & = g_B q_1 \sqrt{{Q_5^- \over Q_5^+}} \cr
\hat L^2 & = q_1 g_B \ .
}}
Note especially that in the near-horizon geometry obtained
from the intersecting
branes, the coordinate $\hat\theta$ is non-compact;
the near-horizon geometry is $AdS_3\times\S^3\times\S^3\times\R$.

This near-horizon geometry can formally be further
compactified using a new isometry that only appears
after taking the near-horizon limit.
Namely, following \deBoerRH, we observe that
$x \to e^{-h} x, y\to e^h y$, is a symmetry of
\metriclimita\ for any real number $h$.
If we take a quotient by $\IZ$ with the generator acting
as $x \to e^{-h_*} x, y\to e^{h_*} y$ with
\eqn\hbstr{
h_*  = {(2\pi L)^2\over g_B} {1\over Q_5^+ Q_5^-}
}
and make a Weyl rescaling of
\iibmet\ by $(2\pi)^2 q_1/\sqrt{Q_5^+ Q_5^-}$
then we obtain $AdS_3\times\S^3\times\S^3\times\S^1$.
Note, however, that this orbifold action relates points
nonperturbatively far apart.  Thus, the relevance of
this orbifold action is open to question.  It is certainly not
a symmetry of the full string theory of the intersecting
branes \bkgdconfig.

The $AdS_3 \times \S^3 \times \S^3 \times \S^1$
supergravity solution of the previous section
is more conventionally related to the brane construction above,
if in the former we take the limit
$Q_1,L\to\infty$, with $g_B$, $Q_5^\pm$ and $q_1 \sim Q_1/L$ fixed;
then we obtain the near-horizon geometry \metriclimitc.
In order to find the relation between various parameters,
one can {\it e.g.} compute the Brown-Henneaux central
charge \bhcf\ - \thrdee\ using \rella. This gives
\eqn\vqq{ q_1 = {1 \over 4\pi^2} \sqrt{Q_5^+ Q_5^-} }
which, after substituting it back into \rella, leads to
the expressions \lngths\ found in the previous section.


While useful for illustrating geometrically the appearance of large $\CN=4$
supersymmetry in the near-horizon limit of branes,
the above intersecting brane configuration is somewhat less
useful for illuminating the nature of the dual CFT,
since the dynamics of fivebranes intersecting over a string
is poorly understood.  To shed some light on this side of
the duality, we can deform the above brane configuration,
simplifying the brane dynamics at the cost of breaking
some of the symmetry.  The deformation involved is expected
to be irrelevant, so that we should still recover large $\CN=4$
supersymmetry in the infrared.

The deformation we wish to consider is only
allowed for $Q_5^+ = Q_5^- \equiv Q_5$; we will also take
the angle $\vartheta$ to have the value
$\vartheta = \pi/4$.
Then, the D5-branes and D5'-branes can join together to form
a single set of $Q_5$ D5-branes along a smooth 4-manifold $\CM\subset\R^8$.
In order to preserve supersymmetry, $\CM$ must be a calibrated
submanifold inside $\R^8$.
Namely, the 4-manifold $\CM$ must be a special Lagrangian
submanifold inside $\IC^4 \cong \R^8$.
Fortunately, the explicit geometry of a special Lagrangian
submanifold in $\IC^4$ with the right properties was found
by Harvey and Lawson \HL.


\bigskip
{\vbox{{\epsfxsize=3.5in
        \nobreak
    \centerline{\epsfbox{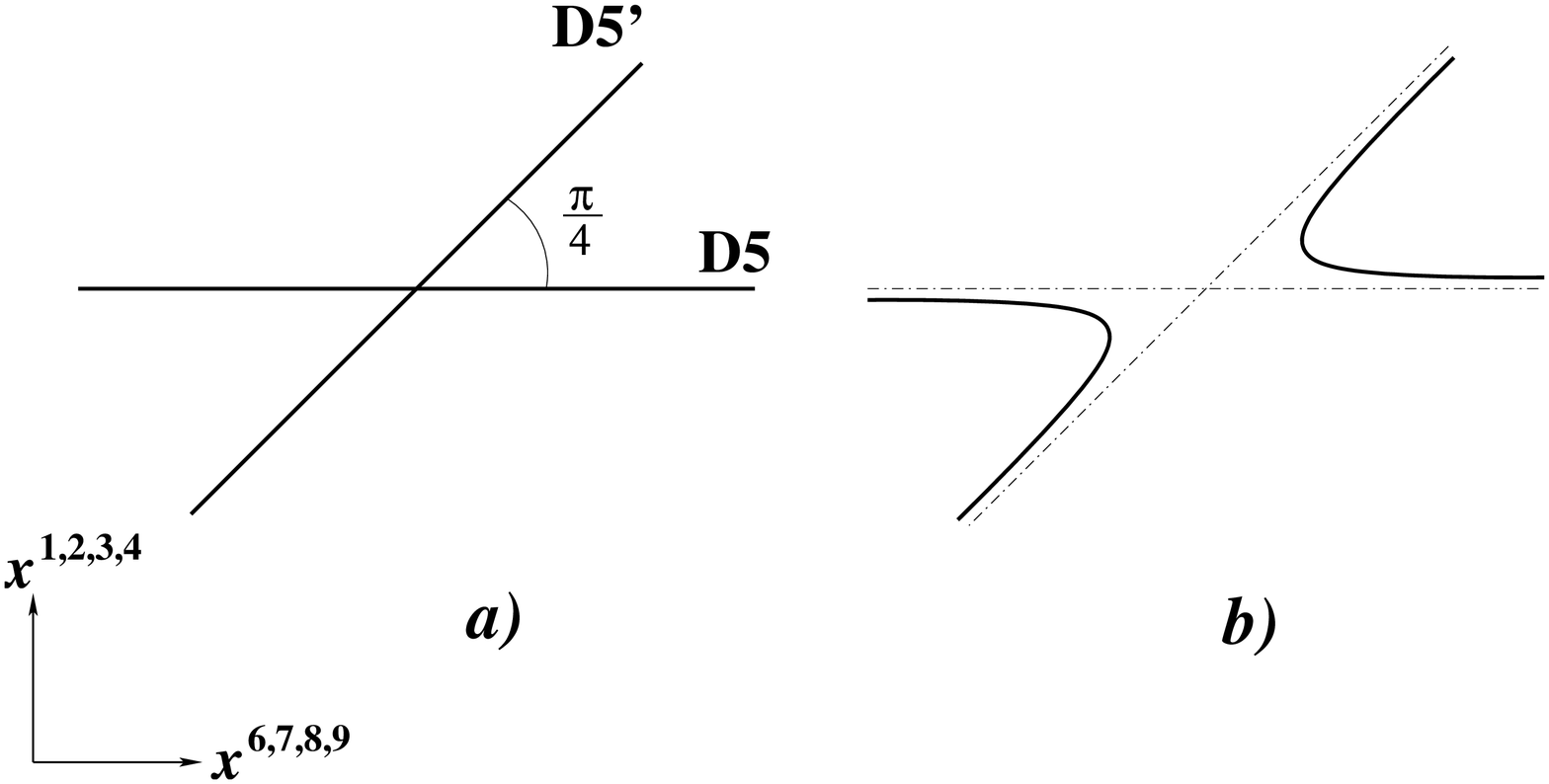}}
        \nobreak\bigskip
    {\raggedright\it \vbox{
{\bf Figure \branefig.}
{\it Intersection of special Lagrangian D5-branes (a)
and its non-singular deformation (b).
} }}}}
\bigskip}

As before, let us represent
\eqn\casrr{\IC^4 \cong \R^4 \times \R^4}
where each copy of $\R^4$ is parametrized by
the 4-vectors $\vec x$ and $\vec y$.
Then, the explicit form of the special Lagrangian
4-manifold $\CM$ is given by a set of points in $\IC^4 = \R^4 \times \R^4$,
which satisfy the following conditions \HL:
\eqn\lmfold{
\CM = \{ (\vec x, \vec y) \in \IC^4 ~~\vert~~
{\vec x \over \vert \vec x \vert} = {\vec y \over \vert \vec y \vert},
\quad xy(x^2 - y^2) = \rho \}
}
where $x\equiv|\vec x|$, $y\equiv|\vec y|$.
In other words, $\CM$ is a cohomogeneity one submanifold
in $\IC^4 = \R^4 \times \R^4$, represented by a graph of the function
\eqn\xyequation{ xy(x^2 - y^2) = \rho }
%
Note that the Lagrangian submanifold $\CM$ has topology
\eqn\ltopology{  \CM \cong \R \times \S^3  }
where the radius of the 3-sphere is determined by
the (real) deformation parameter $\rho$, {\it cf.} \lmfold.

In the limit $\rho \to 0$ the Lagrangian submanifold \lmfold\
degenerates into a union of two 4-planes
\eqn\lsingmfold{ \CM_{{\rm sing}} = \R^4 \cup \R^4 }
and we recover the geometry of intersecting fivebranes \rotatedbranes,
see Figure \branefig a.
On the other hand, when $\rho \ne 0$, $\CM$ is a smooth 4-manifold
with topology \ltopology, which is asymptotic
to the union of planes \lsingmfold.

For generic values of $\rho$, the Lagrangian submanifold $\CM$
is invariant under the symmetry group,
\eqn\sutwod{[SU(2)_L \times SU(2)_R]_D
\subset
\left[ SU(2)_L \times SU(2)_R \right]_{1234}
\times \left[ SU(2)_L \times SU(2)_R \right]_{6789}
}
which is a diagonal subgroup of the $R$-symmetry group \lorentz.
The undeformed rotated brane source \rotatedbranes\
has the same symmetry.
Nevertheless, the near-horizon geometry \metriclimitc\
of the latter is clearly invariant under the full symmetry group
on the RHS of \sutwod; the symmetry breaking to
the diagonal is an irrelevant perturbation in the infrared limit.
We expect the curved geometry of fivebranes located along $\CM$
also to flow to one with the $SU(2)^4$ isometry of \adsssr.
The RG flows considered in \MBerg\ might be relevant to
a further study of this issue.

Such a configuration, with D5 and D5' branes
joined in a single smooth manifold $\CM$, admits
a Higgs branch where D1-branes are realized as
instantons in the D5-brane.
In fact, this branch is very similar to the Higgs branch
in the ordinary D1-D5 system, where D5-branes are wrapped
on a 4-manifold $\CM = T^4$ or $K3$.
In the present case, on a single fivebrane
the vevs of the scalar fields
in the 1-5 string sector parametrize a 4-manifold
$\CM$ with the topology of $\S^3 \times \R$, so that the geometry
of the Higgs branch is given by
the symmetric product of this space,
\eqn\mkrs{ Sym^{N} (\CM) }
where, roughly speaking, one can interpret the coordinates
on this moduli space as parameters of the D1-brane instantons
on the D5-brane.  Unfortunately, because the space
wrapped by the fivebrane is non-compact, a duality
between this Higgs branch sigma model and supergravity
can only take place at $N=\infty$.


As mentioned in the beginning of this section, a rather different
approach using intersecting branes considers the
onebrane-fivebrane system wrapping a special Lagrangian $\S^3$.
This approach does allow us to compactify $\IR$ to $\S^1$. The
low-energy gauge theory of $N=Q_5$ fivebranes wrapping a special
Lagrangian $\S^3$ was considered in
\refs{\AcharyaMU\SchvellingerIB\MaldacenaPB-\GauntlettUR,\GS}. In
this theory, the $k=Q'_5$ units of three-form flux through the
$\S^3$ wrapped by the fivebranes appears in the effective gauge
dynamics through a Chern-Simons term, which is most easily seen
using the RR background frame \eqn\csfive{
  \frac{1}{16\pi^3}\int_{\IR^{\!1,2}\times\S^3_{\parallel}}
    \!\!\! C_2\wedge\Tr[F\wedge F]
    =-\frac{1}{16\pi^3}\int_{\IR^{\!1,2}\times\S_{\parallel}^3}
        \!\!\! F_3\wedge\Tr[AdA+\coeff23 A^3]
    =-\frac{k}{4\pi}\int_{\IR^{\!1,2}}\!\!\!\!  \Tr[AdA+\coeff23 A^3]\ .
}
Dual supergravity solutions
\refs{\AcharyaMU,\SchvellingerIB,\MaldacenaPB,\GauntlettUR,\GS}
have been considered in the NS background frame
(appropriate to the strong coupling gauge theory
that appears in the IR of the D-brane gauge theory).  A solution with
$Q_5=Q'_5$, which preserves 1/16 supersymmetry
and $SU(2)^3$ symmetry, was found in \AcharyaMU:
\eqn\sugrasoln{\eqalign{
  ds^2 & = ds^2_{\IR^{\!1,2}}+dr^2 +\coeff12 r\, d\Omega_{3,\parallel}^{\,2}
    +\coeff14\, d\Omega_{3,\perp}^{\,2} \cr
  e^{2\phi} & = g_s^2\,e^{-2r}r^{3/4} \cr
  H & = \coeff1{32}[\sigma_2\wedge\sigma_3\wedge\nu_1
    +\sigma_3\wedge\sigma_1\wedge\nu_2
    +\sigma_1\wedge\sigma_2\wedge\nu_3]
    +\coeff18\nu_1\wedge\nu_2\wedge\nu_3 \ .
}}
Here $\sigma_a$ ($\omega_a$) are the left-invariant one-forms
on $\S_{\parallel}^3$ ($\S_\perp^3$), and $\nu_a\equiv\omega_a-\hf\sigma_a$.

It was demonstrated in \WittenDS\ that the effective 2+1d gauge theory
obtained by KK reduction on $\S^3_{\parallel}$ spontaneously breaks
supersymmetry unless $|k|\ge N$.%
\foot{This phenomenon is familiar in the context of
supersymetric gauge theories, such as $\CN=1$ super-Yang-Mills in
four dimensions, where a similar argument can be used to show that
the number of BPS domain walls in a $U(N)$ gauge theory is
conserved modulo $N$ \AV.}
In \MaldacenaPB\ it was argued that for $k\ne N$
(\ie\ $Q_5\ne Q'_5$),
one needs to introduce explicit sources for the
three-form field strength corresponding to the
fivebranes wrapping $\S^3_\perp$, and their effects
are crucial for determining the IR dynamics of the theory.
Once again, we suffer from our lack of understanding of
the dynamics of intersecting fivebranes.
However, for equal fivebrane charges it appears that
the effects of one set of fivebranes is taken into account
through the background three-form flux, and the Chern-Simons
term \csfive\ it induces on the other set of fivebranes.

The geometry \sugrasoln\ is quite similar to
the throat geometry of NS5-branes in flat space;
an $\R^3$ parallel to the brane has been replaced by
an $\S^3$ whose warp factor is power law in $r$.
By the UV/IR relation of fivebrane holography,
this variation in the size of the $\S^3_\parallel$
is logarithmic in the energy scale.
Thus, we expect the addition of onebranes to the background
to be essentially the same as adding them to
fivebranes in flat space, up to additional logarithmic warping.
Let us compactify $\R^{1,2}$ to $\R^{1,1}\times\S^1$,
and put onebrane sources along $\IR^{1,1}$,
parametrized by $(t,x_5)$ in keeping with previous notation.
The expected form of the metric is then
\eqn\expmet{
 ds^2 = h_1(r) e^{2r}(-dt^2+dx_5^2)+h_2(r)d\theta^2
    +h_3(r) dr^2 + h_4(r) d\Omega_{3,\parallel}^{\,2}
    +h_5(r) d\Omega_{3,\perp}^{\,2}
}
with $h_i(r)$ having at most polynomial growth at
large $r$ (similarly, one expects the dilaton
$\phi$ to vary logarithmically in $r$).
Constant $h_i$ and dilaton $\phi$ corresponds to
$AdS_3\times\S^3\times\S^3\times\S^1$,
and logarithmic (in energy) dependence
of the geometry would correspond to the RG
flow of the dual sigma model toward an infrared CFT.

One might then look for instanton solutions
to the gauge theory compactified on $\S^3\times\S^1$,
and propose that the CFT
we are interested in is a sigma model on the
instanton moduli space, that represents the
small fluctuations around these configurations.%
\foot{The Chern-Simons term will not affect the solution
of the Yang-Mills equations on $\S^3\times\S^1$;
it will, however, generate additional couplings
in the sigma model obtained by expanding around
the instanton solutions.}
This is the standard logic by which one motivates the sigma
models in the hyperkahler cases of fivebranes
wrapping $T^4$ or $K3$
(see for examples \refs{\DijkgraafGF,\SeibergXZ,\LarsenUK,\LarsenDH,\emillectures} 
 for a discussion),
and we propose that it can be adapted to
the case of $\S^3\times \S^1$.


\newsec{The superconformal algebra $\CA_{\g}$ and its representations}

The superconformal symmetry of the spacetime CFT
for $AdS_3 \times \S^3 \times \S^3 \times \S^1$
consists of left and right copies of
the two-dimensional large $\CN=4$ supersymmetry algebra,
denoted $\cag$ \STVP.
In this section
we review some of the properties of this algebra
and its representations.


\subsec{The superconformal algebra $\CA_{\g}$}

Apart from the usual Virasoro algebra,
the large $\CN=4$ superconformal algebra $\CA_{\g}$
contains two copies of the affine $\widehat{SU(2)}$ Lie algebras,
at the levels $k^+$ and $k^-$, respectively.
The relation between $k^{\pm}$ and the parameter $\g$ is
\eqn\gviakk{ \g = {k^- \over k^+ + k^-} \ .}
Unitarity implies that the Virasoro central charge is:
\eqn\cviakk{ c= {6 k^+ k^- \over k^+ + k^-}\ . }

The superconformal algebra $\CA_{\g}$ is generated by six
affine $\widehat{SU(2)}$ generators $A^{\pm,i} (z)$, four dimension
$3/2$ supersymmetry generators $G^a (z)$, four dimension $1/2$
fields $Q^a (z)$, a dimension 1 field $U(z)$, and the Virasoro
current $T(z)$.
The OPEs with the Virasoro generators, $T_m$, have the usual form.
The remaining OPEs are \refs{\STVP,\Schoutens}:
\eqn\agope{\eqalign{ & G^a (z) G^b (w) = {2c \over 3} {\d^{ab}
\over (z-w)^3} - {8 \g \a_{ab}^{+,i} A^{+,i} (w) + 8 (1-\g)
\a_{ab}^{-,i} A^{-,i}(w) \over (z-w)^2 } - \cr &~~~~~~~~~~~~~~~ -
{4\g \a_{ab}^{+,i} \p A^{+,i} (w) + 4 (1-\g)\a_{ab}^{-,i} \p A^{-,i}
(w) \over z-w } + {2\d^{ab} L(w) \over z-w} + \ldots, \cr & A^{\pm,
i} (z) A^{\pm ,j} (w) = -{k^{\pm} \d^{ij} \over 2 (z-w)^2} +
{\epsilon^{ijk} A^{\pm,k} (w) \over z-w} + \ldots, \cr & Q^a (z)
Q^b (w) = - {(k^+ + k^-) \d^{ab} \over 2 (z-w)} + \ldots, \cr &
U(z) U(w) = - {k^+ + k^- \over 2 (z-w)^2} + \ldots, \cr & A^{\pm,
i} (z) G^a (w) = \mp {2 k^{\pm} \a_{ab}^{\pm,i} Q^b (w) \over (k^+
+ k^-) (z-w)^2 } + {\a_{ab}^{\pm,i} G^b (w) \over z-w } + \ldots,
\cr & A^{\pm,i} (z) Q^a (w) = {\a_{ab}^{\pm,i} Q^b (w) \over z-w }
+ \ldots, \cr & Q^a (z) G^b (w) = {2 \a_{ab}^{+,i} A^{+,i} (w) - 2
\a_{ab}^{-,i} A^{-,i} (w) \over z-w} + {\d^{ab} U(w) \over z-w} +
\ldots, \cr & U(z) G^a(w) = {Q^a (w) \over (z-w)^2}  + \ldots. }}
$\a_{ab}^{\pm,i}$ here are $4 \times 4$ matrices, which project
onto (anti)self-dual tensors.  Explicitly,
\eqn\thoofteta{ \a^{\pm,i}_{ab} = {1 \over 2} \Big( \pm \d_{ia}
\d_{b0} \mp \d_{ib} \d_{a0} + \epsilon_{iab} \Big) \ .}
They obey $SO(4)$ commutation relations:
\eqn\sofourcr{
[\a^{\pm,i} , \a^{\pm,j}] = - \epsilon^{ijk} \a^{\pm k} \quad ,
\quad [\a^{+,i} , \a^{-,j}] = 0 \quad , \quad \{ \a^{\pm,i} ,
\a^{\pm, j} \} = - {1 \over 2} \d^{ij}\ .
}
It is sometimes useful to employ spinor notation,
where for instance
$G^a\to G^{\alpha\dot\alpha}=\gamma_a^{\alpha\dot\alpha}G^a$
(and $\gamma_a^{\alpha\dot\alpha}$ are Dirac matrices);
$A^{+,i}\to A^{\alpha\beta}= \tau^{\alpha\beta}_i A^{+,i}$
(where $\tau^i$ are Pauli matrices);
$A^{-,i}\to A^{\dot\alpha\dot\beta}= \tau^{\dot\alpha\dot\beta}_i A^{-,i}$;
and so on. 
Our conventions are spelled out in appendix B. 


An important subalgebra of $\CA_\gamma$ is denoted $D(2,1|\alpha)$;
here $\alpha=k^-/k^+=\frac{\gamma}{1-\gamma}$.
It is generated (in the NS sector)
by $L_0$, $L_{\pm 1}$, $G^a_{\pm 1/2}$, and $A^{\pm, i}_0$.
The superalgebra $D(2,1|\alpha)\times D(2,1|\alpha)$
constitutes the super-isometries of $AdS_3\times \S^3\times\S^3$.

Yet another useful subalgebra of $\cag$ is the $\CN=2$
subalgebra generated by
\eqn\jntwo{
T\ , \quad
\CG^+=i\sqrt2\,G^{+\dot +}\ ,\quad
\CG^-=i\sqrt2\,G^{-\dot -}\ ,\quad
  J=2i\,[\gamma A^{+,3} - (1-\gamma) A^{-,3}]\
}
where the supercurrents are written in spinor notation.
%
For instance, it will be useful to consider the
states that are chiral with respect to this $\CN=2$.

\subsec{Examples of large $\CN=4$ SCFT's}

The simplest example of a large $\CN=4$ theory
can be realized as a theory of a free boson, $\phi$, and four Majorana
fermions, $\psi_a$, $a=0,\ldots,3$. Specifically, we have \Srep:
\eqn\cthree{\eqalign{
& T = - {1 \over 2} (\p \phi)^2 - {1 \over 2} \psi^a \p \psi^a \cr
& G^a  = - {1 \over 6} i \epsilon^{abcd} \psi^b  \psi^c  \psi^d
- i \psi^a \p \phi \cr
& A^{\pm,i} = {i \over 2} \a^{\pm,i}_{ab} \psi^a \psi^b \cr
& Q^a = \psi^a \cr
& U = i \p \phi\ .
}}
This theory was called the $\CT_3$ theory in \ElitzurMM, but we shall
herein use the notation $\CS$ for simple.
In \ElitzurMM\ it was conjectured that in the case
$k^+ = k^-$ the boundary SCFT dual to type IIB string theory on
$AdS_3 \times \S^3 \times \S^3 \times \S^1$ is a sigma-model based
on the symmetric product orbifold of this $c=3$ theory.

The CFT $\CS$ belongs to a family of large $\CN=4$ theories,
labeled by a non-negative integer number $\kappa$ \refs{\Ivanov,\STVP}:
\eqn\algeone{ \eqalign{ T &  = - J^0 J^0 - {J^a J^a \over \kappa+2} -
\p \psi^a \psi^a \cr G^a & =2 J^0 \psi^a + {4 \over \sqrt{\kappa+2}}
\a^{+,i}_{ab} J^i \psi^b - {2 \over 3 \sqrt{\kappa+2}} \epsilon_{abcd}
\psi^b \psi^c \psi^d\cr A^{-,i} & = \a_{ab}^{-,i} \psi^a \psi^b
\cr A^{+,i} & = \a_{ab}^{+,i} \psi^a \psi^b + J^i \cr U & =
-\sqrt{\kappa+2} J^0 \cr Q^a & = \sqrt{\kappa+2} \psi^a \cr} }
%
where $J^i$ denote $SU(2)$ currents at level $\kappa$ and 
$J^0(z)J^0(w) \sim -\half (z-w)^{-2}$.  We shall denote these
theories $\CS_\kappa$.
It is easy to check that \algeone\ indeed generate
the large $\CN=4$ algebra with $k^+ = \kappa+1$ and $k^- = 1$.
In fact, the $U(2)$ level $\kappa$ theory of \refs{\Ivanov,\STVP}
admits {\it two} distinct large $\CN=4$ algebras.
The second algebra is obtained by the outer automorphism and has
$(k^+=1, k^- = \kappa+1)$:
\eqn\algtwo{ \eqalign{ T &  = - J^0 J^0 - {J^a J^a \over \kappa+2} - \p
\psi^a \psi^a \cr G^a & =2 J^0 \psi^a + {4 \over \sqrt{\kappa+2}}
\a^{+,i}_{ab} J^i \psi^b - {2 \over 3 \sqrt{\kappa+2}} \epsilon_{abcd}
\psi^b \psi^c \psi^d\cr A^{-,i} & = \a_{ab}^{-,i} \psi^a \psi^b
+J^i \cr A^{+,i} & = \a_{ab}^{+,i} \psi^a \psi^b   \cr U & =
+\sqrt{\kappa+2} J^0 \cr Q^a & = -\sqrt{\kappa+2} \psi^a \cr} }
The $c=3$ CFT $\CS=\CS_0$ appears as a special case, $\kappa=0$.
We will consider these simple large $\CN=4$ theories below
in the context of symmetric product orbifolds
as candidates for the spacetime CFT.

Additional examples of large $\CN=4$ are provided by
WZW coset models $\CW\times U(1)$,
where $\CW$ is a gauged WZW model associated to
a quaternionic (Wolf) space.
Examples based on classical groups are
$\CW=G/H=\frac{SU(n)}{SU(n-2)\times U(1)}$,
$\frac{SO(n)}{SO(n-4)\times SU(2)}$,
and $\frac{Sp(2n)}{Sp(2n-2)}$.
These theories carry large $\CN=4$ supersymmetry,
with $k^+=\kappa+1$ and $k^-=\check c_G$;
here $\kappa$ is the level of the bosonic current algebra
for the group $G$ and $\check c_G$ its dual Coxeter number.
However, they are unsuitable as building blocks for a symmetric
product orbifold dual to supergravity.  For example,
any modulus associated to the RR axion would generically
come from the component theory and not the twisted
sector of the symmetric product, and would thus not
deform the spectrum in the appropriate way
as one moves from the orbifold locus to the supergravity regime.%
\foot{
Also, the BPS spectrum of these theories
does not seem to have the requisite properties.
The BPS states are associated to the cohomology of $\CW$,
which is in turn related to the elements
of the Weyl group of affine $G$ (related to the symmetric group).
Instead, in order to match the structure of supergravity,
one typically would want the cohomology to be associated
to the conjugacy classes of the symmetric group,
as in the orbifold cohomology of the symmetric product,
whose cohomology matches supergravity in for example
the D1-D5 system.}


\subsec{Unitary representations}

The unitary representations of the superconformal algebra $\CA_{\g}$
are labeled by the conformal dimension $h$,
by the $SU(2)$ spins $\l^{\pm}$,
and by the $U(1)$ charge $u$.
The generic {\it long} or {\it massive}
representation has no null vectors
under the raising operators of the algebra.
On the other hand, the highest weight states
$\ket{\Omega}^{~}_{\!\cag}$ of
{\it short} or {\it massless} representations
have the null vector \GPTVP
\eqn\nullvec{
  \Bigl(G^{+\dot+}_{-1/2}-\frac{2u}{k^++k^-}\,
    Q^{+\dot+}_{-1/2}-{2i(\ell^+-\ell^-)\over k^++k^-}\,Q^{+\dot+}_{-1/2}\Bigr)
    \ket{\Omega}^{~}_{\!\cag}=0\ .
}
(We have used the property that $ \ket{\Omega}^{~}_{\!\cag}$ is a
highest weight state for the $SU(2)$ current algebras.)
Squaring this null vector leads to a relation among
the spins $\ell^\pm$ and the conformal dimension $h$
\refs{\GPTVP,\PT,\PTtwo,\deBoerRH}
\eqn\unitary{
   h_{\rm short} = {1 \over k^+ + k^-}
    \left( k^- \l^+ + k^+ \l^- + (\l^+ - \l^-)^2 + u^2 \right)\ .
}
Unitarity demands that all representations, short or long,
lie at or above this bound: $h\ge h_{\rm short}$;
and that the spins lie in the range
\hbox{$\ell^\pm=0,\hf,...,\hf(k^\pm-1)$}.
When we consider $U(1)$ singlets, we shall
denote representations by their labels
$(h,\ell^+,\ell^-)$; for short representations with $u=0$
it is sufficient to specify them simply by $(\ell^+,\ell^-)$.
The representations of the spacetime SCFT can be obtained
by combining left and right sectors. Following \deBoerRH,
we shall label such (short)
representations by $(\l^+, \l^-; \bar \l^+ , \bar \l^-)$.

The conformal dimension of short representations is protected,
as long as they do not combine into long ones.
The highest weight components of operators in
short representations with $\l^+=\l^-$ form a ring.
Their dimensions are additive, since $h=\ell^+=\ell^-$.
This ring is the chiral ring of the $\CN=2$ subalgebra
of $\cag$ introduced in subsection 4.1.

We will also be interested in the representations of
the super-isometry group
$D(2,1\vert \alpha)$; for example, the normalizable
wavefunctions on $AdS^3\times \S^3\times \S^3$
lie in representations of $D(2,1\vert \alpha)\times D(2,1\vert \alpha)$.
A {\it short} $\dto$ representation $(\l^+,\l^-)_\sd$
of $\dto$ has a highest weight vector $\ket{\Omega}_D$
which obeys the condition
\eqn\shrttwo{
G^{\ppd}_{-1/2} \vert \Omega \rangle_D =0
}
Long representations $(\l^+,\l^-)_\ld$
have no such null vector in the action of $G^a_{-1/2}$.

The base of a short representation
$(\l^+,\l^-)_\sd$ of $\dto$
can be obtained by acting with $G^a_{-1/2}$:
\eqn\shortrep{\matrix{
h &\quad&  & (\l^+,\l^-) & \cr
h + {1 \over 2} &
& (\l^+ - {1 \over 2},\l^- - {1 \over 2})
& (\l^+ + {1 \over 2},\l^- - {1 \over 2})
& (\l^+ - {1 \over 2},\l^- + {1 \over 2}) \cr
h + 1 &
& (\l^+ , \l^- - 1)
& (\l^+ - 1 , \l^- )
& (\l^+ , \l^- ) \cr
h + {3 \over 2} &&  & (\l^+ - {1 \over 2},\l^- - {1 \over 2}) &
}}
with the rest of the representation filled out
by the action of $L_{-1}$.
We denote these short representations of $\dto$ by $(\l^+,\l^-)_\sd$
with lower case subscript $\sd$ to distinguish them
from short representations $(\l^+,\l^-)_S$ of $\cag$.
These representations are smaller than the
generic large $\dto$ representation, due to
the absence of a spin $(\ell^++\hf,\ell^-+\hf)$
component in $G^a_{-1/2}\ket{h,\ell^+,\ell^-}$.
Note that since $\ell^{\pm}$ take only non-negative values,
short representations with $\l^{\pm} < 1$
have some of the components missing, see {\it e.g.} \deBoerRH\
and below.

Squaring the null vector
\shrttwo\
and using the algebra gives the BPS bound
$h =\frac{k^+ \ell^- + k^- \ell^+}{k^++k^-}$.
Note that {\it unless $u=0$ and $\ell^+ = \ell^-$ the
$\cag$ and $\dto$ BPS conditions are different}.
In particular, if $u\not=0$ or $\ell^+ \not= \ell^-$,
then a BPS state in the $\CA_{\gamma}$ sense
is {\it not} a BPS state in the $\dto$ sense.
In fact, by unitarity, if $\l^+ \not=\l^-$ then
the representation $(\l^+,\l^-)$ of $\cag$ cannot contain
any BPS representations of $\dto$!

The representations of $\cag$ can however be decomposed into
representations of $\dto$.
Let $\rho(\l^+,\l^-,u)$ be a short representation
of $\CA_{\gamma}$. Then, as a representation of $D(2,1;\alpha)$
$\rho$ contains:

\item{a.)} All long $\dto$ representations for $u\not=0$ or for
$\l^+ \not=\l^-$.
\item{b.)} Exactly two short $\dto$ representations for $\l^+ = \l^-$
and $u=0$. That is,
\eqn\decomp{
\rho(\l,\l,0) = (\l,\l)_\sd +
(\l + \coeff12, \l+\coeff12)_\sd + \cdots}
where all representations in $\cdots$ satisfy
$h>\frac{ k^+ \l^- + k^- \l^+}{k^++k^-}$.

\noindent
Part {\sl (a)} is trivial.  From the $\CA_{\gamma}$ bound we get the
inequality:
\eqn\agmm{
(k^++k^-)h =  k^+ \l^- + k^- \l^+ +
    (\l^+ - \l^-)^2 +u^2 > k^+ \l^- + k^- \l^+
}
We have also explained this in detail in comparing the highest weight
conditions above.
For part {\sl (b)} we take the BPS highest weight state
$\vert \Omega\rangle_{\cag}$ for
$\CA_{\gamma}$.
Under the conditions of part {\sl (b)} this is also a
BPS highest weight state for $\dto$.
We also have the state
\eqn\newstate{
Q^{\ppd}_{-1/2}\vert \Omega \rangle_\cag\ .
}
This is a descendent in the $\CA_{\gamma}$ representation,
but since
\eqn\twocomms{
\eqalign{
[A_0^{\pm, +}, Q^{\ppd}_{-1/2}] & = 0 \cr
\{G^{\ppd}_{-1/2} , Q^{\ppd}_{-1/2} \} & = 0 \cr}
}
the state \newstate\ is a BPS highest weight state for the
$D(2,1;\alpha)$ subalgebra.
It generates the representation $(\l + 1/2, \l + 1/2)_\sd$.
Finally, we must show there are no other short $\dto$ highest weight vectors.
The BPS bound is linear in $h,\ell$ and must be obtained from the
$\CA_{\gamma}$ highest weight state by applying $G_{+,-1/2}$
and $Q_{+,-1/2}$. Using \twocomms\ above we see that the only state
we can generate is the second one we have already accounted for.
The two short representations
in \decomp\ are distinct from a long
representation of $D(2,1|\alpha)$, even though they
have the same spin content.


\subsec{The general structure of marginal deformations}

Our goal in this subsection
is to identify the states in the spacetime CFT
which correspond to moduli of the type IIB string theory on
$AdS_3 \times \S^3 \times \S^3 \times \S^1$.
Such states should have conformal dimensions $(h,\bar h)=(1,1)$
and must be $SU(2)\times SU(2)$ singlet components of short multiplets.
Inspection of \shortrep\ shows that
marginal deformations in the large $\CN=4$
superconformal field theory come from upper components
of the $u=0$ short multiplets $({1 \over 2}, {1 \over 2})_S$.
These representations have the $\dto$ structure
\eqn\hfshort{
  \matrix{h=\hf&\quad& &(\half,\half)& \cr
  h=1&\quad& (0,0)&(1,0)&(0,1) \cr
  h=\coeff32&\quad& &(\hf,\hf)& \cr
  h=2&\quad& &(0,0)& }
}
and so are even more truncated than the generic
short representation.
The spin $(0,0)$ state on the second level is dimension one
and invariant under the $SU(2)\times SU(2)$ $R$-symmetry;
acting by the raising operators
$G^{\alpha\dot\alpha}_{-1/2}$ gives only $L_{-1}$ descendants,
as we will now show momentarily.%

First we note a very interesting consequence of the structure \hfshort:
there are no constraints on the number of moduli.
This result should be compared with a similar situation
in superconformal theories based on the small $\CN=4$ algebra,
where marginal deformations are also upper components of
chiral primary states $\Phi_{\alpha\bar\beta}$ 
with $(j,\bar j)=({1 \over 2}, {1 \over 2})$. 
However, in that case every short multiplet contains four singlet
states with $(h,\bar h)=(1,1)$, namely, 
$\CT^{a\bar b}=G_{-\hf}^{a\alpha}\bar {G_{-\hf}^{b\beta}}
    \Phi_{\alpha\bar\beta}$, where $a,\bar b$ are custodial $SU(2)$ 
indices. In particular, the number
of massless moduli has to be a multiple of 4.

A large $\CN=4$ chiral primary with $\ell^+=\ell^-$ has the null vector
\nullvec, which may be written more invariantly,
\eqn\shortnullvec{
  G_{{\shf}(\dot\alpha}^{~~~(\alpha}
        \;\Phi^{\alpha_1...\alpha_n)}_{\dot\alpha_1...\dot\alpha_n)} = 0\ .
}
The candidate modulus operator is
\eqn\modcand{
  \CT = G_{\shf}^{\beta\dot\beta}
        \bar G_{\shf}^{\bar\alpha\dot{\bar\alpha}}
        \Phi_{\beta\dot\beta;\bar\alpha\dot{\bar\alpha}} \ ,
}
where $\Phi_{\beta\dot\beta;\bar\alpha\dot{\bar\alpha}} $
has $(\ell^+,\ell^-;{\bar\ell}^+,{\bar\ell}^-)=(\hf,\hf;\hf,\hf)$.
We will for the remainder of the discussion suppress the anti-holomorphic
structure, which will not be needed explicitly.
Expanding in components,
\eqn\modcomp{
  G_{\shf}^{\beta\dot\beta} \Phi_{\beta\dot\beta}
        = G_{\shf}^\ppd\Phi^\mmd + G_{\shf}^\mmd\Phi^\ppd
        - G_{\shf}^\pmd\Phi^\mpd - G_{\shf}^\mpd\Phi^\pmd\quad .
}
The supercharge anticommutation relations and the nullvector condition
\shortnullvec\ then imply
\eqn\stepb{
  G_{\shf}^\ppd(G_{\shf}^\pmd\Phi^\mpd+G_{\shf}^\mpd\Phi^\pmd)
        = \{G_{\shf}^\pmd,G_{\shf}^\mpd\}\Phi^\ppd
        = -L_{-1}\Phi^\ppd\ .
}
Similarly, we have
\eqn\stepc{
  G_{\shf}^\ppd(G_{\shf}^\ppd\Phi^\mmd+G_{\shf}^\mmd\Phi^\ppd)
        = \{G_{\shf}^\ppd,G_{\shf}^\mmd\}\Phi^\ppd
        = L_{-1}\Phi^\ppd\ ;
}
putting it all together, we have%
\foot{Similarly, one can show that
\eqn\afterstep{
  G_{\sst+\frac12}^{\alpha\dot\alpha}(G_{\shf}^{\beta\dot\beta}
        \Phi_{\beta\dot\beta})
        = 2 \Phi^{\alpha\dot\alpha}\ .
}}
\eqn\laststep{
  G_{\shf}^{\alpha\dot\alpha}(G_{\shf}^{\beta\dot\beta}
        \Phi_{\beta\dot\beta})
        = 2 \partial\Phi^{\alpha\dot\alpha}\ .
}
Thus, while the candidate modulus is not
the highest component of the supermultiplet
based on $\Phi^{\alpha\dot\alpha}$, it nevertheless
varies into a total derivative under the action
of the supercharges and so its integral preserves
all the supersymmetries.  All that remains to be checked
is that it preserves conformal invariance.
A proof of conformal invariance to all orders in
conformal perturbation theory, following \DixonBG,
is given in Appendix A.

As an aside, it is curious that it appears not to be
possible to write this candidate modulus operator
as an integral over even $\CN=1$ superspace!
In particular, we cannot directly use the results of Dixon \DixonBG\
on the marginality of $h=1$, $\CN=2$ chiral operators,
even though the lowest component $\Phi_{\alpha\dot\alpha}$
of the multiplet is a chiral operator under the
canonical $\CN=2$ subalgebra of large $\CN=4$.
The argument of \DixonBG\ uses the structure of $\CN=2$ chiral
superspace integrals in an essential way.
Fortunately, it is possible to adapt the analysis
to fit the structure of large $\CN=4$.

A key ingredient of the analysis of Appendix A
is the demonstration that, in the partition function,
one can replace the operator \modcand\ by the operator
\eqn\altmod{
  \tilde\CT = \Bigl(G_{\shf}^\ppd\bar G_{\shf}^{+\dot{+}}
            \Phi_{\ppd,+\dot{+}}
    + G_{\shf}^\mmd\bar G_{\shf}^{-\dot{-}}
            \Phi_{\mmd,-\dot{-}}\Bigr)
    + \Bigl(G_{\shf}^\mmd\bar G_{\shf}^{+\dot{+}}
            \Phi_{\mmd,+\dot{+}}
    + G_{\shf}^\ppd\bar G_{\shf}^{-\dot{-}}
            \Phi_{\ppd,-\dot{-}}\Bigr)
}
which is a sum, in equal proportion, of a chiral and a twisted chiral
modulus under the canonical $\CN=2$ algebra \jntwo;
moreover, the chiral and twisted chiral moduli are real.
If we were to give each term in \altmod\ a different coefficient
(compatible with hermiticity), we would explore the
moduli space of an $\CN=2$ superconformal theory.
This theory is manifestly self-mirror.
The large $\CN=4$ locus on this moduli space is thus
the fixed point set under both the mirror map, and also
the antiholomorphic involution of the $\CN=2$ algebra.%
\foot{Note that this reduces a $4n$ dimensional
moduli space to an $n$ dimensional one.  Again there is
no constraint on $n$.}
This rather constrains the geometry of the moduli space;
it would be interesting if the structure of large $\CN=4$
could yield further information about this geometry.


There is a universal $({1 \over 2}, {1 \over 2})_S$
representation that canonically appears in the theory --
the singleton bilinear $U\bar U$.
In the application to $AdS_3\times\S^3\times\S^3\times\S^1$,
it implements (among other things) a change in the
boundary condition on the corresponding bulk
gauge field \AharonyDP.
There is also a second modulus associated to the $\S^1$,
the mode which corresponds
to changing the $\S^1$ radius in supergravity,
the combination of the metric and dilaton found in
equation \newlgths.
In general the structure is rather complicated,
since these two moduli mix non-trivially.
Conventionally, the singleton bilinear modulus
is turned off, and only the supergraviton mode
is considered.
%
We will make the same restriction here.

We are also expecting that the spacetime CFT contains
another modulus, corresponding to the RR axion,
as discussed in section 2.
In symmetric products of the $U(2)$ WZW model,
we will find the corresponding
marginal deformation in twisted sectors.


\subsec{Spectral flow}

Since the superconformal algebra \agope\ contains two copies
of $SU(2)$, there are several types of spectral flow
one can consider\foot{In the case $k^+ = k^-$, the superconformal
algebra $\CA_{\g}$ has additional automorphisms,
which we are not going to discuss here.}.
Following \DST, let us call the corresponding parameters $\rho$ and $\eta$.
Then, the relation between the generators looks like \DST,
\eqn\sflow{\eqalign{
& L_m^{\rho, \eta}
= L_m - i (\rho A_m^{+3} + \eta A^{-3})
+ {1 \over 4} ( k^+ \rho^2 + k^- \eta^2 ) \d_{0,m} , \cr
& A_m^{\rho, \eta;+3}
= A_m^{+3} + {i \over 2} \rho k^+ \d_{m,0} , \cr & A_m^{\rho, \eta;-3}
= A_m^{-3} + {i \over 2} \eta k^- \d_{m,0} , \cr & U_m^{\rho, \eta} = U_m
}}

The Neveu-Schwarz sector corresponds to $(\rho,\eta)=(0,0)$,
whereas the Ramond sector can be obtained by a spectral flow
with $\rho=1$, $\eta=0$ or $\rho=0$, $\eta=1$.
{}From \sflow\ one finds the following relation between
the conformal dimensions and other quantum numbers
in the Ramond and Neveu-Schwarz sectors (note, that our
transformations of $\l^{\pm}$ differ from those given in \GPTVP),
\eqn\sflowplus{\eqalign{
& h_R = h_{NS} - \l^+_{NS} + {1 \over 4} k^+ \cr
& \l^+_R =  \l^+_{NS} - {1 \over 2} k^+ \cr
& \l^-_R = \l^-_{NS} \cr
& u_R = u_{NS}
}}
for the spectral flow in the $SU(2)^+$.
Similarly, for the spectral flow in the $SU(2)^-$, we have
\eqn\sflowminus{\eqalign{
& h_R = h_{NS} - \l^-_{NS} + {1 \over 4} k^- \cr
& \l^+_R = \l^+_{NS} \cr
& \l^-_R = \l^-_{NS} - {1 \over 2} k^- \cr
& u_R = u_{NS}\ .
}}
In particular, NS states saturating the BPS bound \unitary\
flow to R states with
\eqn\Rwt{
  h_R-\frac{c}{24} = \frac{(\ell^++\ell^-)^2+u^2}{k^++k^-}\ .
}
Note the rather peculiar fact that the right-hand side is nonzero.
Here again we see an important qualitative difference between
the large $\CN=4$ algebra and other superconformal algebras.

The $\CN=2$ subalgebra \jntwo\ leads to yet another
version of spectral flow to a Ramond sector, with $\rho=\eta=1/2$.
This leads to Ramond boundary conditions for the
$\CN=2$ currents $\CG^\pm=i\sqrt2 G^\pmpmd$,
but the boundary conditions on $G^{\pm\dot\mp}$ become
fractionally moded
(as do the raising and lowering operators $A^{\pm,i}$,
$i=\pm$, of the two $SU(2)$'s).


\subsec{An index for theories with $\CA_\gamma$ symmetry}

When one is working with families of theories with $\cag$ symmetry,
as we are in the present paper, it is useful to know quantities
which are invariant under deformations. The traditional
elliptic genus does not provide useful information in the present
context, but one can nevertheless define an index which summarizes
some important information about the BPS spectrum of the theory and
which remains invariant under deformations.
In this section we
briefly define such an index.\foot{For a related
discussion see also \OPT.} Further details and comments can be found
in  a  companion paper \indexsummary\ where we investigate
this large $\CN=4$ index in some detail.

The representation content of a theory with $\cag$ symmetry  is summarized by
the  RR sector   supercharacter:
\eqn\defnewzee{
Z(\tau, \omega_+,\omega_-; \bar \tau, \tl \omega_+,\tl \omega_-):=
{\Tr}_{\CH_{RR} } q^{L_0-c/24}\tilde q^{\tilde L_0-c/24}
 z_+^{2T_0^{+,3}} (-z_-)^{2T_0^{-,3}}
\tilde z_+^{2\tilde T_0^{+,3}} (-\tilde z_-)^{2\tilde T_0^{-,3}}
}
Here and hereafter we denote  $z_\pm = e^{2\pi i \omega_\pm }$ for left-movers
and   $\tl z_\pm = e^{2\pi i \tl \omega_\pm }$ for right-movers.
The spectrum in other sectors can be obtained from \defnewzee\ by
spectral flow.


Now \defnewzee\ can be expanded in the supercharacters of the irreducible
representations, defined by
\eqn\defchrs{
\SCh(\rho)(\tau,\omega_+,\omega_- )
= {\Tr}_{\rho} q^{L_0-c/24} z_+^{2T_0^{+,3}} z_-^{2T_0^{-,3}}
(-1)^{2T_0^{-,3}}
}
we just write $\SCh(\rho)$ when the arguments are understood.
Explicit formulae for these characters have been derived by Peterson and
Taormina.
Using the formulae of \PTtwo\ one finds that
short representations have a character
with a first order zero at $z_+ = z_-$, while long representations have a
character with a second order zero at $z_+ = z_-$.%
\foot{The fact that all characters vanish at $z_+ = z_-$ is a reflection of
the fact that one can always make a GKO coset construction factoring out
the free $\CS$-theory defined by $U, Q^{A\dot A}$.
The character of this theory
has a first order zero. The characters of the quotient $\tilde \cag$
$W$-algebra are nonvanishing for short representations, and have a first
order zero for long representations. }

Thanks to the second order vanishing of the characters of long representations
we can define the {\it left-index} of the CFT $\CC$  by
\eqn\soperat{
I_1(\CC):= - z_+
{d\over dz_-}\biggl\vert_{z_- =  z_+}Z
}
Only short  representations can contribute on the left. On the
right, long representations might contribute. However, due to the constraint
$h-\bar h = 0 \mod 1$ the right-moving conformal weights which do contribute
are rigid, and hence $I_1$ is a deformation invariant.

Of course, one could also define a right-index.
Since we will consider left-right
symmetric theories here this is redundant information. Nevertheless, it is
often useful to define the left-right index:
\eqn\indxone{
I_2(\CC)  := z_+ \tl z_+ {d\over dz_-}{d\over d\tilde z_-} Z
}
where one evaluates at $ z_- =  z_+$, $\tl z_- =  \tl z_+$.


\subsec{Digression: Taking the tensor product of two large $\CN=4$ algebras}

Although it is not directly used in the present paper, we would like
to mention in this section on $\cag$ symmetry a curious behavior of these
theories under the tensor product operation.
Since the Virasoro central charge \cviakk\ is nonlinear
it is therefore not obvious how to take a tensor product of
algebras $\CA(k_1^+,k_1^-)$ with $\CA(k_2^+,k_2^-)$.

The tensor product formula is given as follows. Denote the
generators of the two {\it commuting} $\CN=4$ algebras by $G_1^a,
G_2^a$, etc. Then we form:
\eqn\tensprd{ \eqalign{ T & = T_1 + T_2 + \half \p (p U_1 + q
U_2) \cr G^a & = G_1^a + G_2^a + \p (p Q^a_1 + q Q^a_2) \cr
A^{\pm,i} & = A^{\pm,i}_1+ A^{\pm,i}_2\cr Q^a & = Q_1^a + Q_2^a
\cr U & = U_1 + U_2 \cr} }
with
\eqn\eqfrpa{ p = 2 {k_1^+ k_2^- - k_1^- k_2^+ \over k_1 ( k_1 +
k_2) } }
\eqn\eqfrpb{ q = 2 {k_2^+ k_1^- - k_2^- k_1^+ \over k_2 ( k_1 +
k_2) } }
where $k_i=k_i^++k_i^-$, $i=1,2$.
Moreover, this is the unique way of combining the generators to
form a large $\CN=4$ algebra.

{\bf Remarks}

\item{1.} The computation of the $AG$ commutator shows
that one {\it cannot} give a Feigin-Fuks
deformation of a single copy of the large $\CN=4$ algebra, it is too
rigid. This is actually a special case of \eqfrpa\eqfrpb\ with
$k_2^\pm =0$.

\item{2.} Note that $p=q=0$ when $k_1^\pm =\lambda k_2^\pm$. Thus, for example, in
symmetric products the generators are simply made by direct sum.

\item{3.}  Given a large $\CN=4$ algebra one can form \STVP\  a small
$\CN=4$ algebra $\hat \CA = (\hat L, \hat G^a, A^+)$
with $c= 6 k^+$ and
\eqn\subaone{ \eqalign{ \hat T & = T + {k^+\over k} \p U \cr \hat
G^a & = G^a  + 2 {k^+\over k} \p Q^a . \cr} }
On the other hand, one can also form a small $\CN=4$ algebra $\check
\CA= (\check L, \check G^a, A^-)$ with   $c=6 k^-$ and
\eqn\subatwo{ \eqalign{ \check T & = T - {k^-\over k} \p U \cr
\check G^a & = G^a  - 2 {k^-\over k} \p Q^a . \cr} }
We find that $\hat \CA = \hat \CA_1 \oplus \hat \CA_2$ and
$\check \CA = \check \CA_1 \oplus \check \CA_2$ are small $\CN=4$
algebras; and now note that $p$ and $q$ in \tensprd\
have opposite signs.

\item{4.} It is useful to state the combination rule in terms of
an effective bosonizing field defined by
\eqn\bosonfld{
U := \sqrt{k\over 2} \p \phi
}
Then when combining two algebras we have, by \tensprd\
\eqn\tensprd{ \phi_{12}: = \sqrt{k_1\over k_1+ k_2} \phi_1 +
\sqrt{k_2\over k_1+ k_2} \phi_2 }
The orthogonal linear combination is a linear-dilaton field,
\eqn\tensprd{
\phi_{L}: = \sqrt{k_2\over k_1+ k_2} \phi_1 -  \sqrt{k_1\over k_1+ k_2} \phi_2
}
contributing to the
stress tensor as
\eqn\strsstnsr{
T = - \half (\p \phi_L)^2 + {Q_{12}\over 2} \p^2 \phi_L + \cdots
}
where
\eqn\qfactor{ Q_{12} =\sqrt{2} {k_1^+k_2^--k_2^+k_1^- \over
\sqrt{k_1 k_2 ( k_1+k_2) } } }
Note the interesting fact that if we combine three theories then
\eqn\nonassoci{
Q_{(12)3} \not= Q_{1 (23) }
}
So this operation of combining large $N=4$ theories is {\it nonassociative}!


\newsec{Symmetric product CFT's with large $\CN=4$}

As we have mentioned, a natural candidate for the spacetime CFT dual is
the symmetric product of a simple CFT $\CS_\kappa$ with large $\CN=4$
\eqn\symmt{ {\rm Sym}^N (\CS_\kappa) = (\CS_\kappa)^N / S_N \ .}
More precisely, we would like to explore the possibility
that this symmetric product CFT is on the same
moduli space as the supergravity regime of string
theory on $AdS_3\times \S^3\times\S^3\times\S^1$;
the perturbative CFT regime and the supergravity
regime are typically well-separated in the moduli
space, as befits a strong-weak coupling duality.

\subsec{General structure of symmetric product orbifolds}

To begin, let us recall some of the features
of symmetric product orbifolds that suggest
their relation to supergravity.  First and foremost
is the match between the BPS spectra.
We will discuss in detail this matching below,
for specific examples related to $AdS_3\times\S^3\times\S^3\times\S^1$.
First let us discuss the common features of all such orbifolds.

BPS states of an orbifold come from the ground states
of twisted sectors.  Twisted sectors are in one-to-one correspondence
with the conjugacy classes of the orbifold group.
In the case of the symmetric product orbifold,
the conjugacy classes $[g]$ of $g\in S_N$
can be decomposed into combinations of cyclic
permutations, $[g] = \prod (n_i)^{m_i}$,
where $(n)$ is a cycle of length $n$ in $S_N$.
This carries the structure of a Fock space of identical
particles, in that cycles of the same length
are symmetrized over, and represent identical objects.
One is thus led to the idea that twist operators for
single cycles create one-particle states from the CFT
vacuum, and that twist operators containing several
cycles correspond to multiparticle states.

Of course, the notion of Fock space only makes sense
at weak coupling, \ie\ large $N$.  Consider the twist
operator for a cycle of length $n$:%
\foot{The discussion here parallels \LuninYV.}
\eqn\cyctw{
  \CO_n = \frac{\lambda_n}{N!}\sum_{h\in S_N} \sigma_{h(1...n)h^{-1}}
}
where $\sigma_{(1...n)}$ is the normalized twist operator
permuting the first $n$ copies of $\CS$,
\eqn\twnorm{
  \vev{\,\sigma_{(1...n)}^\dagger\,\sigma^{~}_{(1...n)}\,}=1\ .
}
We will abbreviate $\sigma_{(1...n)}\equiv\sigma_n$.
If we then demand that $\CO_n$ is unit normalized,
we find
\eqn\twnormcoeff{
  \lambda_n= \left[\frac{n\;(N-n)!}{N!}\right]^{1/2}
    \sim \sqrt{n\,N^{-n}}\quad,\qquad N\to\infty
}
by elementary combinatorics.
The operator product of twists
obeys the selection rules
\eqn\selrule{
  \sigma_m\sigma_n\sim \sum_{p}C_{mn}^{p}[\sigma_p]\quad,\qquad
    p\in\left\{\,|m-n|+1\,,\,|m-n|+3\,,\,\dots,m+n-1\,\right\}\ ,
}
and one can readily see that at large $N$ one has
\eqn\threetw{
  \vev{\,\CO_m\,\CO_n\,\CO_p\,}
    \sim \sqrt{\frac{mnp}{N}}~
        C_{mnp}
}
where the leading behavior of $C_{mnp}$
is $N$-independent.
The $C_{mnp}$ can be calculated \LuninYV\
using an application of the covering space method of \DixonQV.
This scaling is consistent with that of the string coupling
in the NS background, $g_B^2\sim 1/Q_1\propto 1/c$.
Note the similarity to the large $N$ scaling of
operator products in the $AdS_5$/SYM duality \BanksDD.

When we apply an operator to one of the
states of the symmetric product,
say for instance the ground state of a cyclic twist,
at large $N$ the result will be predominantly states
in twisted sectors with two cycles
(we assume that the second operator does not simply
annihilate the first one) with coefficient $O(N^0)$;
there will also be a small admixture at order $N^{-1/2}$
of twist sectors of single cycles
according to the interaction \threetw.
At large $N$, the mixing of various twisted sectors
is suppressed by $N^{-1/2}$.
Consequently, it is natural to identify the
cycles of the symmetric orbifold as the analogue
of single trace operators in gauge theory,
which realize the single particle excitations of supergravity;
we may regard the
twist operators for cycles as the creation/annihilation
operators for single particles.
This structure will be important below in understanding
the Hilbert space.


\subsec{Twist operators and moduli of the symmetric product}

As explained in section 4.4, the moduli are BPS states
of the form $(\hf,\hf)_S$. Any large $\CN=4$ theory contains at
least one modulus, $U\bar U$, which changes the radius of the
$U(1)$ in the algebra. In a symmetric product orbifold, this
yields two moduli in the untwisted sector: $\sum_i U_i\bar U_i$,
and $|\sum_i U_i|^2$.  The latter `bi-singleton' perturbation does
not correspond to a single particle
operator in supergravity.%
\foot{It is the analogue of a double-trace operator
in the $\CN=4$ SYM/$AdS_5\times \S^5$ correspondence
\AharonyPA.
Such perturbations also exist in the small $\CN=4$
theory on $AdS_3\times \S^3\times T^4$ \AharonyDP;
one has the 20 moduli from supergravity deformations of the
background, and in addition an $8\times 8=64$
dimensional moduli space from the eight left- and right-moving
currents coupling to the charges of wrapped branes on $T^4$.}
We then wish to identify the former with the combination of the
dilaton operator and the $\S^1$ radius deformation which is the
modulus $\Im\tau$ of supergravity (see section 2.5). If a
symmetric product is related to the spacetime CFT dual to
supergravity, we need to find the second modulus corresponding to
$\Re\tau$ in supergravity, equation \dxc. A simple theory, such as
any of the $\CS_\kappa $ theories \cthree, \algtwo, has no
additional moduli beyond the universal one; any second modulus
must come from twisted sectors of the symmetric product orbifold.

In this subsection, we will construct not only this marginal
twist operator, but also all the chiral twist operators
of the symmetric product (that is, all the operators
with $h=\l^+=\l^-$ and $u=0$ which are chiral
under the $\CN=2$ subalgebra of large $\CN=4$).
As discussed above,
the single-particle chiral operators
are built out of the more basic chiral twist operators
for $\Z_n$ cyclic twists.

In order to construct cyclic chiral twist fields in the symmetric
product CFT \symmt, it is convenient to recall the properties of a
generic symmetric product CFT based on $\sym^N(\CS_\kappa)$, where
$\CS_\kappa$ has central charge $c$.\foot{The following analysis
can be made for any symmetric product CFT. Those based on
$\CS_\kappa$ are of special interest as effective theories for the
GKS long strings discussed in section 8.} Given an operator in
$\CS_\kappa$ with dimension $h_0$ and $R$-charge $R_0$, there is
an operator in the $\Z_n$ twisted sector of $\CS_\kappa^n / \Z_n$ with
dimension and $R$-charge given by \KS:
\eqn\hnfla{
h_n = {h_0 \over n} + {c \over 24} {n^2 - 1 \over n}
\quad , \quad R_n = R_0 }
For example, if we apply this formula to the ground state $h_0=R_0=0$
of the Neveu-Schwarz sector, we obtain a singlet state in
the $\Z_n$ twisted sector with conformal dimension
\eqn\snh{ h_{n,{\rm gd}}
= {c \over 24} \left( n - {1 \over n} \right) }
This state corresponds to a non-chiral twist operator $\s_n$ which
permutes the copies of $\CS_\kappa$. The dimension \snh\ of the
twist operator $\s_n$ can be understood as a difference between
the vacuum energy in a theory based on $n$ separate copies of
$\CS_\kappa$ and a theory on a single copy of $\CS_\kappa$, defined on
the covering space defined by the map $t\sim z^n$ of the parameter
space of the CFT. 

In order to build the chiral twist spectrum, we must use
nontrivial operators of $\CS_\kappa$ carrying the appropriate
$R$-charges. 
%
Recall the
$\CS_\kappa$ theory consists of a bosonic SU(2) WZW model at level
$\kappa=k^--1$, a free boson, and four free fermions.  The scale
dimensions of the conformal highest weight states of these
respective factors, and their contributions to the various
$R$-charges (the $SU(2)$ spins $\ell^\pm$ and the $U(1)$ charge
$u$), are as follows: The bosonic contributions are \eqn\bhwdims{
\eqalign{
  h_b &= \Bigl[\frac{j(j+1)}{k^-+1} +
        jw +\frac{w^2(k^--1)}{4}\Bigr] + u^2 \cr
  \ell_b^- &= j + \coeff12 w(k^--1) \ ,\quad
        j=0,\coeff12,...,\coeff12(k^--1)\ ,\quad w=0,1,2,...\cr
  \ell_b^+ &= 0
}} (here $j$ is the spin of an $SU(2)$ level $k^--1$ highest
weight representation, and $w$ is a spectral flow index), while
the fermionic contributions are \eqn\fhwdims{ \eqalign{
  h_f &= (\ell_f^+)^2 + (\ell_f^-)^2 \cr
  \ell_f^\pm &= 0,\coeff12,1,\dots
}}
Then using $h_0=h_b+h_f$ in \hnfla\ with the choices
\eqn\chsp{
  \ell_b^- = j+\coeff12 w(k^--1)\ ,\quad
  u = 0 \ ,\quad
  \ell_f^- = \coeff12 w\ ,\quad
  \ell_f^+ = j+\coeff12 w k^-
}
leads to a spectrum of chiral operators with
\eqn\chspec{
\eqalign{
  h_n &= \ell^- = \ell^+ = j+\coeff12 wk^-  \cr
  n &= 2j+1+w(k^-+1)
}} where again $w=0,1,...$, and
$j=0,\coeff12,...,\coeff12(k^--1)$.

Some other important properties of the chiral spectrum are that
the chiral spectrum for $k^-=1$ appears only in the sectors with
odd twist, since $n=2w+1$. On the other hand, for $k^->1$ all
twist sectors contribute to the chiral spectrum. Note also that
there are no gaps in the chiral spectrum. All values of $\ell^\pm$
occur, up to the bound set by the stringy exclusion principle; the
maximum twist $n\le N$ implicitly restricts $\ell^\pm \lessapprox
N/2$ via \chspec.

In addition to the chiral spectrum of these twist ground states,
one can construct chiral operators by applying the fermionic
operator $Q$, which can raise both the spin and the dimension by
one half. We can now see explicitly what part of the action of $Q$
is `one-particle', and what part `two-particle'. Consider the
twist $(n)(1)^{N-n}$.  The operator $Q$ decomposes into
\eqn\Qn{
  Q_n^a \equiv \sum_{i=1}^n \psi_i^a
}
and the remainder, $Q-Q_n$.  The two-particle component
is $(Q-Q_n)\sigma_n$, while the one-particle component
is $Q_n\sigma_n$.  Summing over the symmetric group as
in \cyctw\ and normalizing as in \twnorm-\twnormcoeff,
we see indeed that the single-particle component
is suppressed by a factor of $g_s\sim N^{-1/2}$ at large $N$.
Restricting consideration to the single-particle BPS spectrum,
we see that the twisted sector of order $n=2\ell+1$
($n=4\ell+1$ for $\kappa=0$)
gives rise to a quartet of chiral operators,
with
\eqn\chiralwts{
  (h,\bar h)=(\ell^\pm,\bar\ell^\pm)=
    (\ell,\ell)\ ,\ \
    (\ell+\coeff12,\ell)\ ,\ \
    (\ell+\coeff12,\ell)\ ,\ \
    (\ell+\coeff12,\ell+\coeff12)\ .
}
Two are fermionic and two are bosonic,
and so their contribution to the index \soperat\ cancels.

Note that there is always a chiral twist operator with $h=\ell^- =
\ell^+ =\hf$, which we identify with the second modulus
corresponding to $\Re\tau$ in supergravity. Generically this
modulus is in the $\IZ_2$ twisted sector, with $j=\hf$ and $w=0$;
however, in the special case $k^-=1$ we find the modulus in the
$\IZ_3$ twisted sector, with $j=0$ and $w=1$
\refs{\ElitzurMM,\AGS}.

One can also see that this modulus is a RR operator.
RR fields are odd under $(-1)^{F_{\!L}}$; this operation
maps to parity of the spacetime CFT.  The perturbative
regime of the spacetime CFT is the weak coupling
limit of the theory in the RR duality frame; thus
we should identify the Wess-Zumino term of the
SU(2) WZW model with the background RR three-form flux
through $\S^3_-$; the distinguishing characteristic
of this term is its odd parity.%
\foot{Similarly, in the D1-D5 system on \eg\ $T^4$,
the (parity-even) metric moduli of the $T^4$ in the spacetime CFT
map to NS moduli -- the shape moduli map onto one another,
and the $T^4$ volume of the spacetime CFT
maps to the six dimensional string coupling $g_6^2=g_s^2/V_{T^4}$
in supergravity.  On the other hand, the parity-odd
moduli (the antisymmetric tensor $B^{\rm\sst (cft)}_{ij}$)
maps to the RR deformation $C^{\rm\sst(sugra)}_{ij}$.
Furthermore, the twisted sector moduli
$\CT^{a\bar b}=G_{-\hf}^{a\alpha}\bar {G_{-\hf}^{b\beta}}
    \Phi^{tw}_{\alpha\bar\beta}$
of $\sym^N(T^4)$ (here $\Phi^{tw}_{\alpha\bar\beta}$
is the $h=\ell=\hf$ highest weight twist field)
can be seen to decompose into a parity-odd singlet,
which is the RR axion; and a parity-even triplet,
which comprises the self-dual NS B-field moduli of $T^4$.}
Indeed, the (parity-even) radius modulus of $\S^1$ in the spacetime
CFT is identified with the (NS sector) dilaton.
The twisted sector modulus is
\eqn\twm{
  \CT = G_{-\hf}^{\alpha\dot\alpha}\bar {G_{-\hf}^{\beta\dot\beta}}
        \Phi^{tw}_{\alpha\dot\alpha;\bar{\beta\dot\beta}}\ ;
}
this operator is parity-odd, since the lowest component
$\Phi^{tw}_{\alpha\dot\alpha;\bar{\beta\dot\beta}}$
of the twist field multiplet is parity-even,
and parity interchanges the two supercharges
$G$, $\bar G$, thus introducing a fermion minus sign.
Hence it is natural to identify the twist modulus
with the RR axion.

For $k^->1$ we also get a geometrical picture of the twist
modulus.  The bosonic target space $U(2)$ is four dimensional, so
as usual the $\IZ_2$ twist blows up the diagonal in $U(2)\times
U(2)$, which locally looks like $\IR^4/\IZ_2$.  It would appear
that the modulus is a B-flux through the ${\bf P}^1$ of the
resolution, as is familiar from other contexts \AspinwallEV,
and turning off this B-flux results in a singular CFT.%
\foot{The triplet of geometrical blowups of $\IR^4/\IZ_2$ are not
moduli in the present context, since $U(2)$ is not hyperkahler .}

Thus the twisted sector modulus acts as a kind of B-flux turned on
by a finite amount at the orbifold point, that resolves the
geometrical singularities of the orbifold.  This B-flux is a
periodic modulus; we have argued that it is the RR axion and has
period in the given (RR) background is
$\Re\tau \sim \Re\tau + d$. Symmetry considerations analogous
to those discussed in \WittenYU\ lead one to suspect that the
orbifold locus is the line $\Re\tau=\hf d$. The twist
modulus is parity odd, hence at generic points on the moduli
space, the spacetime CFT does not respect parity.  There are
however two points, $\Re\tau=0$ and $\Re\tau= d/2$
(\ie\ the half-period points) at which parity is conserved.
The line $\Re\tau=0$ is the singular locus, thus the (non-singular)
symmetric orbifold CFT could lie on the line $\Re\tau=\hf d$.

Are there other BPS multiplets in the cyclic twist spectrum?
Apart from the $N^{\rm th}$ twisted sector,
the answer is no.
Potential BPS multiplets in the symmetric product with
$\ell^+\ne \ell^-$ will not have a contribution
$\frac{(\ell^+-\ell^-)^2}{N(k^++k^-)}$
to their energy unless we are in the $N^{\rm th}$ twisted sector,
where these states come from applying $\kappa/N$-moded
fermion oscillators to the chiral twist ground state.%
\foot{Below, we will exhibit these BPS states with
$\ell^+\ne\ell^-$ in the $N^{\rm th}$ twisted sector, in the
special case $k^-=1$.} This means that as one perturbs across the
moduli space from the supergravity regime to the symmetric
orbifold regime, states with $\ell^+\ne\ell^-$ are not protected
and move off the BPS bound. This was observed for $k^-=1$ in
\ElitzurMM.


As an aside, the spectrum \hnfla\
makes it clear why the orbifold locus is outside
the geometrical regime of the spacetime CFT.
Suppose we are taking the symmetric orbifold of some theory
$W$ of central charge $c_w$.
Consider the theory at some fixed energy $E$.
The way to partition this energy
that maximizes the entropy is to take half
of it to make a long string, that is to go to the twist sector of
a single cyle of order $E/c_w$; then populate that long string
with oscillators using the remaining energy (the oscillator gap
will be $c_w/E$ and so the string will be thermalized if $c_w$
is not too big).  The entropy is thus of order
\eqn\hagent{
  S \sim \sqrt{c_w  (E/c_w) E} \sim E
}
(here $c_w (E/c_w)$ is the effective central charge of the long string),
\ie\ the symmetric product has a Hagedorn spectrum as soon as
the long strings can be thermalized \BanksDD.
For this we need the temperature to be larger than the gap.
The temperature is determined by \eg\
$ S = c_w LT^2 $ where $L$ is the length of the long string,
which is $E/c_w$.  Since $S\sim E$ we have $T\sim 1$.
So as soon as $E> c_w$ we are in the Hagedorn regime --
there is no gap parametrically large in the order
of the symmetric product between the $AdS$ scale
(order one in our conventions) and the string scale.


\subsec{Explicit construction}

One can give an explicit construction of the cyclic twist
operators in symmetric product orbifolds, in the case where the
component theory is $\CS=\CS_0$. This theory consists of one free
boson $\phi_r$ and four free fermions $\psi_r^a$, and so the
cyclic twist operators can be built out of standard orbifold twist
operators, see for example \DixonQV. Here $r=1,...,N$ labels the
copies of $\CS_0$ of the symmetric product.

The cyclic twist of order $n$ permutes $n$ copies of $\CS_0$
labelled by $r=1,...,n$ via $r\to r+1$ with $r+n\equiv r$.
A discrete Fourier transform diagonalizes the twist:
\eqn\twdiag{
  \phi_\nu = \frac1n \sum_{r=1}^{n}  exp[2\pi i r \nu/n]\, \phi_r
}
and similarly for the fermions.
The action of the twist on $\phi_\nu$ is then
rotation by $\omega^\nu$, where $\omega=exp[2\pi i/n]$; similarly
for the fermions.
To keep explicit the $SU(2)\times SU(2)$ content,
it is convenient to bosonize the fermions.
Define bosons $H_\nu$, $H'_\nu$ with corresponding exponentials
representing the fermions $exp[\pm i H_\nu]$ and $exp[\pm iH'_\nu]$.
Note that $H$, $H'$ are {\it not} the bosons corresponding to the
Cartan subalgebra of $SU(2)\times SU(2)$;
the latter are $\half(H\pm H')$.

The standard $\Z_n$ twist operator $\sigma_\nu$ for $\phi_\nu$
has dimension
\eqn\hkappa{
  h_\nu = \frac14\frac\nu n\Bigl(1-\frac\nu n\Bigr)\ ,
}
and the full bosonic twist operator is the product
of the twist operators for each $\phi_\nu$, $\nu=1,...,n-1$:
\eqn\bostw{
  \sigma_n^{\rm bos} = \prod_{\nu=1}^{n-1}\sigma_{\nu}\quad,\qquad
    h_{n}^{\rm bos} = \sum_{\nu=1}^{n-1}h_\nu=\frac{n^2-1}{24 n}\ .
}
A fermionic twist operator with the appropriate monodromy is
\eqn\fermtw{
  \sigma_n^{\rm ferm} = \prod_{\nu=1}^{n-1}
    exp[i\coeff{\nu}n(H_\nu+H'_\nu)]\quad,\qquad
    h_n^{\rm ferm} = \frac{(n-1)(2n-1)}{6n}\ .
}
The full twist operator is then
$\sigma_n = \sigma_n^{\rm bos}\sigma_n^{\rm ferm}$,
whose quantum numbers are
\eqn\twqnos{
\eqalign{
  h &= (n-1)(3n-1)/8n\cr
  \ell^+ &= (n-1)/2\cr
  \ell^- &= 0\ .
}}
Note that this operator is on the unitarity bound
\eqn\ubd{
   h = \frac{k^-\ell^+ + k^+\ell^- + (\ell^+ - \ell^-)^2}{k^+ + k^-}
}
if we take $k^+=k^-=n$ for the $n$ copies being wound together;
however, this lies above the unitarity bound for
$k^+=k^-=N$ of the full symmetric product.
Successive operator products with the antifermions
$exp[-iH'_\nu]$, $\nu=n-1,n-2,\ldots$,
lowers $\ell^+$ by 1/2 and raises $\ell^-$ by 1/2
for each applied antifermion, while staying on
the bound \ubd\ for $k^+=k^-=n$.
When $n$ is odd, applying the $\half(n-1)$ antifermions for
$\nu=\hf(n+1),\dots,n-1$ yields a BPS twist operator
with quantum numbers
\eqn\bpstw{
  h = \ell^+ = \ell^- = \coeff14(n-1)\ .
}
This is the operator whose existence was inferred from
spectral flow arguments in the previous subsection.

To summarize, for the symmetric product $\sym^N(\CS_\kappa)$
there are `single-particle' BPS states in twisted sectors
for each $\ell^+=\ell^-=0,\hf,...,\hf[\frac12(N-1)]$;
in addition, for $\CS_\kappa$ at level $\kappa=0$,
we have exhibited BPS states
with $\ell^++\ell^-=\hf(N-1)$ for $\ell^-=0,1,...,\hf(N-1)$.


\subsec{A conjectural geometrical interpretation of the chiral spectrum}

It is very important to understand what part of the spectrum of the
theory is invariant under perturbations by the modulus \twm.
 In the next section we will
examine the large $\CN=4$ index for these theories. While this
detects some invariant states, it turns out not to detect all the important
ones. In this section we argue for the existence of some protected states,
which turn out not to be detected by the index.

The spectrum of chiral operators we have found above bears some similarity
to that of the small $\CN=4$ symmetric product $\sym^N(T^4)$,
which points to a possible geometrical interpretation.
So let us recall the single-particle chiral twist
spectrum of $\sym^N(T^4)$
\refs{\MaldacenaBW,\deBoerIP,\LarsenUK}.
There is again a chiral twist field
$\Phi^{\alpha_1\cdots\alpha_{n}}_{{\bar\alpha}_1\cdots{\bar\alpha}_{n}}$
for every $(n+1)$-cycle in the symmetric group,
$n=1,...,N-1$, with quantum numbers
$(h,\bar h)=(\ell,\bar\ell)=(\frac{n}2,\frac{n}2)$.

The $\Z_{n+1}$ cyclic twist highest weight states of the symmetric product
can be given a cohomological interpretation
in terms of the hyperkahler resolution of the singularities
along the $n+1$-fold diagonal of $\sym^N(T^4)$.
This resolution blows up the diagonal, such that the
(orbifold) cohomology of the symmetric product has
a representative in dimension $2n$.

Furthermore, the isometries of $T^4$ lead to four
$U(1)$ currents $J^{\dot a a}$,
and their superpartners $\psi^{\dot a\alpha}$
(and similarly for right-movers).  Here $\alpha$ is an
$SU(2)$ doublet index under the small $\CN=4$ algebra,
$a$ is a doublet index for the custodial $SU(2)$,
and $\dot a=1,2$.
In the symmetric product, the diagonal U(1) fermion field
acts much as in \Qn\ to generate a collection of
single-particle operators built on $\Phi$; starting
with the highest weight state, we can act with
$\psi^{\dot a+}$ to make two additional states with $\ell=\hf(n+1)$,
and act again to make one more state with $\ell=\hf(n+2)$.
Combined with the action of the right-moving
$\bar\psi^{\dot{\bar a}\bar\alpha}$,
there are all told $16=8_B+8_F$ states built on $\Phi$,
with a spectrum analogous to \chiralwts.

{}From the geometrical viewpoint,
the  chiral operators in $\sym^N(T^4)$ can   be interpreted
in terms of the cohomology of the (hyperkahler) target.
The twist highest weight ground states are identified
with the cohomology of the resolution of diagonals
in the symmetric product, and
the action of the fermions can be identified with
the product in cohomology with the eight even and eight odd
cohomology classes of $T^4$
\refs{\MaldacenaBW,\deBoerIP,\MaldacenaBP}.

Similarly, we would like to identify the chiral twist fields
$\Phi$
of the large $\CN=4$ theory $\sym^N(\CS_\kappa)$
with even cohomology elements of some  resolution of
diagonals of  the complex orbifold $\sym^N(\S^3\times\S^1)$.%
\foot{In this context, note that $\S^3\times \S^1$
is a rather special target.  It is the unique WZW model
whose left and right complex structures commute \RocekVK;
in fact it has a quaternionic structure, with
two commuting triplets of complex structures
\refs{\STVP,\SpindelSR,\RocekVK}.
Its Dolbeault cohomology is
$H^{p,q}=\IC$ for $(p,q) = (0,0), (0,1), (2,1)$, and $(2,2)$, and trivial
otherwise.  Additional interesting facts
about $\S^3\times\S^1$ may be found in \refs{\braam,\IvanovAI}.}
We saw an example of this above, when we argued that
the fixed locus of
the $\IZ_2$ twist is resolved by a $B$-flux through
a string-sized $\IP^1$.
We also wish to identify the action of the fermion \Qn\ that makes the two
bosonic and two fermionic states \chiralwts, with
the action of tensoring with the
two even and two odd cohomology classes of $\S^3\times\S^1$.
This would account for all the chiral cohomology states
exhibited above.

We expect that just as there is a smooth metric on a small $\CN=4$
resolution of ${\rm Sym}^N(T^4)$ and ${\rm Sym}^N(K3)$, there is
also a smooth metric on a large $\CN=4$ resolution $\tilde X \to
{\rm Sym}^N(\S^3 \times \S^1)$ which can be used to define an
$\CN=2$ sigma model. The chiral primaries of this model will be
given by the cohomology of $\tilde X$, and will be invariant under
smooth deformations of $\tilde X$, which we suppose to include the
perturbations inherited from \twm. If this interpretation is
correct, it would go a long way to explaining why (as we will see
below in section 7) the chiral twist spectrum is seen both in the
symmetric product and in the supergravity limit, whereas the BPS
states with $\ell^+\ne\ell^-$ are seen in supergravity but not in
the symmetric product.  The latter states would not be associated
to any particular cohomology of the target, and being paired up
into long representations, nothing prevents them from being lifted
as we move around the moduli space.  On the other hand, the chiral
states are, according to the above proposal, associated to
cohomology; even though they are invisible to the index,
nonetheless they are not lifted as we cross the moduli space
unless we move to a singular point where the cohomology disappears
(such as the singular locus at $C_0=0$).

If the chiral ring is preserved across moduli space,
then we can rule out iterated symmetric products
such as $\sym^{Q_1}[\sym^{Q_5}(\CS)]$, as candidate
duals.%
\foot{Or $\sym^{Q_5}[\sym^{Q_1}(\CS)]$; note that this is distinct
from $\sym^{Q_1}[\sym^{Q_5}(\CS)]$.}
The chiral ring in this situation differs in the states
we would call multiparticle BPS states; for example,
the two-particle states correspond to words in the symmetric
group that are products of two cycles.  In the iterated
symmetric product, one has a choice
of whether these two cycles come from the same
$\sym^{Q_5}(\CS)$ component or different ones.
In the single symmetric product, there is only
one state of this type.  One readily sees that
the growth of states is much faster than that
of the Fock space of BPS supergravity states.
Note that, apart from this problem, the iterated
symmetric product appears to pass the other
tests of a duality; the central charge is correct,
there is a long string sector
with gap of order $\frac{1}{Q_1Q_5}$, and one can
show that the index is the same as that of the
single symmetric product of order $Q_1Q_5$
whenever $Q_1$ and $Q_5$ are relatively prime \indexsummary.


\newsec{The index for the theory $\syms$}

In this section we summarize briefly the result of some
computations of the index defined
 in section 4.6 above for the
theory $\syms$. Details of the computations can be found in  
our companion paper \indexsummary.

We consider the theory $\CC = \syms$ with $\cag$ symmetry.
The formula for $I_2(\CC)$ is:
\eqn\itwoindx{ \eqalign{ I_2(\CC) &  = \sum_{ad=N}
\sum_{n_0,m_0=0}^{d-1} a \Theta_{a(4n_0+1),k}^-(\omega,\tau)
\overline{ \Theta_{a(4m_0+1),k}^-(\tl \omega,\tl \tau) }\cdot \cr
&\qquad \qquad  \cdot {1\over d} \sum_{b=0}^{d-1}e^{2\pi i {b
\over d}(n_0-m_0)(2n_0+2m_0+1)} Z_{\Gamma}({a\tau+b \over d}) \cr}
}
where
\eqn\narain{
Z_{\Gamma}
=
\sum_{\Gamma^{1,1}} q^{\half p_L^2} \bar q^{\half p_R^2}
}
is the standard Siegel-Narain theta function
for the  compact scalar of radius $R$ in
the theory $\CS$. In \itwoindx\ we sum over factorizations $N=ad$. 
As discussed at length in \indexsummary\ the $(a=N,d=1)$ term 
should be identified as a 
``short string'' and the $(a=1, d=N)$ term as a  ``long string'' 
contribution. 

To simplify
matters we assume that $N$ is prime and we restrict
attention to the charge zero sector.
The result is
\eqn\itoosym{ I_2^0(\CC) = (N+1) \vert \Theta^-_{N,k}\vert^2 +
\sum_{\mu >0, {\rm odd}}^{N-2} \biggl\vert \Theta^-_{\mu,k} +
\Theta^-_{2N-\mu,k}\biggr\vert^2 }
Here and below the conjugation operation implied in $\vert \Theta \vert^2$
takes $\omega_\pm\to \tilde \omega_\pm $ and acts as complex conjugation.

Turning to $I_1$, we find  the simplest RR spectrum
consistent with this index is
\eqn\simplest{
\eqalign{
& \oplus_{\l^-=1/2}^{N/2} \vert ({N+1\over 2} - \l^-, \l^-) \vert^2 \cr
&
\oplus_{\l=1/2}^{(N-1)/4} \biggl\vert (\l,\l)
+ ({N+1\over 2}-\l, {N+1\over 2}-\l) \biggr\vert^2
\oplus \biggl\vert   ({N+1\over 4} , {N+1\over 4} ) \biggr\vert^2 \cr}
}
where the first line comes from the short string   and the
second from the long string  contribution.  The short string states have
$h= {N\over 4} + {u^2\over 2N}$ for all states, and the gap to the next excited state is
order $1$. The long string states   have
\eqn\lngwt{
h= {N\over 8} + {(4\l -1)^2\over 8N}
}
and have small gaps $\sim 1/N$ to the first excited state.

Applying spectral flow to the representation \simplest\ gives:
\eqn\simplestns{
\eqalign{
& \oplus_{\l=0}^{(N-1)/2} \vert (\l,\l)_{NS} \vert^2 \cr
& \oplus_{\l=0}^{(N-3)/4} \biggl\vert ({N-1\over 2}-\l,\l)_{NS}
+ (\l, {N-1\over 2}-\l)_{NS} \biggr\vert^2
\oplus \biggl\vert   ({N-1\over 4} , {N-1\over 4} )_{NS} \biggr\vert^2 \cr}
}
where the first line is from the short string contribution $a=N, d=1$
and the second from the long string  contribution $a=1,d=N$.
The short string states have
\eqn\shrtwt{
h= \l
}
while the long string states have:
\eqn\lngwt{
h= {N-1\over 4} + {(N-1-4\l)^2\over 8N}  }

In the case of the general $\sym^N(\CS_\kappa)$ theory,
where the $U(2)$ is at level $\kappa$,
we have not managed to evaluate the index completely.
However, the short-string contribution is amenable to analysis
and the simplest (BPS,BPS) spectrum consistent with the index is
\eqn\rmadni{
\oplus_{j=0}^{\kappa/2} \oplus_{a=0}^{N-1}
\vert (  Nj + (a+1)/2,  (N-a)/2 )\vert^2
}
Upon spectral flow to the NS sector this is
\eqn\rmadnii{
\oplus_{j_1=0}^{\kappa/2} \oplus_{j_2=0}^{\half( N-1)}
\vert \bigl(Nj_1 + j_2, j_2\bigr)_{NS}\vert^2
}

The true BPS spectrum of the $\syms$ theory, which can in principle be directly
examined on the orbifold line, differs from that above
by short representations with cancelling indices.
A detailed examination of the states shows that it is most natural to
account for the above spectrum in terms of multiparticle states
of the singleton $Q$ and the modulus operator for $\Im\tau$,
the size of the $\S^1$; these are all operators in
the untwisted sector of the symmetric product.
The twisted sector BPS states constructed in the previous
section always come with partner representations with cancelling index,
as argued around equation \chiralwts.


\newsec{Comparison of BPS spectra}

\subsec{The supergravity spectrum }

Let us ask how the spectrum of supergravity
single-particle states fits into the representations
described in section 4.

Reference  \deBoerRH\ derives the particle content of the
KK reduction of the 10-dimensional type II supergravity
multiplet decomposed in terms of representations of the
$\dto\times \dto$ super-isometries of spacetime.
The KK spectrum  is perhaps most
clearly written as
\eqn\newkksc{
\bigoplus_{\l^+,\l^-\geq 0\, ;\;u}~
\rho(\l^+,\l^-,u)\otimes \overline{\rho(\l^+,\l^-,u)}
}
%
where the highest weight state in $(0,0;0,0)$
corresponds to the vacuum, and not
to a single-particle state.

This result can be understood intuitively as follows.
We should be looking for two things:
(a) 256 polarization states, and
(b) all the on-shell Fourier modes on
$AdS_3\times \S^3 \times \S^3 \times \S^1$.
Vertex operators will be products of left-moving and right-moving
states. Under the $SO(4)_L  = SU(2)_{+,L} \times SU(2)_{-,L}$
isometry of $\S^3_+ \times \S^3_-$, a scalar operator in 10-dimensions
gives rise to a tower of vertex operators
\eqn\towera{
\Phi^{\alpha_1\cdots \alpha_{n^+} }_{\dot \alpha_1\cdots \dot \alpha_{n^-} }
}
where $n^\pm $ are integers and the tensor is totally symmetric.
Since the decomposition of normalizable functions on $\S^3_+ \times \S^3_-$
under $SO(4)_L \times SO(4)_R$ is
\eqn\towerb{
L^2(\S^3_+ \times \S^3_-)  =
\bigoplus_{\l^\pm\geq 0} (\ell^+,\ell^-;\ell^+,\ell^-)\ ,
}
all tensors in \towera\ occur with degeneracy 1,
where $n^\pm = 2 \ell^\pm $.
Now, for $u\ne0$ or $\l^+\ne\l^-$,
there are no vanishings in the application of $G_{-1/2}$
(the only candidate, equation \nullvec, simply relates
$G_{-1/2}$ to $Q_{-1/2}$).  Raising with the four $G$'s
fills out a 16 component base of the representation, leading to
\eqn\sxtr{
  16 (2 \l^+ + 1)(2 \l^- +1)
}
states that are not descendants under $L_{-1}$.
Combining left and right quantum numbers,
we identify this $\cag$ multiplet with the supergravity
multiplet of states carrying angular momentum
$(\l^+,\l^-;\l^+,\l^-)$ and momentum $u$ on $\S^1$.

For $\l^+=\l^-$ and $u=0$, the action of $G_{-1/2}^{\ppd}$ vanishes
according to \nullvec, and we must be more careful. Previously,
\nullvec\ related $G_{-1/2}^\ppd$ to $Q_{-1/2}$; in the present case
$G_{-1/2}^\ppd$ vanishes, however we can still act with $Q_{-1/2}$.
Thus we can also make the vertex operator
\eqn\towerc{
\Phi^{(\alpha_1\cdots \alpha_{n} }_{(\dot \alpha_1\cdots\dot\alpha_{n}}
    Q^{\alpha)}_{\dot \alpha)}
}
(related to the state \newstate),
leading to a $\dto$ short multiplet
$(\l + \half , \l + \half)_\sd$.%
\foot{More precisely, in the regime of supergravity
weak coupling the product \towerc\ decomposes into
an operator creating a two-particle state and
(with a coefficient $g_s$)
an operator creating a one-particle state.
We focus on the single-particle operator component.}
Thus we find the $\dto$ representation content
\eqn\comb{
(\l,\l)_\sd\oplus (\l + \coeff12 ,\l + \coeff12)_\sd
}
obtained by combining \towera\ and \towerc.
The number of states in the $\dto$ representation \comb\
which are not descendents of $L_{-1}$ is
\eqn\sytr{
16 (2 \l + 1)(2 \l +1)
}
(this remains true for the special short
representations having $\l^\pm=\hf$).
Combining left and right quantum numbers,
we identify the multiplet \comb\
formed by \towera\ and \towerc\ with the supergravity
multiplet of states carrying angular momentum
$\l^+=\l^-$ and $u=0$.

We now come to the question of whether the
BPS condition \nullvec\ is sufficient to protect
the conformal dimensions of such states
as we move along the moduli space.
Here we encounter the distinction between $\cag$
and the super-isometry algebra $\dto$;
as mentioned above, their BPS conditions are different
unless $\l^+=\l^-$ and $u=0$.
This is a situation not encountered in other contexts,
such as $AdS_3$ backgrounds with $\CN=2,3$ or
small $\CN=4$.  However, this distinction disappears
in the classical limit $k^++k^-\to\infty$,
where the $\cag$ unitarity bound \unitary\ degenerates
to the $\dto$ bound $h = \frac{k^+ \l^- + k^- \l^+}{k^++k^-}$.
Nevertheless, since this classical dimension {\it violates}
the $\CA_{\gamma}$ BPS bound, and since
$\CA_{\gamma}$ is the true symmetry of the theory,
we know that supergravity states with $\l^+ \not=\l^-$ or $u\ne 0$
{\it must} get a quantum correction to their mass.
Moreover, there must be a corrections to the
$\dto$ BPS condition
$G^{\ppd}_{-1/2}\vert \l^+,\l^-\rangle =0$
for such states.  This is a novel situation in which
{\it states which appear to be BPS in the classical
approximation, in fact can receive quantum corrections.}
This is perhaps an important cautionary tale.

Since we have not computed the corrections
$\sim \frac{(\l^+ - \l^-)^2}{k^++k^-}$
to the masses in string theory on
$AdS_3 \times \S^3 \times \S^3 \times \S^1$,
we do not know if particle states with $\l^+ \not= \l^-$
or $u\ne 0$ arrange themselves into
long or short representations of $\CA_{\gamma}$.
There is no reason that prevents the various single-particle
and multiparticle states in the supergravity Fock space
from combining to form massive representations
that leave the bound \unitary\ in this case.
The same mechanism that arranges the BPS states with
$\ell^+=\ell^-$ into long multiplets
(namely, acting with $Q^a$)
applies also to the states with $\ell^+\ne\ell^-$.
Indeed, in the symmetric product, we saw
that typical single-particle states do not
contribute to the index, and indeed we also saw
that there were no BPS states with $\ell^+\ne\ell^-$ small.

If such states were to remain BPS then the string theory
corrections would have to be  {\it exactly}
\eqn\exctmss{
h = \frac{k^+ \l^- + k^- \l^+ + (\l^+ - \l^-)^2+u^2}{k^++k^-}
}
If true it would be very striking and
would suggest some kind of integrability.

Are there additional BPS objects 
we can consider?  The topological 
classification of D-brane sources is given by the 
(twisted) K-theory of spatial infinity, modulo classes which 
extend to the interior \refs{\MooreGB,\MaldacenaSS}. For both IIA
and IIB theories this group is $\IZ_{Q_5^+} \otimes \IZ_{Q_5^-}$, 
where the two torsion factors come from the twisted $K$-theory 
of $\S^3$ and we are working in the NS flux picture. These are 
classes representing D1-branes wrapping the $\S^1$. 
However, in the familiar way
(reviewed, for example,  in \refs{\SchomerusDC,\MooreVF})
the D-objects blow up into $\S^2$ spheres in each of the $\S^3$ 
factors, so the strings blow up into 5-branes of topology 
$\S^2 \times \S^2 \times \S^1$.  One novelty of the present 
context is that the mathematical identity 
$\IZ_{Q_5^+} \otimes \IZ_{Q_5^-} \cong \IZ_{gcd(Q_5^+,Q_5^-)}$
implies interesting instabilities 
of the Chan-Paton degrees of freedom.%
\foot{Note that if either fivebrane charge is equal to one,
the K-theory is trivial.}


\subsec{Comparison of (BPS,BPS) states with the symmetric product}

One of the key tests of any proposal for a duality is the matching
of the BPS spectrum. Here we focus on the left and right BPS
states and make several loosely connected remarks concerning the
comparison between the supergravity background and the
proposed dual $\syms$.

First, comparison  with the simplest spectrum suggested by
the index \simplestns\ strongly suggests that  the
spectrum of short representations of $\CA_{\gamma}$
associated with  supergravity particles is in fact precisely
\eqn\sugrasp{
\oplus_{\l \geq 0 }^{(N-1)/2} \rho(\l,\l,0) \otimes  \rho(\l,\l,0)
}
where the upper bound is imposed by hand, in supergravity,
as part of the ``stringy exclusion principle'' \MaldacenaBW.

Now, we have actually argued for {\it two}
towers of BPS states in the representations $\vert (\l,\l)\vert^2$
in the symmetric product CFT. On the one hand there are the multiparticle
states made of untwisted sector states, which contribute to the index.
On the other hand, there are the twisted sector states
constructed in section 5.  The latter states are more naturally
identified with the supergravity one-particle states
carrying momentum on $\S^3_\pm$.
The companion representations which cancel in the index may
be understood in terms of boundstates with singletons
(\cf\ the discussion surrounding equation \chiralwts).
We also gave a conjectural cohomological interpretation
to these twisted sector states
which suggests that, even though they cancel in the index,
they might nevertheless be preserved along the moduli space.
%

As we have stressed, there are generically {\it no} BPS states
with $\ell^+\ne\ell^-$ with small $\ell^\pm$
in the symmetric product.%
For instance, in the $\sym^N(\CS_0)$ theory,
the one-particle supergravity states with these quantum numbers
get corrections to their mass of order
$\delta h\sim\frac{(\ell^+-\ell^-)^2}{\ell^++\ell^-}$
at large $N$, under the assumption that the states in \ubd\
should be identified with the supergravity
one-particle states with the corresponding quantum numbers.
On the one hand, one might take this result as
a cautionary tale regarding the extent to which the BPS
property as seen in the supergravity approximation actually
extends to a property of the full theory;
on the other hand, if the symmetric product orbifolds
only describe situations where one of $Q_5^\pm=1$,
supergravity calculations are suspect and there
might not be any contradiction.

It should also be noted that  the spectrum \newkksc\ does not
depend on whether $N= Q_1 Q_5$ is prime, nor
on whether $Q_5$  is equal to $ Q_5'$. On the other
hand, we show in \indexsummary\ that the BPS spectrum of
${\rm Sym}^N(\CS_\kappa)$ depends on the prime factorization
of $N$.  Moreover, the conjectural holographic dual
for $Q_5/Q_5' = \kappa+1$ has a BPS spectrum
which depends on $\kappa$.   For the case $\kappa>0$
we found some BPS states in
\rmadni.
The states with
$j_1>0$ are ``new'' in comparison to the spectrum at
$\kappa=0$.
Note that  the conformal weight of these states is very simple:
\eqn\confdim{
h  = j_2  + N (j_1 (j_1 +1))/(\kappa+2)
}
 It follows that  particles with $j_1>0$  are heavy - parametrically
of order $N$. These states should probably not
be identified with supergravity particles. It is possible they
can be identified with ``conical defect geometries'' or  smooth
versions thereof. Thus, the light spectrum remains
$(j_2,j_2) $  for all the $U(2)_\kappa $ theories
and is insensitive to $ \kappa$,
This is 
at least consistent with the idea that   ${\rm Sym}^N(w(\kappa+1,1))$
is the holographic dual for $Q_1 Q_5 = N(\kappa+1) $, $Q_1 Q_5' = N$ .

On top of the above considerations we are left with  
 the ``long string BPS states'' contributing to line 2 in 
equations \simplest\simplestns. These are
 unaccounted for on the sugra side.
It is possible that these states  correspond
to conical defect geometries smoothed out into
supertubes, along the lines described in \LuninIZ.
The solutions of \LuninIZ\ are specific to the $T^4$ case,
but perhaps could be generalized to $\S^3 \times \S^1$.
Moreover, when   $N$ is not prime,
there will be many further BPS states \indexsummary.
They will be heavy, parametrically having mass of order $N$, but
still, they must  have supergravity duals, since they are BPS.
It should be very interesting to see this structure arising
in the supergravity side.
For N nonprime the construction of \MartinecCF\ might provide some duals to
the new BPS states which are associated with nontrivial divisors of $N$.

To summarize, even in the most promising case where $Q_5^+=1$ or $Q_5^-=1$ there 
are discrepancies in the BPS spectrum between supergravity and the CFT dual. 
However, for reasons discussed above one cannot rule out the $\syms$ theory 
as a CFT dual solely on this basis. 


\newsec{The near-BPS spectrum of
$AdS_3 \times \S^3 \times \S^3 \times \S^1$}

The perturbative ``long string'' spectrum for
$AdS_3\times \S^3\times \S^3\times \S^1 $
provides a point of comparison between the boundary
conformal field theory and supergravity that goes
beyond the BPS spectrum.  States of high spin on
the $\S^3$'s, specifically with $j'=j''\gg 1$ and $h\sim j'$,
are near-BPS states, the so-called BMN states. 
Their dimensions are expected to be slowly varying functions 
along the moduli space.
Thus, we might expect that this portion of the spectrum
to remain intact as we move from the orbifold locus
to the singular locus in moduli space, where
the perturbative GKS description of \GiveonNS\
can be applied.


\subsec{The spectrum of GKS long strings}

In the worldsheet formalism of \GiveonNS,
$AdS_3$ is described by an $SL(2,R)$ WZW algebra of level $k$,
two $SU(2)$ current WZW models of levels $k'=Q_5^+$ and $k''=Q_5^-$
(and a free field theory on $\S^1$); recall that
the levels are related by $1/k = 1/k' + 1/k''$.
Long strings pulled out of the background ensemble,
that wind some number $w$ of times around
the angular direction of $AdS_3$,
are obtained by $w$ units of spectral flow
from primary states in the $SL(2,R)$ WZW model \MaldacenaHW.
The standard worldsheet formalism requires the
absence of RR backgrounds, and so describes the
NS background duality frame with all RR potentials
vanishing.  This is the singular locus of the
spacetime CFT, where the Coulomb branch of separated
onebranes and fivebranes meets the Higgs branch
of onebrane/fivebrane bound states described
by the $AdS_3\times \S^3\times \S^3\times \S^1$ background.

The worldsheet formalism of \GiveonNS\
is a perturbative approximation to
the structure of the exact spacetime CFT.
So for example one builds a Fock space of strings,
ignoring back reaction, stringy exclusion, black holes, \etc.
Back-reaction is a higher loop (and/or nonperturbative/collective) effect.
To the extent that one can ignore these latter effects,
the long string states describe a subspace of the Hilbert space.
However, we are on the singular locus of the spacetime CFT
and one could wonder whether such a subspace or Hilbert space 
is even well-defined. 
We believe that the answer is yes when sugra is weakly
coupled, but then the spacetime CFT
is strongly coupled and hard to analyze.
A rough physical picture is that the long strings
are in a corner of the configuration space of the symmetric
product sigma model
(where a Coulomb branch meets a Higgs branch);
if this is a sufficiently deep pocket in the sigma model
target space, a state can get trapped there for a long time and we can
usefully think of it as a separate entity (like a resonance).
In 1+1 dimensions, the sigma model fields cannot have
expectation values due to infrared fluctutations;
instead they have wavefunctions.
We continue to employ the usual terminology `Higgs' and `Coulomb'
applied to moduli spaces of scalar vevs in higher dimensions;
but these are now regions of configuration space
of the theory where the wavefunctions may have support.
The support of the wavefunctions of long string states
is predominantly on the Coulomb branch.
In that region of configuration space,
it is energetically cheaper for excitations
to be carried by the long string -- \eg\ a
U(1) quantum by itself costs energy going like $n/R$,
while on the long string of winding $w$ it costs
$(1/w)(n/R)^2$ which is smaller
for small enough $n$ and large enough $w$.

Let the $SL(2)\times SU(2)\times SU(2)$ spins
of the worldsheet primaries be denoted $j,j',j''$,
and let the spectral flow winding be $w,w',w''$
(as usual, $j'<k'/2$, the total $SU(2)$ spin of a state
is $\ell'=\half k'w'+j'$, \etc).
Then the formula
for the spacetime energy and spin of a long string is
(\cf\ \MaldacenaHW, eq. 75, for the bosonic string,
and \MartinecCF, eq. 97, for the superstring)%
\foot{To derive this expression, one solves
the worldsheet Virasoro condition $L_0-1=0$
for the spacetime energy $h=m+\half kw$ in
the $SL(2)$ sector of spectral flow winding $w$,
with $m$ the unflowed $J_3$ of $SL(2)$.}
\eqn\longstr{
\eqalign{
  h &= \frac{ kw}{4}
    + \frac1w\Bigl[-\frac{j(j-1)}{k}
    + \Bigl(\frac{j'(j'+1)}{k'} + j'w' + \frac{k'w'^2}{4} \Bigr)\cr
    &\hskip 3cm
    + \Bigl(\frac{j''(j''+1)}{k''} + j''w'' + \frac{k''w''^2}{4} \Bigr)
    + \Delta_{\rm int}-\coeff12\Bigr]\cr
  \ell' &= \hf k'w'+j'\cr
  \ell'' &= \hf k''w''+j''\quad.
}}
It is implicit in these formulae that $w\ge 1$;
the winding number zero sector is the supergravity spectrum
(which is given by a different expression).
Adding $U(1)$ charge $u$ simply puts $\Delta_{\rm int}=u^2$
inside the square bracket
(we use $h$ to denote spacetime energy,
$\Delta$ to denote worldsheet conformal dimension).
The normalization of this term
is set by the winding number zero sector, which is the
supergravity spectrum.  States that satisfy the GSO projection
will need at least one fermion excitation, which we choose
orthogonal to $\S^3\times \S^3$ in order not to deal
with multiple cases according to the addition of angular momenta
(since the fermions along $\S^3\times \S^3$ are vectors
of $SU(2)\times SU(2)$).  Henceforth we will add such
an orthogonal fermion excitation, and
drop the $-1/2$ in the square brackets of \longstr.

The near-BPS states have large $w$ and sufficiently
small $\Delta_{\rm int}$ that the fractional excess of energy
above the BPS bound is tiny.  For simplicity,
let us restrict to states with $\ell'=\ell''$,
for which the BPS bound is particularly simple:
\eqn\bpsbd{
  h\ge\frac{k''\ell'+k'\ell''}{k'+k''} +
        \frac{(\ell'-\ell'')^2 + u^2}{Q_1(k'+k'')}
    = \ell'+\frac{u^2}{Q_1(k'+k'')}\ .
}
All the states in \longstr\ satisfy this bound.


\subsec{Comparison with the symmetric product orbifold $\sym^N(\CS)$}

The symmetric product orbifold $\sym^N(\CS)$
is a candidate for the boundary CFT dual to
the above supergravity, for $k'|k''$.
The orbifold is a non-singular CFT, and
so if it is at all related to supergravity
on $AdS_3\times\S^3\times\S^3\times \S^1$,
it is at a different point in moduli space.
Our working hypothesis is that the deformation of $\Re\tau$
from the orbifold line to the singular locus
does not drastically change the near-BPS spectrum,
so that a comparison is possible between
the computation of the previous subsection and
the near-BPS spectrum of the symmetric product.

For simplicity, consider first the special case $k'=k''$
(\ie\ $Q_5^+=Q_5^-$),
for which a candidate dual is $Sym^{Q_1Q_5}(\CS_0)$
with $\CS_0$ the U(2) WZW model at level $\kappa=0$.
In the $n^{\rm th}$ twisted sector of the symmetric product
(\ie\ the twisted sector for a single cycle of length $n$
in the symmetric group), the spectrum is
\eqn\twsp{
\eqalign{
  h &= \frac{n-1}4 + \frac{h_{\rm int}}n\cr
  \ell'&=\ell''=\frac{n-1}4\ .
}}
Here $n$ is necessarily odd, $n=2r+1$.

For supergravity long strings \longstr\ with $k'=k''$,
in order to have $\ell'=\ell''$ we must set $w'=w''$ and $j'=j''$.
The simplest way to satisfy
the BPS bound is to set $w=w'+w''=2w'$,
and put $j-1=j'=j''$ to cancel the $SU(2)$ and $SL(2)$ Casimir terms
in \longstr.
Then one finds a spectrum of BPS states
with $\ell'=\ell''$, one for each value of the spin.
The states in the zero-winding sector in $SL(2)$ are BPS supergravity
states, whose spins are bounded by $k'/2$;
once we add the long string sectors, we can get
arbitrary spin.  The states are grouped according
to the $SL(2)$ winding $w$ in blocks of size $k'/2$.
Back-reaction is supposed to lead to the upper cutoff
(due to stringy exclusion) of spin less than of order $k Q_1$,
but in the GKS formalism this restriction is nonperturbative
and therefore invisible.
If we now add U(1) charge, we get a spectrum of BMN type states with
\eqn\equalk{
 h = \hf k'w'+j'  + \frac{{u'}^2}{2w'}
}
with $j'=0,1/2,...,k'/2$ and $w'=1,2,3,...~$.  These states are BPS if $u'=0$,
and near-BPS in the BMN sense if $\ell'$ is large and $u'$ is small.

Let us compare \equalk\ to the spectrum \twsp\
of the symmetric product of the $\CS_0$ theory.
The latter has BMN type states with
\eqn\symbmn{
  h =\frac{n-1}4 + \frac{u^2}{n} \ .
}
Each increment of $SU(2)$ spin is accompanied by an increment
in the winding sector.  The order of the winding is deduced
from the (assumed large) first term on the RHS:
$n=2(k'w'+2j'+2)$.  Identifying $u'=u$,
we see that the second terms differ by a factor $k'$.
If we think in terms of the ``invariant mass'' of the state,
which is highly boosted along the $\S^3$'s, we have
\eqn\invmi{
  m_{\rm inv}^2 \sim (h-\ell)\ell \sim \coeff14 u^2
}
for the symmetric product, and
\eqn\invmii{
  m_{\rm inv}^2 \sim (h-\ell)\ell \sim \coeff14 k'{u'}^2
}
for super(string)gravity.

It is tempting to identify $k'=Q_5^\pm=1$ from this result;
however, the spectra are being compared across a distance
in moduli space proportional to $gcd(Q_5^+,Q_5^-)$, {\it cf.} section 2.
If $Q_5^+ = Q_5^-$ then this distance is order $1$ for values
of $Q_5^\pm$ for which supergravity is valid. In this case the
deviation from the BPS bound might well vary significantly.
If $Q_5^+= Q_5^-=1$ the distance is order $g_B$,  and the deviation
from the BPS bound should be controllable. However, in this case
the supergravity approximation is not valid.
In spite of all these cautionary remarks, we cannot
help noting that the best match is for $Q_5^\pm=1$,
reminiscent of the fact \LarsenUK\ that in the D1-D5 system
on $T^4$, the symmetric orbifold was determined to lie in
the cusp of the moduli space related to $Q_5=1$.


\subsec{Spectra for $k'\ne k''$}

An analysis of the more general backgrounds with $k' \ne k''$
indicates again a discrepancy in the BMN spectrum
between supergravity and the symmetric
product of $U(2)$.

Consider the special case $j'=j''=0$.
Then the special BPS states $\ell'=\ell''$
will have $2\ell'=k'w'=k''w''$.
Consider the further specialization $w'=pk''$, $w''=pk'$.
The BPS condition is satisfied for $j=0$, $w=p(k'+k'')$;
then the energy of long BMN-type strings with these particular
quantum numbers is
\eqn\kpnotkpp{
  h = \frac{kw}{2}+\frac{u^2}{w}
    = \frac{pk'k''}{2}+\frac{u^2}{p(k'+k'')}\ .
}
Let us compare this answer to the symmetric product of
$U(2)$.  The BMN spectrum is easily
determined by the analysis of section 5:
\eqn\utobmn{
  h = j+\coeff12 \hat wk^- +\frac{u^2}{2j+1+\hat w(k^-+1)}\ .
}

We again fix the order of the twisted sector by
comparing the large first terms.  For $j$ small, we determine
\eqn\minvi{
  m_{\rm inv}^2\sim \frac{k_-}{2(k_-+1)}\, u^2
}
for the symmetric product, and
\eqn\minvii{
  m_{\rm inv}^2\sim \frac{k'k''}{2(k'+k'')}\, u^2
}
for super(string)gravity (recall $k'=Q_5^+$, $k''=Q_5^-$).
Again the best match is for one of $Q_5^\pm$
equal to one, but we cannot exclude other possibilities
given the  considerations mentioned at the end of
the previous subsection.


\subsec{Comparison with the PP-wave Limit
of $AdS_3 \times \S^3 \times \S^3 \times \S^1$}

As a check on the near-BPS spectrum derived using the GKS formalism 
above, we reproduce that spectrum by taking the Penrose limit 
of $AdS_3 \times \S^3\times \S^3 \times \S^1$ and analyzing the 
spectrum along the lines of \BMN. 
%
For this purpose, it is convenient to write
the space-time metric \iibmet\ as
\eqn\adsmetric{\eqalign{
ds^2 =
& \ell^2 \left( - \cosh^2 \rho dt^2 + d \rho^2 + \sinh^2 \rho d \phi^2 \right)
+ \cr
& + \sum_{i = \pm} R_{i}^2
\left( d \th_{i}^2 + \cos^2 \th_{i} d \psi_{i}^2
+ \sin^2 \th_{i} d \varphi_{i}^2 \right)
+ L^2 d \th^2
}}

The plane wave limit of the geometry \adsmetric\
is obtained by boosting along a null geodesic
in $AdS_3 \times \S_{+}^3 \times \S_{-}^3 \times \S^1$ \BMN.
Specifically, we consider a limit where some of the radii
in \adsmetric\ are taken to infinity, with $\a'$ and $g_B$ kept fixed.
In the boundary theory, this corresponds to focusing on the sector
of the theory spanned by operators with large values of spin.
There are many choices of boost; one can associate these choices
with a choice of direction inside  $SU(2)_+ \times SU(2)_- \times U(1)$.
Recently, one particular choice of the Penrose limit was
considered in \Sommov, 
but the PP-wave limit considered there does not describe states which are near BPS.
Here, we shall consider another limit,
given by the rescaling, {\it cf.} \refs{\Blauetal,\LuPoritz},
\eqn\bmnlimitx{\eqalign{
t & = \mu_0 x^+ \cr
\psi_{\pm} & = \mu_{\pm} x^+ - {x^- \over 2 \mu_{\pm} R_{\pm}^2}
\pm \left( \mu_- R_- \over \mu_+ R_+ \right)^{\pm 1/2} {y_1 \over R_{\pm}} \cr
\th & = {y_2 \over L} \cr
\rho & = {r \over \ell} \cr
\th_{\pm} & = {y_{\pm} \over R_{\pm}}
}}
where $\mu_0$ and $\mu_{\pm}$ are some parameters.
In CFT, this limit corresponds to $\l^{\pm} \to \infty$.

Substituting \bmnlimitx\ into \adsmetric,
and taking the limit $R \to \infty$, we obtain
\eqn\dsppx{
ds^2 = - 2 dx^+ dx^-
- {1 \over 2} (\mu_0^2 r^2 + \mu_+^2 y_+^2 + \mu_-^2 y_-^2) dx^+ dx^+
+ d \vec r^2 + d \vec y_+^2 + d \vec y_-^2 + d \vec y^2
}
where, in order to cancel the terms of order $R_{\pm}^2$,
we need to take
\eqn\nux{
\mu_0^2 \ell^2 = R_+^2 \mu_+^2 + R_-^2 \mu_-^2
}
where $\ell$ denotes the radius of AdS$_3$ (not to be confused with $SU(2)$ spin).
Notice, that the terms $dx^+ dy_1$ cancel automatically due
to a particular choice of the coefficients in \bmnlimitx.
The last four terms in \dsppx\ describe
the usual flat metric on $\R^8$ written in polar coordinates,
$$
\eqalign{
ds_6^2
& = \left( dr^2 + r^2 d \phi^2 \right)
+ \left( dy_+^2 + y_+^2 d \varphi_+^2 \right)
+ \left( dy_-^2 + y_-^2 d \varphi_-^2 \right)
+ dy_1^2 + dy_2^2 \cr
& = d \vec r^2 + d \vec y_+^2 + d \vec y_-^2 + d \vec y^2
}
$$
Similarly, the following components of the 3-form flux \hthree\
remain non-zero in the pp-wave limit \bmnlimitx:
%
%
\eqn\hppa{H_{+ 12} = 2 \mu_0 \quad , \quad
H_{+ 34} = 2 \mu_+  \quad , \quad H_{+ 56} = 2 \mu_- }

Using \bmnlimitx, we find the relation between charges in the pp-wave
geometry and the charges in the dual CFT:
\eqn\ppchargesx{\eqalign{
p^- & = i \p_{x^+} = i \mu_0 \p_t + i \mu_+ \p_{\psi_+}
= \mu_0 h - \mu_+ \l^+ - \mu_- \l^- \cr
p^+ & = i \p_{x^-} = - {i \over \mu_+ R_+^2} \p_{\psi_+}
- {i \over \mu_- R_-^2} \p_{\psi_-}
= {\l^+ \over \mu_+ R_+^2} + {\l^- \over \mu_- R_-^2}
}}

In the light-cone gauge, $x^+ = p^+ \tau$, the string world-sheet
theory is Gaussian (hence, solvable). The bosonic excitations are
described by the Hamiltonian
\eqn\hlc{2p^- = - p_+ = H_{{\rm l.c.}}
= \sum_{n=-\infty}^{\infty} \sum_{I=1}^8 (a_n^I)^{\dagger}
a_n^I \sqrt{\mu_I^2 + \left( { 4\pi^2 n \over p^+} \right)^2 }
}
where, for different values of the space-time index $I$, we have
$$
\mu_I = \cases{\mu_0 & \cr \mu_{\pm} & \cr 0 & }
$$
Substituting \ppchargesx\ into \hlc, we find that the string spectrum
in the plane wave background \dsppx\ looks like
\eqn\bmnrrspecx{
h - {\mu_+ \over \mu_0} \l^+ - {\mu_- \over \mu_0} \l^-
= \sum_n N_n \sqrt{
\left( {\mu_I \over \mu_0} \right)^2
+ \left( { 4\pi^2 n \over \mu_0 p^+} \right)^2}
+ { 4\pi^2 h_{{\rm int}} \over \mu_0 p^+}
}
in the RR case. Similarly, in the NS frame, we obtain, {\cf} \BMN:
\eqn\bmnnsspecx{
h - {\mu_+ \over \mu_0} \l^+ - {\mu_- \over \mu_0} \l^-
= \sum_n N_n
\left( {\mu_I \over \mu_0}  + {4\pi^2 n \over \mu_0 p^+} \right)
+ { 4\pi^2 h_{{\rm int}} \over \mu_0 p^+}
}

Let us now consider in more detail the symmetric case,
where $R_+ = R_- = \sqrt{2} \ell$.
In this case, the constraint \nux\ implies $\mu_+ = \mu_- = \mu_0 /2$.
The momenta \ppchargesx\ take a simple form
\eqn\ppchgsx{
p^- = \mu_0 \left( h - \half (\l^+ + \l^-) \right)
\quad , \quad p^+ = {2 (\l^+ + \l^-) \over \mu_0 R_+^2}
}
Correspondingly, the string spectrum \bmnrrspecx\ becomes
\eqn\specllrr{
h - \half (\l^+ + \l^-) = \sum_n N_n
\sqrt{1 + \left( { 2\pi^2 R_+^2 n \over \l^+ + \l^-} \right)^2 }
+ { 2\pi^2 R_+^2 h_{{\rm int}} \over \l^+ + \l^-}
}
On the other hand, in the NS-NS case the spectrum \bmnnsspecx\
takes the form
\eqn\specllns{
h - \half (\l^+ + \l^-) = \sum_n N_n
\left(1 +  {2\pi^2 R_+^2 n \over \l^+ + \l^-} \right)
+ { 2\pi^2 R_+^2 h_{{\rm int}} \over \l^+ + \l^-}
}

In order to compare this with the spectrum of the GKS long strings,
we need to write \specllns\ in terms of $k' = Q_5^+ = 4 \pi^2 R_+^2$.
For $\l^+ = \l^- = \ell'$ and $N_n=0$, we obtain
\eqn\speckx{
h = \ell' + {k' h_{{\rm int}} \over 4 \ell'}
}
This agrees with the GKS long string spectrum
\equalk\ in the limit of large $w'$,
and also agrees with the spectrum \symbmn\
of the symmetric product orbifold provided that $k'=1$.

Finally, let us briefly describe interactions in the pp-wave
geometry \dsppx\ when $Q_5^+ = Q_5^- = Q_5$.
Since the transverse string fluctuations are confined in this geometry,
the strings are effectively two-dimensional.
Six transverse directions in \dsppx\ are massive with
a characteristic scale $(\mu_I p^+)^{-1/2}$, whereas the other
two transverse directions have sizes $R_+$ and $L$, respectively.
Therefore, the effective two-dimensional string coupling constant
is given by
\eqn\gtwodpp{
g_2^2 = {g_B^2 (\mu p^+)^3 \over R_+ L} \sim {(\ell')^3 \over N}
}
where $N=Q_1 Q_5$, and in the last equality we expressed
$p^+$, $R_+$, and $L$ in terms of the background charges.
The result \gtwodpp\ has to be compared with the genus-counting
parameter in the pp-wave limit of AdS$_3 \times \S^3 \times K3$.
Since the latter geometry has only four massive transverse
directions, the effective two-dimensional string coupling
in this case scales with the $SU(2)$ spin $\ell'$ as \Gomis:
\eqn\cupscl{
g_2^2 = {g_B^2 (\mu p^+)^2 \over {\rm Vol} (K3)} \sim {(\ell')^2 \over N}
}
It is tempting to speculate that the cubic power in \gtwodpp\
is related to the four-string interaction in
$AdS_3 \times \S^3 \times \S^3 \times \S^1$
suggested by the structure of the twisted sectors
in the dual symmetric product CFT (see section 5.2).



\newsec{The $U(1)\times U(1)$ gauge theory}

The low energy supergravity contains a $U(1)\times U(1)$ gauge
theory with Chern-Simons term. The study of the associated
topological field theory provides further information on the
holographic dual of the theory. Indeed, it leads to our strongest
argument that $\syms$ can only be the holographic dual for $Q_5=1$.

\subsec{Actions}

\def\l{\ell}

In the NS flux picture with IIA on  $AdS_3 \times \S^3 \times \S^3 \times \S^1$
we find a $U(1)\times U(1)$ massive gauge theory  for two U(1) gauge fields
in $AdS_3$ by dimensional reduction of the metric and NS
B-field on the $\S^1$.  The relevant ansatz is 
\eqn\iibmetagain{\eqalign{
  ds^2 &= \frac{\ell^2}{x_2^{~2}}
    \Bigl(-dt^2 + (dx^1)^2 + (dx^2)^2\Bigr) +
    {Q_5^+ \over 4 \pi^2} ds^2(\S^3_+) +{Q_5^-\over 4 \pi^2} ds^2(\S^3_-)
    + L^2 (d\theta+ a)^2 \cr
  H &= \lambda_0 \omega_0 + \lambda_+ \omega_+ +
    \lambda_- \omega_- + db \wedge d\theta
}}
where $a$ and $b$ are gauge fields on $AdS_3$ and $da$ and $db$
have integral periods. 
%
%
%
%
The relevant part of the action for the $H$-flux
is proportional to 
\eqn\hstrh{ H\wedge * H =
  {1\over L } db \wedge (*_{\sst\rm AdS_3} db)\wedge
    \omega_+\wedge \omega_- \wedge d\theta
  - 2L \lambda_0\, db \wedge a \wedge
    \omega_+\wedge \omega_- \wedge d\theta \ .}
%
The
second term in \hstrh\ gives a Chern-Simons term in $AdS_3$.
{}From \iialag, \hstrh\ we get 
\eqn\cstrm{ {16 \pi^5 \over g_A^2} \frac{L R_+^3 R_-^3}{\ell} \int
db \wedge a }
Substituting \lengths\ gives
\eqn\cstrmi{ 2\pi Q_1 \int db \wedge a }
On topologically nontrivial three-manifolds we would define this
term by $2\pi Q_1 \int_{M_4} f_b \wedge f_a $. Since our field
strengths have integer periods, having integral $Q_1$ is precisely
the right topological quantization.

Including the kinetic terms  we have an action of the form
\eqn\abthry{
S_a =
\int {-1\over 2e_A^2} da*da + {-1\over 2e_B^2} db*db + 2\pi Q_1  a db
}
It is very useful to introduce $\mu:= \vert e_B/e_A\vert$
and the linear combinations
\eqn\linecombs{
\eqalign{
A^{(+)} & :={1\over \sqrt{2}}\biggl( \mu^{-1/2} b +  \mu^{1/2}  a\biggr) \cr
A^{(-)} & := {1\over \sqrt{2}}\biggl( \mu^{-1/2}b- \mu^{1/2} a \biggr) \cr}
}
%
%
%
In terms of these fields we may write
\eqn\apamthry{
S_s = \int \biggl[ {-1\over 2 \vert e_A e_B\vert}
d\ap* d \ap + \pi Q_1   \ap d \ap \biggr]
+
\int \biggl[ {-1\over 2 \vert e_A e_B\vert}
d\am* d \am - \pi Q_1   \am d \am \biggr]
}
The equation of motion is:
\eqn\diagn{
\eqalign{
d*d A^{(+)} & =  2\pi Q_1 \vert e_A e_B \vert d A^{(+)} \cr
d*d A^{(-)} & = - 2\pi Q_1 \vert e_A e_B \vert d A^{(+)} \cr}
}
and therefore there  are two  propagating vector fields of
$m^2 = (2\pi Q_1 e_A e_B)^2$.

%
%
{}From straightforward Kaluza-Klein reduction we find   
\eqn\arsii{
e_B^2 = {g_A^2 L \over 8\pi^5 R_+^3 R_-^3 }
}
(Note that it is important to work at $C_0=0$ here.
Otherwise $R_3 = - C_0 H$ and
the term in the action $\sim \int R_3*R_3$
lead to a correction $\sim C_0^2$ to $1/e_B^2$.)
%
%
%
%
%
%
%
For the Kaluza-Klein gauge field we obtain: 
\eqn\arsi{
e_A^2 = {g_A^2 \over 8 \pi^5 R_+^3 R_-^3 L^3}
}
Note that this means that 
\eqn\muparam{
\mu^2 := {e_B^2\over e_A^2 } = L^4
}
The gauge group must be $U(1)\times U(1)$ and not $\IR \times \IR$
because we know there are KK monopoles and H-monopoles.

The gauge fields \linecombs\ are 
\eqn\lindom{
A^\pm = \pm \sqrt{\mu\over 2}  ( a\pm {1 \over L^2 } b)
}
and these are indeed the combinations
which appear in the covariant derivatives
for left- and right-moving supersymmetry transformations.
Moreover, equation \diagn\ above shows
that $A^\pm$ have mass-squared in AdS units:
\eqn\msssq{
m^2 \l^2 = (2\pi Q_1 e_A e_B)^2 \l^2 = 4
}
where we used the fixed-values for the radii.
 There is a nice check on our formulae.
Equation   41 of \LarsenXM\ says that in the AdS/CFT correspondence
a vector field satisfying
\eqn\confwt{
\l *dA = \mp (h+\bar h-1) A
}
corresponds to a primary of dimension $(h,\bar h)$, where
$h-\bar h = \pm 1$.
This is to be compared with equation  \diagn. Using the above values
for $e_A, e_B$, that  equation reads:
\eqn\diagnii{
\ell *dA^\pm = \pm 2 A^\pm
}
So,  the massive scalar mode of $A^+$ is dual to a primary field
of dimension $(1,2)$ and $A^-$ is dual to a primary field of dimension $(2,1)$.
Meanwhile $(1,0)$ and $(0,1)$ primaries, i.e., the currents,
correspond to $dA=0$, i.e.  the flat fields.

{\bf Remark}:
Very similar considerations apply for $AdS_3\times \S^3 \times T^4$.
If we choose the background 
\eqn\tfi{
ds^2 = \l^2 ds^2_{AdS} + R^2 ds^2(\S^3)
+ \sum_{i=1}^4 L_i^2 (d\theta_i + a_i)^2
}
\eqn\tfii{
H = \lambda_0 \omega_0 + \lambda_1 \omega_1
+ \sum_{i=1}^4 db_i d\theta_i
}
then the Einstein equations give $\lambda_0^2 = \lambda_1^2$, and charge
quantization gives
$
2\pi^2 \lambda_1 R^3 = Q_5
$
The Chern-Simons interaction is
\eqn\abextra{
2\pi Q_1 \int \sum db_i \wedge a_i
}
and 
$(e_{a_i}/e_{b_i})^2 = L_i^4$. 
Meanwhile, the RR fields give another set of $4+4$ $U(1)$ gauge
fields $\beta_i, \alpha_i$ with Chern-Simons term
\eqn\rrcstrm{
2\pi Q_5 \int \sum d\beta_i \wedge \alpha_i
}
The formulae for the charges change when we turn on the background but the
Chern-Simons terms are quantized. For the general $U$-duality invariant
formula, valid for all backgrounds, see sec. 7 of \MaldacenaSS.

\subsec{Path integral on the torus}

Imagine doing the path integral on the solid torus  for the theory \abthry\
with hyperbolic metric.
In the topology $D \times \S^1$
let $\rho$ be the radial coordinate on the disk.
Consider the path integral where we just integrate over fields
for $\rho\leq \rho_1$.
The path integral over $a,b$ defines some state
$\Psi(a,b;\rho_1)$ in the Hilbert space
of the massive Chern-Simons theory, as a function of the
gauge fields $a,b$ on the boundary torus at $\rho=\rho_1$.
Now consider the path integral
at $\rho_2> \rho_1$. How is the new state
$\Psi(a,b;\rho_2)$ related to the old state? We view evolution in $\rho$ as
a Euclidean time evolution.
Since the hyperbolic metric is of the form
\eqn\hypermet{
ds^2 \sim  d \rho^2 + {e^{2\rho}\over 4} \vert d \phi + \tau d t\vert^2
}
for large $\rho$, and since the Hamiltonian is conformally invariant  (for the
flat gauge fields) we find that
$$
\Psi(a,b;\rho_2) = e^{- (\rho_2-\rho_1) H } \Psi(a,b;\rho_1)
$$
Thus, if we let $\rho_2 \to \infty$ the wavefunction
$\Psi(a,b;\rho_2) $ is projected onto the lowest
energy level of the Hamiltonian.  It is therefore a linear combination
of the gauge invariant wavefunctions for quantization on the torus.

In the companion paper \massivecs\ we work through
the exercise of implementing the above
procedure in detail for the theory \abthry.
The result is that the gauge invariant wavefunctions may be understood
in terms of two   Gaussian models
with radius 
\eqn\leftright{
R_A^2 = {1 \over 4\pi^2} Q_1\mu = {1 \over 4\pi^2} Q_1 L^2
}
and 
\eqn\leftright{
R_B^2 = {Q_1 \over 4\pi^2 \mu} = {Q_1 \over 4\pi^2 L^2}
}

The partition function can be
written  in terms of ``higher level Siegel-Narain theta functions.''
It takes the form:
\eqn\asymwvfn{
\sum_{\beta \in \Lambda^*/\Lambda} \Psi_\beta(A) \Psi_{\bar \beta}(\lambda)
}
where
\def\zb{{\bar z}}
\eqn\asymvfw{
\Psi_\beta(A) = \sqrt{1\over Q_1} {1\over \vert \eta(\tau)\vert^2}
e^{-2\pi Q_1 \imt [ A_z^+ A_\zb^+
+ A_z^- A_\zb^- + \ap_z \am_\zb - \am_z \ap_\zb ]}
\Theta_{\Lambda}(\tau,0,\beta;P;\xi(A))
}
\eqn\asymvfw{
\Psi_{\bar\beta}(\lambda) = \sqrt{2\tau_2\over Q_1}
 \Theta_{\Lambda}(\tau,0,\bar\beta;P;\xi(\lambda))
}
Here
\eqn\lamlat{
\Lambda = e_1 \IZ + f_1 \IZ \cong \sqrt{Q_1} II^{11}
}
 is a lattice with hyperbolic
metric: $e_1^2=f_1^2 =0$, $e_1\cdot f_1 = Q_1$.
$\Theta_\Lambda$ is a Siegel-Narain theta function for
the embedding $P$ of $\Lambda\otimes \IR$
into $\IR^{1,1}$ defined as usual by
left- and right-moving momenta $(p_L;p_R)$,
with metric $p_L^2- p_R^2$. Also,
$\beta = \rho/Q_1 e_1 - \tl \rho/Q_1 f_1$ and $
\bar \beta = \rho/Q_1 e_1 + \tl \rho/Q_1 f_1 $
are representatives of the dual
quotient group $ \Lambda^*/\Lambda\cong (\IZ/Q_1 \IZ)^2$ while
\eqn\xises{
\eqalign{
\xi(A) & = (\sqrt{Q_1} 2i \tau_2 \am_\zb ; - \sqrt{Q_1} 2i \tau_2 \ap_z) \cr
\xi(\lambda) & = ( - {\bar \lambda \over 2\pi i \sqrt{Q_1}};
{ \lambda \over 2\pi i \sqrt{Q_1}}) \cr}
}
Here $\lambda,\bar \lambda$ are arbitrary constants
that depend, for example, on
what kind of operators have been inserted in the solid torus.
See \massivecs\ for further details.

The ``conformal blocks'' $\Psi_\beta(A)$ in
\asymvfw\ predict conformal weights
that give an explicit realization to the
``level $Q_1$ $U(1)$ current algebra'' in
the sense of \refs{\KutasovXU,\LarsenUK}.

\subsec{Comparison to $I_2(\syms)$}

The path integral for the gauge fields $A,B$ that we have discussed is only
part of the bulk superstring  path integral dual  to - say -
the index $I_2$ of the boundary CFT. First, the gauge fields couple to the
charged supergravity modes, and hence
perturbative string interactions should be
taken into account.   One might naively think that
since there   are no couplings between
the $SU(2)^4$ gauge fields and the   $U(1)\times U(1)$ then
\asymwvfn\ would have to be an overall factor in the partition function.
This is not true when one takes into account instanton effects such as
NS5-brane instantons and KK monopole instantons. These effects can lead to a
correlation between the $U(1)$ and $SU(2)$ quantum numbers of the spectrum
computed from the supergravity viewpoint.

Nevertheless, since the topological theory is expected to dominate
at long distance it is very natural to   conjecture that
the $\ap,\am$-dependent wavefunctions are valid in the full AdS/CFT
duality of string theory.
That is, if $Z^{ab}$ is the partition function on the solid torus
and is written as in \asymvfw\ as
a linear combination of some finite dimensional space
of ``conformal blocks'':
\eqn\zabfrct{
Z^{ab} = \sum_\beta  \zeta^\beta \Psi_\beta (A)
}
where $\zeta^\beta$ are  constants (the $\lambda$-dependent terms in
\asymvfw)
then the full string theory partition function is of the form:
\eqn\zabfrcti{
Z^{string} = \sum_\beta  Z^\beta(\Phi_\infty ) \Psi_\beta (A)
}
where $\Phi_{\infty}$ are the boundary values
of the other fields in the supergravity theory.
That is, the exact $\ap,\am$ dependence is given
by a linear combination of the same
``conformal blocks'' as in the massive gauge theory.

If we accept the above conjecture, then we can compare to
the proposed  holographic dual $Z({\rm Sym}^N({\cal S}))$.
Let us consider the index $I_2$,
for simplicity. Then from \itwoindx\ we can deduce that the dependence on
$\ap,\am$ - defined to be the coordinates $(\chi_L;\chi_R)$ dual
to the charges $u,\tilde u$ -  is given by
higher level Siegel-Narain theta functions for
$\Lambda_{cft} = \sqrt{N} II^{1,1}$.
On the other hand, the wavefunctions appearing in \zabfrcti\ are
Siegel-Narain theta functions for $\Lambda_{sg}= \sqrt{Q_1}II^{1,1}$.
Comparing with the ``conformal blocks'' \asymvfw\ of the theory \abthry\
  suggests that we must identify   $N=Q_1$.
Since $N= Q_1 Q_5$, this conjecture supports
the idea that the orbifold  theory ${\rm Sym}^N({\cal S})$ is
only on the moduli space of the supergravity theory for  $Q_5=1$.

Let us comment on possible subtleties that could invalidate the conclusion
that the partition functions can only match for $Q_5=1$.
First, it is possible that one loop determinants associated with
charged fermions on $AdS_3$ induce a renormalization of the
Chern-Simons \cstrmi. We think this unlikely, but it bears further
thought.
Second, it is
possible to change the level of a theta function by summing over
certain vectors $\beta\in \Lambda^*/\Lambda$. In this way, one can
express a theta function of level $k$ in terms of a theta function of
level $k\Delta^2$, where $\Delta$ is an integer. In our present example
we would require
\eqn\includelatt{
\sqrt{Q_1}II^{1,1} \subset {1\over \sqrt{Q_1Q_5} } II^{1,1}
}
This, in turn, is true iff  $Q_5$ is a perfect square.
Thus, when $Q_5$ is not a perfect square, we cannot evade the conclusion.
It is conceivable that some unknown physical mechanism changes the
basic periodicity of the large gauge transformations of the $a,b$ fields
to be multiplied by $Q_5$. In this case, the supergravity partition function
would be expressed in terms of level $Q_1 Q_5$ theta functions. However,
we cannot see any justification for this. Thus, we conclude that $Q_5=1$ is
necessary to match to the simplest proposal for the holographic dual $\syms$.

It is worth stressing that the above argument does {\it not} apply to the 
case of $AdS_3 \times \S^3 \times T^4$. Here the enlarged $U$-duality group 
allows one to redefine a basis of gauge fields so that \abextra\ and \rrcstrm\ are 
rearranged into level $1$ and level $Q_1Q_5$ Chern-Simons theories. This redefinition, 
of course, depends on the  cusp in moduli space. 


\vskip 2cm \noindent{\bf Acknowledgments:}
We would like to thank J. de Boer, R. Dijkgraaf,
J. Maldacena, B. Mazur, H. Ooguri and K. Skenderis
for useful conversations.
This work was conducted during the period S.G. served as a Clay
Mathematics Institute Long-Term Prize Fellow.  G.M. thanks the 
LPTHE theory group at Jussieu, and the KITP at Santa Barbara 
for hospitality during the course of part of this work. The work of E.M. is
supported in part by DOE grant DE-FG02-90ER40560, that of G.M. by
DOE grant DE-FG02-96ER40949 and that of A.S. by DE-FG02-90ER40654.


\appendix{A}{Proof of marginality}

As mentioned in section 4.4,
although the candidate modulus operator
preserves large $\CN=4$ supersymmetry,
it cannot be written as a superspace integral,
and so the proof of \DixonBG\ that this
operator is truly a modulus must be reconsidered.
Since we will be using the methodology of \DixonBG\
in an essential way, and this article may
not be readily available to the reader,
let us reproduce (more or less verbatim) the proof there
of marginality for $\CN=2$ massless perturbations.
We will then adapt this proof to our modified circumstances,
and show that the candidate modulus $\CT$
preserves conformal invariance to all orders
in conformal perturbation theory.

\subsec{Dixon's proof for $\CN=2$}

The reasoning of \DixonBG\ runs as follows.  Consider an $\CN=2$
theory with an $h=\ell=\hf$ chiral primary field with lower
component $\Phi_0^+$ and upper component $\Phi_1^+=G_\shf^-\Phi_0^+$,
and and antichiral field $\Phi_{0}^-$ 
with upper component$\Phi_1^-=G_\shf^+\Phi_0^-$.
The $k^{\rm th}$ term in conformal perturbation theory
involves a correlation function
\eqn\corfn{
  \Bigl\langle \prod_{i=1}^k \CT(z_i) \Bigr\rangle
}
of the modulus deformation $\CT=\Phi_1^++\Phi_1^-$,
integrated over $k-3$ arguments.

Consider the term with $m$ operators $\Phi_1^+$ and
$n$ operators $\Phi_1^-$, $m+n=k$.
We suppress the right-moving structure except as needed.
Embed this CFT correlator in a string scattering amplitude%
\foot{One could worry that this restricts the CFT
to have $\hat c=\frac23 c\le 9$,
but at least in tree level string amplitudes
one can admit larger $\hat c$ together with a compensating
timelike linear dilaton.  The tree level scattering amplitudes
are unlikely to exhibit any pathology.}
\eqn\scattamp{\eqalign{
  \int\prod_{i,j}d^2 z_i d^2 w_j
    \Bigl(\prod_i\bigl(\Phi_1^+(z_i)+ik_i\cdot\psi_i\Phi_0^+(z_i)\bigr)
        e^{ik_i\cdot X(z_i)} \cr
    \prod_j\bigl(\Phi_1^-(w_j)+ik_j\cdot\psi_j\Phi_0^-(w_j)\bigr)
        e^{ik_j\cdot X(w_j)} \Bigr)\quad .
}}
The leading term as $k_i\to 0$ comes from taking $\Phi_1^\pm$
in each factor, leading to the correlation function
\eqn\Fterm{
  F(z_i,w_j) = \Bigl\langle
    \prod_{i=1}^m \Phi_1^+(z_i) \;
    \prod_{j=1}^n \Phi_1^-(w_j) \Bigr\rangle
}
where, despite the slightly confusing notation,
$\Phi_1^\pm$ carry zero $R$-charge so the correlators
are non-vanishing even when $m\ne n$.  It will turn out
that $F$ is a total derivative; integration by parts
brings down factors of $k_i\cdot k_j$ from the correlator
of the exponentials, and one can choose the kinematics
such that the surface terms vanish in the integration by parts
\GreenQU.
Replacing pairs of $\Phi_1^\pm$ by
pairs of $ik_i\cdot\psi\Phi_0^\pm$ also leads to terms
with at least two powers of momenta.

The scattering amplitude can develop poles $\frac{1}{k_i\cdot k_j}$
from on-shell intermediate states.  This would lead to
contact terms at zero momentum and a non-vanishing effective
potential for the candidate modulus, as the pole cancels
the quadratic vanishing of the numerator.
But fortunately $F$ also picks up a total derivative
in $(\bar z_i,\bar w_j)$ from its right-moving
superstructure.  The amplitude behaves as $k^4/k^2\to 0$
as the momenta are uniformly scaled to zero,
and no effective potential is generated for $\CN=(2,2)$ supersymmetry.

To complete the proof, one must show that $F$ is indeed
a total derivative with three of the coordinates
$(z_i,w_j)$ held fixed.  Consider the expression
for the upper component
\eqn\upp{
  \Phi_1^\pm(z_1) = \oint_{z_1} dz\; 2G^\mp(z)\Phi_0^\pm(z_1)
    = \frac1{z_1}\oint_{z_1} dz\;z\;2G^\mp(z)\Phi_0^\pm(z_1)
}
and deform the integration contour so that it
surrounds the other vertices in the correlator
(these pick out the only two modes of $G(z)$ that are regular
at $z=0,\infty$).
The relevant operator products are
\eqn\opeso{
\eqalign{
G^{+}(z) \Phi_0^+(w) & \sim 0 \cr
G^{+}(z) \Phi_1^+(w) & \sim 2 {\p \over \p w}( {1\over z-w} \Phi_0^+(w))  \cr
G^{-}(z) \Phi_0^+(w) & \sim {1\over z-w} \Phi_1^+(w) \cr
G^{-}(z) \Phi_1^+(w) & \sim 0 \ ,}
}
and similarly for $\Phi^-_{0,1}$.
One finds
\eqn\twoexpr{\eqalign{
  F(z_i,w_j) &= -\sum_{r=1}^n\partial_{w_r} F_r(z_i,w_j) \cr
  z_1 F(z_i,w_j) &= -\sum_{r=1}^n\partial_{w_r}[w_r F_r(z_i,w_j)]\ ,
}}
where
\eqn\Fr{
  F_r \equiv \vev{\Phi_0^+(z_1)\Phi_1^+(z_2)\cdots\Phi_1^+(z_m)
    \Phi_1^-(w_1)\cdots\Phi_0^-(w_r)\cdots\Phi_1^-(w_n)}\ .
}
If $m\ge 3$, we can fix three of the $z_i$ and then $F$
is a total derivative with respect to the $w_j$, which are integrated.
Similarly for $n\ge 3$.  Thus one need only examine
the cases $(m,n)=(2,1)$, $(1,2)$, and $(2,2)$.
Without loss of generality we can assume $m=2$
(otherwise just interchange the roles of chiral
and antichiral in the following).  Use $SL(2,C)$ invariance
to fix $z_1$, $z_2$, and $w_1$.  Multiply the first of \twoexpr\
by $w_1$ and subtract from the second to obtain
\eqn\fone{
  F(z_i,w_j) = \frac{1}{w_1-z_1} F_1(z_i,w_j)
    +\frac{1}{w_1-z_1} \partial_{w_2}[(w_2-w_1)F_1]
}
(we have assumed the most complicated case $n=2$; if
$n=1$, replace $\Phi_{0,1}^-(w_2)$ by the identity operator).
The problem boils down to showing that $F_1$ is a total derivative.
Apply the expressions \upp\ to $\Phi_1^+(z_2)$
and deform contours to obtain
\eqn\ftwo{\eqalign{
  F_1(z_i,w_j) &= - H - \partial_{w_2}H_2 \cr
  z_2F_1(z_i,w_j) &= -z_1 H - \partial_{w_2}[w_2 H_2]\ ,
}}
where
\eqn\Hr{\eqalign{
  H(z_i,w_j) &= \vev{\Phi_1^+(z_1)\Phi_0^+(z_2)\Phi_0^-(w_1)\Phi_1^-(w_2)}\cr
  H_2(z_i,w_j) &= \vev{\Phi_0^+(z_1)\Phi_0^+(z_2)\Phi_0^-(w_1)\Phi_0^-(w_2)}\ .
}}
Now eliminate $H$ from \ftwo\ to obtain
\eqn\fthree{
  F_1(z_i,w_j) = \frac{1}{z_1-z_2}\; \partial_{w_2}[(w_2-z_1)H_2]\ ,
}
a total derivative with respect to the integrated variable $w_2$
if $n=2$, or vanishing if $n=1$.  Thus the effective
potential for $\CN=2$ massless chiral fields vanishes,
and they are moduli.

Note that the key here is the last OPE in \opeso,
which says that the contour deformation of $G^{-}$ does not
act on any $\Phi_1^+$.  Then in \Fr, there are no
derivatives with respect to any of the $z_i$, in
particular the unintegrated ones; if there were,
the expression could not be reduced further and
we could not show that the correlator is a total
derivative with respect to integrated variables.
This is for instance why the argument does not
apply to $\CN=1$ supersymmetry, where all the
operator insertions are on the same footing.


\subsec{Application to large $\CN=4$}

The above proof relied essentially on the properties \opeso\
of chiral superderivatives.
The modulus appeared in the combination
$\CT = g_+\Phi_1^++g_-\Phi_1^-$
(where $g_-=g_+^*$),
and the terms with $m$ chiral operators
$\Phi_1^+=G_{\shf}^-\Phi_0^+$ and $n$ antichiral operators
$\Phi_1^-=G_{\shf}^+\Phi_0^-$ were analyzed separately.

For large $\CN=4$, the modulus deformation has the form
\eqn\modop{  \CT = G_{\shf}^{\beta\dot\beta}
    \bar G_{\shf}^{\bar\alpha\dot{\bar\alpha}}
    \Phi_{\beta\dot\beta;\bar\alpha\dot{\bar\alpha}} \ ;
}
expanding in components,
one has two canonical $\CN=2$
substructures
\eqn\substr{
  \CT= \CT_1+\CT_2\ ,
}
where (again suppressing the anti-holomorphic structure)
\eqn\tonetwo{\eqalign{
  \CT_1 &= \Bigl(G^\ppd\Phi^\mmd+G^\mmd\Phi^\ppd\Bigr)\cr
  \CT_2 &= -\Bigl(G^\pmd\Phi^\mpd+G^\mpd\Phi^\pmd\Bigr) \ ;
}}
the problem is that $\Phi^\pmmpd$ are neither chiral
nor anti-chiral under the $\CN=2$ generated by $G^\pmpmd$,
and similarly $\Phi^\pmmpd$ are nonchiral under the
$\CN=2$ generated by $G^\mppmd$;
so we cannot directly apply Dixon's argument.
Fortunately, we will be able to find an appropriate modification.

The correlation function \corfn\ of conformal perturbation
theory can be broken apart into contributions
\eqn\contribs{
  \Bigl\langle \prod_{i=1}^m \CT_1(z_i)
    \prod_{j=1}^n\CT_2(w_j) \Bigr\rangle\ .
}
Embedding the problem again in string theory via
an expression of the sort in \scattamp,
the problem again boils down to showing that all
of these contributions
are total derivatives with respect to the integrated
variables.

We now claim that there is a rearrangement lemma, namely:
\eqn\newward{
  \Bigl\langle
    \prod_{j=1}^n\CT_2(w_j)\prod_{i=1}^m \CT_1(z_i) \Bigr\rangle\  =
\Bigl\langle  \CT_1(w_1)
    \prod_{j=2}^n\CT_2(w_j)\prod_{i=1}^m \CT_1(z_i) \Bigr\rangle\  }
This Ward identity allows us to reduce the large $\CN=4$ moduli problem
to the $\CN=2$ subalgebra, and then we can invoke Dixon's theorem.

In order to prove \newward\ we will need the following OPE's:
\eqn\opeemm{
\eqalign{
G^\ppd(z) \CT_1(w) & = G^\ppd(z)\Bigl(G^\mmd\Phi^\ppd\Bigr)=
- {\p \over \p w} \Bigl( {1\over z-w} \Phi^{\ppd}(w)\Bigr) \cr
G^\ppd(z) \CT_2(w) & =
- {\p \over \p w} \Bigl( {1\over z-w} \Phi^{\ppd}(w)\Bigr) \cr
G^\mmd(z) \CT_1(w) & = G^\ppd(z)\Bigl(G^\ppd\Phi^\mmd\Bigr)=
- {\p \over \p w} \Bigl( {1\over z-w} \Phi^{\mmd}(w)\Bigr) \cr
G^\mmd(z) \CT_2(w) & =
- {\p \over \p w} \Bigl( {1\over z-w} \Phi^{\mmd}(w)\Bigr) \cr}
}
as well as
\eqn\sutwowi{
\eqalign{
A^{+,\mdmd}(z) \CT_1(w) & \sim 0 \cr
A^{+,\mdmd}(z) \CT_2(w) & \sim 0 \cr
A^{-,\mm}(z) \CT_1(w) & \sim 0 \cr
A^{-,\mm}(z) \CT_2(w) & \sim 0 \cr}
}
The equations \sutwowi\ are proved using the identities \shortnullvec.

Now we write:
\eqn\newwardpf{
\eqalign{
&   \Bigl\langle
    \prod_{j=1}^n\CT_2(w_j)\prod_{i=1}^m \CT_1(z_i) \Bigr\rangle\   \cr
& = - \Bigl\langle \bigl(\oint_{w_1} G^{\pmd}\Phi^{\mpd} +
    \oint_{w_1} G^{\mpd}\Phi^{\pmd} \bigr)
    \prod_{j=2}^n\CT_2(w_j)\prod_{i=1}^m \CT_1(z_i) \Bigr\rangle\  \cr
& = - \sum_{r=2}^n {\p\over \p w_r}
\Biggl\{ \Bigl\langle \Phi^{\mpd}(w_1) \Phi^{\pmd}(w_r)
{\prod}''\CT_2 \prod \CT_1 \Bigr\rangle +
\Bigl\langle \Phi^{\pmd}(w_1) \Phi^{\mpd}(w_r)
{\prod}''\CT_2 \prod \CT_1 \Bigr\rangle \Biggr\} \cr
& \hskip 1cm
- \sum_{i=1}^m {\p\over \p z_i}
    \Biggl\{ \Bigl\langle \Phi^{\mpd}(w_1) \Phi^{\pmd}(z_i)
{\prod}'\CT_2 {\prod}' \CT_1 \Bigr\rangle +
    \Bigl\langle \Phi^{\pmd}(w_1) \Phi^{\mpd}(z_i)
{\prod}'\CT_2 {\prod}' \CT_1 \Bigr\rangle \Biggr\} \cr
& = + \sum_{r=2}^n {\p\over \p w_r}
    \Biggl\{ \Bigl\langle \Phi^{\mmd}(w_1) \Phi^{\ppd}(w_r)
{\prod}''\CT_2 \prod \CT_1 \Bigr\rangle +
    \Bigl\langle \Phi^{\ppd}(w_1) \Phi^{\mmd}(w_r)
{\prod}''\CT_2 \prod \CT_1 \Bigr\rangle \Biggr\} \cr
&\hskip1cm + \sum_{i=1}^m {\p\over \p z_i}
    \Biggl\{ \Bigl\langle \Phi^{\mmd}(w_1) \Phi^{\ppd}(z_i)
{\prod}'\CT_2 {\prod}' \CT_1 \Bigr\rangle +
    \Bigl\langle \Phi^{\ppd}(w_1) \Phi^{\mmd}(z_i)
{\prod}'\CT_2 {\prod}' \CT_1 \Bigr\rangle \Biggr\} \cr
& = \Bigl\langle  \CT_1(w_1)
    \prod_{j=2}^n\CT_2(w_j)\prod_{i=1}^m \CT_1(z_i) \Bigr\rangle\ \cr }
}
The primes on the products indicate that the appropriate factor is deleted
from the product. In the first equality,
we have written the definition of $\CT_1, \CT_2$.
In the second we have deformed contour integrals
of $G$ and used \opeemm. In the third
we have used the Ward identity following from contour deformation
of integrals of $A^{\pm, \mm}$ and made use of \sutwowi.
Finally, in last equality, we have used again
a Ward identity following from deformation of contour integrals of $G$.
Thus we can systematically reduce the correlators \contribs\
to correlators of only $\CT_1$.

Now let us separate the product over $\CT_1(z_i)$ in \contribs\
(with $n=0$ now) into its separate contributions from
$m_+$ operators $G^\mmd\Phi^\ppd(z_i)$, and
$m_-$ operators $G^\ppd\Phi^\mmd(z'_i)$.
Since $G^\mmd$ in the OPE \opeemm\ treats
$G^\ppd\Phi^\mmd(z'_i)$ in the same way
that the antichiral operators $\Phi_1^-(w)$
behaved in Dixon's analysis, we are done;
we can simply apply the same analysis with
$G^\mmd\Phi^\ppd$ playing the role of $\Phi_1^+$ and
$G^\ppd\Phi^\mmd$ playing the role of $\Phi_1^-$.


\appendix{B}{$\CN=4$ algebra in bispinor notation}

Spinor conventions: $\epsilon^{+-}= \epsilon_{-+}=1$ raises/lowers spinor
indices from the left. The adjoint of $su(2)$ is a bispinor according to:
\eqn\bisp{
\eqalign{
x^{A\dot B} & = x^j \sigma_j^{A\dot B} \cr
x^{++} & = x^1 + i x^2 \cr
x^{+-} & = - x^3 \cr
x^{--} & = - (x^1-i x^2) \cr}
}
In four dimensions:

\eqn\gmma{
\eqalign{
\sigma^\mu_{A\dot B} & \equiv (i , \vec \sigma) \cr
\bar \sigma^{\mu\dot A B} & \equiv  (i , -\vec \sigma) \cr
v^{\dot B A} = v^{A \dot B} & \equiv
-{1 \over  2} (\bar \sigma^\mu)^{\dot B A} v_\mu\cr
v_\mu & = (\sigma_\mu)_{A\dot B} v^{A \dot B} \cr}
}

\def\db{\dot B}
\def\dc{\dot C}
\def\dd{\dot D}

In terms of $\gamma = k^-/k$, $1-\gamma = k^+/k$ we have
\eqn\spinlarge{
\eqalign{
\{ G_m^{A \db}, G_n^{C \dd}\} & = - \half \epsilon^{\db \dd} \epsilon^{AC}
\biggl[ 2L_{n+m} + {c\over 3} \delta_{n+m,0} (m^2-{1\over 4}) \biggr] \cr
& + i(n-m) \biggl[ -\gamma A^{+,\db \dd}_{n+m} \epsilon^{AC} + (1-\gamma)
A^{-,AC}_{n+m} \epsilon^{\db \dd} \biggr] \cr
[A_m^{+,j}, G_n^{\db A} ] & = {i\over 2} (\sigma^j)^{\db}_{~~\dc}
\biggl( G^{\dc A}_{n+m} - 2(1-\gamma) m Q^{\dc A}_{n+m} \biggr) \cr
[A_m^{-,j}, G_n^{\db A} ] & = -{i\over 2}
\biggl( G^{\db C}_{n+m} +2 \gamma  m Q^{\db C}_{n+m} \biggr)
(\sigma^j)_{C}^{~~A}  \cr}
}

where
\eqn\gees{
G^{A \dot B} = G^{\dot B A} = \half \pmatrix{ G^3 - i G^4 & G^1 - i G^2 \cr
G^1 + i G^2 & - G^3 - i G^4 \cr}
}

\def\ppd{{+\dot +}}
\def\mmd{{-\dot -}}
\def\pmd{{+\dot -}}
\def\mpd{{-\dot +}}
\def\pmpmd{{\pm\dot \pm}}

\def\pmmpd{{\pm\dot \mp}}
\def\mppmd{{\mp\dot \pm}}
\def\mdmd{\dot - \dot -}

\def\mm{--}

\listrefs
\end